\documentclass[fleqn,usenatbib]{mnras}

\usepackage{newtxtext,newtxmath}

\usepackage[T1]{fontenc}

\DeclareRobustCommand{\VAN}[3]{#2}
\let\VANthebibliography\thebibliography
\def\thebibliography{\DeclareRobustCommand{\VAN}[3]{##3}\VANthebibliography}

\usepackage{graphicx}	
\usepackage{amsmath}	
\usepackage{orcidlink}  
\usepackage{siunitx}
\usepackage{booktabs,makecell} 
\usepackage{multirow}   
\usepackage{rotating} 

\newcommand{\cav}{[Ca\,\textsc{v}]}      
\newcommand{\arx}{[Ar\,\textsc{x}]}      
\newcommand{\sxii}{[S\,\textsc{xii}]}    
\newcommand{\fev}{[Fe\,\textsc{v}]}      
\newcommand{\fevi}{[Fe\,\textsc{vi}]}    
\newcommand{\fevii}{[Fe\,\textsc{vii}]}  
\newcommand{\fex}{[Fe\,\textsc{x}]}      
\newcommand{\fexi}{[Fe\,\textsc{xi}]}    
\newcommand{\fexiv}{[Fe\,\textsc{xiv}]}  
\newcommand{\nev}{[Ne\,\textsc{v}]}      
\newcommand{\arxiv}{[Ar\,\textsc{xiv}]}  

\newcommand{\oiii}{[O\,\textsc{iii}]}    
\newcommand{\oii}{[O\,\textsc{ii}]}      
\newcommand{\nii}{[N\,\textsc{ii}]}      
\newcommand{\sii}{[S\,\textsc{ii}]}      
\newcommand{\neiii}{[Ne\,\textsc{iii}]}  

\newcommand{\ha}{H\,$\upalpha$}          
\newcommand{\hb}{H\,$\upbeta$}           
\newcommand{\hei}{He\,\textsc{i}}        
\newcommand{\heii}{He\,\textsc{ii}}      

\newcommand{\mosfit}{\textsc{mosf}{\tiny i}\textsc{t}} 
\newcommand{\tdemass}{\textsc{tdem}{\tiny ass}}        
\newcommand{\bagpipes}{\textsc{bagpipes}}      

\newcommand{\wise}{\textit{WISE}} 

\title[Mapping the nuclear environments of ECLEs]{Mapping the nuclear environments of extreme coronal line emitting galaxies}

\author[D. Kynoch et al.]{
Daniel Kynoch$^{\orcidlink{0000-0001-8638-3687}}$,$^{1}$\thanks{E-mail: danielkynoch@gmail.com}
Or Graur$^{\orcidlink{0000-0002-4391-6137}}$,$^{1,2}$\thanks{E-mail: or.graur@port.ac.uk}
Peter Clark$^{\orcidlink{0000-0002-6576-7400}}$,$^{3}$
J.N.~Aguilar,$^{4}$
S.~Ahlen$^{\orcidlink{0000-0001-6098-7247}}$,$^{5}$
D.~Bianchi$^{\orcidlink{0000-0001-9712-0006}}$,$^{6,7}$
D.~Brooks,$^{8}$
\newauthor
T.~Claybaugh,$^{4}$
A.~de la Macorra$^{\orcidlink{0000-0002-1769-1640}}$,$^{9}$
P.~Doel,$^{8}$
J.E.~Forero-Romero$^{\orcidlink{0000-0002-2890-3725}}$,$^{10,11}$
S.~{Gontcho A Gontcho}$^{\orcidlink{0000-0003-3142-233X}}$,$^{12}$
\newauthor
G.~Gutierrez,$^{13}$
R.~Joyce$^{\orcidlink{0000-0003-0201-5241}}$,$^{14}$
S.~Juneau$^{\orcidlink{0000-0002-0000-2394}}$,$^{14}$
M.~Landriau$^{\orcidlink{0000-0003-1838-8528}}$,$^{4}$
L.~Le Guillou$^{\orcidlink{0000-0001-7178-8868}}$,$^{15}$
A.~Meisner$^{\orcidlink{0000-0002-1125-7384}}$,$^{14}$
\newauthor
R.~Miquel,$^{16,17}$
J.~Moustakas$^{\orcidlink{0000-0002-2733-4559}}$,$^{18}$
S.~Panda$^{\orcidlink{0000-0002-5854-7426}}$,$^{19}$\thanks{Gemini Science Fellow}
W.~J.~Percival$^{\orcidlink{0000-0002-0644-5727}}$,$^{20,21,22}$
F.~Prada$^{\orcidlink{0000-0001-7145-8674}}$,$^{23}$
I.~P\'erez-R\`afols$^{\orcidlink{0000-0001-6979-0125}}$,$^{24}$
\newauthor
G.~Rossi,$^{25}$
E.~Sanchez$^{\orcidlink{0000-0002-9646-8198}}$,$^{26}$
D.~Schlegel,$^{4}$
J.~Silber$^{\orcidlink{0000-0002-3461-0320}}$,$^{4}$
D.~Sprayberry,$^{14}$
G.~Tarl\'{e}$^{\orcidlink{0000-0003-1704-0781}}$,$^{27}$
B.~A.~Weaver,$^{14}$
\newauthor
R.~Zhou$^{\orcidlink{0000-0001-5381-4372}}$,$^{4}$
and H.~Zou$^{\orcidlink{0000-0002-6684-3997}}~^{28}$
\\
$^{1}$Institute of Cosmology and Gravitation, University of Portsmouth, Dennis Sciama Building, Burnaby Road, Portsmouth, PO1 3FX, UK \\
$^{2}$Department of Astrophysics, American Museum of Natural History, New York, NY 10024, USA \\
$^{3}$School of Physics and Astronomy, University of Southampton, Southampton, SO17 1BJ, UK \\
$^{4}$Lawrence Berkeley National Laboratory, 1 Cyclotron Road, Berkeley, CA 94720, USA \\
$^{5}$Department of Physics, Boston University, 590 Commonwealth Avenue, Boston, MA 02215 USA \\
$^{6}$Dipartimento di Fisica ``Aldo Pontremoli'', Universit\`a degli Studi di Milano, Via Celoria 16, I-20133 Milano, Italy \\
$^{7}$INAF-Osservatorio Astronomico di Brera, Via Brera 28, 20122 Milano, Italy \\
$^{8}$Department of Physics \& Astronomy, University College London, Gower Street, London, WC1E 6BT, UK \\
$^{9}$Instituto de F\'{\i}sica, Universidad Nacional Aut\'{o}noma de M\'{e}xico, Circuito de la Investigaci\'{o}n Cient\'{\i}fica, Ciudad Universitaria, Cd. de M\'{e}xico C.~P.~04510, M\'{e}xico \\
$^{10}$Departamento de F\'isica, Universidad de los Andes, Cra. 1 No. 18A-10, Edificio Ip, CP 111711, Bogot\'a, Colombia \\
$^{11}$Observatorio Astron\'omico, Universidad de los Andes, Cra. 1 No. 18A-10, Edificio H, CP 111711 Bogot\'a, Colombia \\
$^{12}$University of Virginia, Department of Astronomy, Charlottesville, VA 22904, USA \\
$^{13}$Fermi National Accelerator Laboratory, PO Box 500, Batavia, IL 60510, USA \\
$^{14}$NSF NOIRLab, 950 N. Cherry Ave., Tucson, AZ 85719, USA \\
$^{15}$Sorbonne Universit\'{e}, CNRS/IN2P3, Laboratoire de Physique Nucl\'{e}aire et de Hautes Energies (LPNHE), FR-75005 Paris, France \\
$^{16}$Instituci\'{o} Catalana de Recerca i Estudis Avan\c{c}ats, Passeig de Llu\'{\i}s Companys, 23, 08010 Barcelona, Spain \\
$^{17}$Institut de F\'{i}sica d’Altes Energies (IFAE), The Barcelona Institute of Science and Technology, Edifici Cn, Campus UAB, 08193, Bellaterra (Barcelona), Spain \\
$^{18}$Department of Physics and Astronomy, Siena University, 515 Loudon Road, Loudonville, NY 12211, USA \\
$^{19}$International Gemini Observatory/NSF NOIRLab, Casilla 603, La Serena, Chile \\
$^{20}$Department of Physics and Astronomy, University of Waterloo, 200 University Ave W, Waterloo, ON N2L 3G1, Canada \\
$^{21}$Perimeter Institute for Theoretical Physics, 31 Caroline St. North, Waterloo, ON N2L 2Y5, Canada \\
$^{22}$Waterloo Centre for Astrophysics, University of Waterloo, 200 University Ave W, Waterloo, ON N2L 3G1, Canada \\
$^{23}$Instituto de Astrof\'{i}sica de Andaluc\'{i}a (CSIC), Glorieta de la Astronom\'{i}a, s/n, E-18008 Granada, Spain \\
$^{24}$Departament de F\'isica, EEBE, Universitat Polit\`ecnica de Catalunya, c/Eduard Maristany 10, 08930 Barcelona, Spain \\
$^{25}$Department of Physics and Astronomy, Sejong University, 209 Neungdong-ro, Gwangjin-gu, Seoul 05006, Republic of Korea \\
$^{26}$CIEMAT, Avenida Complutense 40, E-28040 Madrid, Spain \\
$^{27}$University of Michigan, 500 S. State Street, Ann Arbor, MI 48109, USA \\
$^{28}$National Astronomical Observatories, Chinese Academy of Sciences, A20 Datun Road, Chaoyang District, Beijing, 100101, P.~R.~China \\
}

\date{Accepted XXX. Received YYY; in original form ZZZ}

\pubyear{\the\year{}}

\begin{document}
\label{firstpage}
\pagerange{\pageref{firstpage}--\pageref{lastpage}}
\maketitle

\clearpage
\begin{abstract}
Extreme coronal line emitters (ECLEs) are a rare class of galactic nuclei exhibiting unusually strong high-ionisation forbidden emission lines, and several ECLEs have been linked to tidal disruption events (TDEs). 
In this work, we compile and analyse optical spectra of 33 ECLEs, dividing them into variable, TDE-linked sources and non-variable, AGN-linked systems. 
Using multi-epoch spectroscopy from the Sloan Digital Sky Survey, Dark Energy Spectroscopic Instrument, and other facilities, we investigate the evolution of the emission line spectra and measure emission line profiles. Many variable ECLEs have changing spectra in which the highest-ionisation lines (e.g., \fex--\fexiv) appear and fade first, followed by \fevii, accompanied by brightening of \oiii. 
These changes may reflect a softening ionising continuum, the outward propagation of the ionisation front following the TDE flare, or both. Assuming virial motion, we translate line widths into characteristic radial distances, reconstructing the spatial distribution of line-emitting gas. 
Coronal lines are generally emitted at radii intermediate between the broad line region and the low-ionisation narrow line region. 
This ionisation stratification is seen in many sources, with similar incidence in variable and non-variable ECLEs, suggesting no apparent difference in circumnuclear gas distributions between active and quiescent nuclei. 
We find positive correlations between gas distance and black hole mass for both \oiii\ and \fevii: the log(Distance)–log(Mass) relations have slopes $0.63\pm0.08$ and $0.69\pm0.12$, respectively, broadly consistent with a Mass$^{0.5}$ dependence and with characteristic radii set primarily by photoionisation.
\end{abstract}

\begin{keywords}
transients -- transients: tidal disruption events -- galaxies: active -- galaxies: nuclei
\end{keywords}



\section{Introduction}\label{sec:intro}
`Coronal' emission lines, arising from forbidden transitions of highly-ionised species with ionisation potentials typically exceeding 100 eV, are rare features in the optical spectra of galactic nuclei (\citealt{Rodriguez-Ardila06}; \citealt{MullerSanchez11}; \citealt{Rose15}; \citealt{Reefe22,Reefe23};  \citealt{Rodriguez-Ardila25}). 
The presence of these lines requires exposure of gas to an intense extreme-ultraviolet or soft X-ray radiation field, such as that produced by accretion onto a supermassive black hole (SMBH).
Coronal lines are therefore commonly associated with active galactic nuclei (AGNs: e.g., \citealt{Tadhunter08}; \citealt{Netzer08}; \citealt{Padovani17}). 
As these transitions arise from ions with both high ionisation potentials and high critical densities, they provide valuable probes of dense circumnuclear gas that is directly influenced by the central engine.
\textcolor{black}{Coronal lines in AGNs trace photoionised gas spanning a wide range of scales.
Some of the emission has been shown to arise at an intermediate scale between the broad emission line region (BLR) and the narrow line region (NLR), possibly in an accretion-driven outflow (e.g., \citealt{Penston84}; \citealt{Erkens97}; \citealt{Gelbord09}; \citealt{Kynoch22}).
Further, spatially resolved studies have shown that coronal line emission can extend to hundreds of parsecs or even kiloparsec scales (e.g., \citealt{Prieto05}; \citealt{Mazzalay10}; \citealt{MullerSanchez11}; \citealt{Negus21}; \citealt{Rodriguez-Ardila25}).}
Despite their diagnostic potential, optical coronal lines are typically weak even in AGNs (\citealt{Cerqueira-Campos21}; \citealt{McKaig24}), which limits their usefulness.

A small number of galactic nuclei exhibit unusually strong coronal emission lines, forming the class known as extreme coronal line emitters (ECLEs). 
In these objects, lines from species such as \nev\ and \fevii\ have strengths comparable to traditional low-ionisation ($\mathrm{IP}\lesssim100$~eV) narrow emission lines such as \oiii~$\lambda5007$. 
The first example, SDSS\,J0952$+$2143, was reported by \cite{Komossa08}, who noted exceptionally strong iron coronal lines including \fevii~$\lambda3758$, 5720, and 6087, \fex~$\lambda6374$, and \fexiv~$\lambda5302$. 
Subsequent searches of large spectroscopic survey data have uncovered additional ECLEs (e.g., \citealt{Wang12}; \citealt{Callow24,Callow25}; \citealt{Clark26}), although the overall population remains small. 
The unusual strength of the coronal lines in these sources suggests that large quantities of circumnuclear gas are exposed to powerful ionising radiation fields.
\textcolor{black}{A related class of objects, CLiF (Coronal Line Forest) AGNs, characterised by a rich spectrum of high-ionisation forbidden lines in AGNs, was independently identified by \cite{Rose15}. 
These sources are proposed to be `ordinary' AGNs for which we have a preferential viewing angle onto the coronal line emission region on the inner edge of the dust torus.
The preliminary definition of CLiF AGNs included a detailed set of observational criteria applicable to AGN spectra (\citealt{Rose15}), whereas ECLEs are commonly identified by the single criterion that one coronal line has a flux $\geqslant$20~per cent that of \oiii~$\lambda5007$ (a criterion which was relaxed in recent ECLE surveys; e.g., \citealt{Clark26}).
At least one object (SDSS\,J1241) has been identified as both an ECLE (by \citealt{Wang12}) and a CLiF AGN (by \citealt{Rose15}), although a systematic comparison of the two classes has not been performed.}

\cite{Komossa08}, \cite{Wang11,Wang12}, \cite{Yang13}, \cite{Clark24,Clark26}, and \cite{Callow24,Callow25}, among others, have argued that the subset of ECLEs in which the coronal lines fade over time are caused by tidal disruption events (TDEs: first theorised by \citealt{Hills75} and \citealt{Rees88}; for a recent review see \citealt{Gezari21}).    
During a TDE, a star passing within the tidal radius of the SMBH is disrupted and partially accreted, producing a luminous accretion flare that radiates strongly in the ultraviolet and soft X-ray bands. 
This transient radiation field can illuminate pre-existing circumnuclear gas, producing strong and evolving emission line spectra. 
Optical and ultraviolet observations of TDEs commonly reveal broad hydrogen and helium recombination lines (\citealt{Arcavi14}; \citealt{Leloudas19}), but in some cases high-ionisation forbidden lines also appear, forming the class of coronal-line TDEs (CL-TDEs: e.g., \citealt{Short23}; \citealt{Hinkle24a}).
\cite{Hinkle24a} argue that coronal lines appear in the spectra of TDEs that occur in gas-rich galactic nuclei.
In several events these lines evolve dramatically over time, with the highest-ionisation transitions appearing first and fading before lower-ionisation species.

Because the ionising flare is transient, TDE-associated ECLEs offer a unique opportunity to probe the structure of circumnuclear gas in formerly dark galactic nuclei. 
As the radiation from the flare propagates outward and declines in luminosity, different regions of gas may become visible at different times. 
The widths and profiles of emission lines provide information on the kinematics of the emitting material and therefore its characteristic distance from the SMBH under the assumption of virial motion. 
In this way, emission line spectroscopy of ECLEs can be used to construct approximate `maps' of the circumnuclear environment on sub-parsec scales.

A small number of recent studies have applied this approach to individual ECLEs. 
\cite{Short23} analysed the evolving emission line spectrum of TDE\,2019qiz, while \cite{Newsome24} performed a similar analysis for TDE\,2022upj. 
\cite{Clark25} investigated the ECLE AT\,2018dyk, finding evidence for stratified emission line regions spanning sub-parsec scales. 
In these systems the coronal lines appear to originate at radii intermediate between the BLR and the more extended NLR, similar to AGNs. 
\cite{Clark25} suggested that differences in the inferred gas distances between TDE\,2022upj and AT\,2018dyk been reflected differences in their black hole (BH) masses. 
However, it remains unclear whether such behaviour is typical of ECLEs in general, or whether the circumnuclear gas distributions differ between transient TDE-driven systems and persistent AGNs.

In this work we assemble optical spectra for a sample of $\approx30$ ECLEs drawn from the literature. 
We divide the sample into variable ECLEs (v-ECLEs), likely associated with TDEs, and non-variable ECLEs (nv-ECLEs), likely powered by AGN activity. 
Using emission line profile measurements from multi-epoch spectroscopy obtained by the Sloan Digital Sky Survey (SDSS: \citealt{York00}; \citealt{Eisenstein11}; \citealt{Blanton17}; \citealt{SDSS-DR19}; \citealt{Kollmeier26}), the Dark Energy Spectroscopic Instrument (DESI: \citealt{DESI-I,DESI-II} and \citealt{DESI-KP1-Inst}), and other facilities, we investigate the evolution of the emission line spectra and use the observed line widths to infer characteristic distances of the emitting gas. 
This allows us to explore the spatial distribution of circumnuclear material across the ECLE population and to test whether these distributions depend on BH mass or on the transient or persistent nature of the ionising source.

Most of the galaxies in this work \textcolor{black}{(although not all of them)} have recently been observed as part of the 8-yr DESI survey being conducted with the 4-m Mayall Telescope at Kitt Peak National Observatory, Arizona, which began operations in 2021 (\citealt{Schlafly23}).
The science focus of DESI is primarily on cosmology: cosmological constraints from full-shape modelling of clustering measurements were presented by \cite{DESI-VII}.
Its data releases include the publicly-available Early Data Release (EDR: \citealt{DESI-EDR}) and Data Release 1 (DR1: \citealt{DESI-DR1}), and the forthcoming Data Release 2 (DR2).
Early results from DESI DR2 include baryon acoustic oscillation measurements from the Ly\,$\upalpha$ forest and from galaxies and quasars, respectively (\citealt{DESI-DR2-results-I,DESI-DR2-results-II}).
Further technical details can be found in the following papers: \cite{Miller23} describes the Optical Corrector while \cite{Poppett24} describes the DESI Fiber System.
We make use of optical spectra produced by the DESI spectroscopic pipeline, which is described by \cite{Guy23}.

This paper is organised as follows.
In Section~\ref{sec:sample} we describe the sample and source classifications.
In Section~\ref{sec:methods} we explain the methods used to obtain the emission line measurements and BH mass estimates that lead to the calculations of the line-emitting gas radii.
The emission line measurements of the sources are presented in Section~\ref{sec:spectro}, in which we also describe their spectroscopic evolution over time and look at trends in emission line properties across the sample.
We discuss our findings and present plans for future work in Section~\ref{sec:disc} and we summarise our conclusions in Section~\ref{sec:conc}.
In this work we have assumed a flat $\Lambda$CDM cosmology with $H_0=70$~km\,s$^{-1}$\,Mpc$^{-1}$, $\Omega_\mathrm{m}=0.3$ and $\Omega_\Lambda=0.7$.

\section{The sample}\label{sec:sample}
\begin{figure*}
    \centering
    \includegraphics[width=\linewidth]{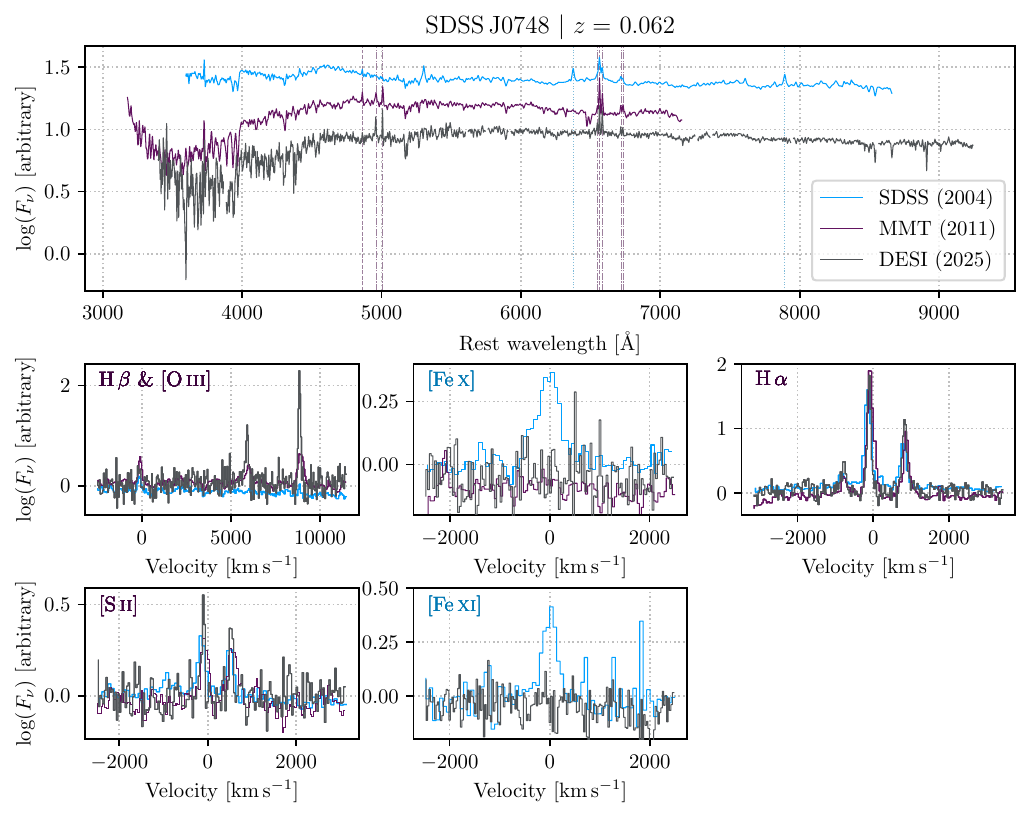}
    \caption{Optical spectra of SDSS\,J0748.  The upper panel shows the full spectra plotted as a function of rest-frame wavelength.
    Key emission lines are indicated with vertical lines: \ha~$\lambda6565$ and \hb~$\lambda4863$ (\textcolor{black}{dashed purple lines}); \oiii~$\lambda\lambda4960, 5007$, \nii~$\lambda\lambda6549, 6585$, and \sii~$\lambda\lambda6718, 6732$ (\textcolor{black}{dot-dashed purple lines}); and \fex~$\lambda6374$ and \fexi~$\lambda7891$ (\textcolor{black}{dotted cyan lines}).
    The lower panels show cut-outs of key emission lines in velocity space.}
    \label{fig:sdssj0748_spectra}
\end{figure*}

We have compiled from the literature a sample of reported optical ECLEs,
which we have divided into v-ECLEs and nv-ECLEs.
In this work, `variable' is meant in the context established by \cite{Yang13} and \cite{Clark24}, referring to ECLEs that display one or more of the following characteristics: 
the appearance and/or fading of coronal lines and low-ionisation forbidden lines on timescales of months to decades (see e.g., Fig.~\ref{fig:sdssj0748_spectra}); and 
substantial ($\gtrsim0.5$~mag), typically monotonically decreasing, mid-infrared (MIR) brightness changes over several years, accompanied by MIR colours evolving away from AGN-like values (\citealt{Clark24,Clark26}).
MIR colours are quantified using the W1$-$W2 colour index from the Wide-field Infrared Survey Explorer \citep[\wise;][]{Wright10}, where W1 and W2 refer to the 3.4 and \SI{4.6}{\micro\metre} bands, respectively.
Following \cite{Stern12}, galaxies with W1$-$W2\,$\geq$\,0.8~mag are considered to display AGN-like MIR colours, whilst those with W1$-$W2\,$<$\,0.8~mag are consistent with quiescent or star-forming host galaxies.
v-ECLEs typically show an initial W1$-$W2 colour at or above this threshold, then declining below it as the source fades, 
whereas nv-ECLEs maintain AGN-like colours throughout (\citealt{Clark24,Clark26}).
The changes observed in v-ECLEs are consistent with a transient event such as a TDE, in contrast to the lower-amplitude, stochastic variability typical of AGNs.
We note that some newly identified v-ECLEs have not yet displayed coronal line fading, but are classified on the basis of their MIR properties alone.
For the variable sources we make the further distinction between (a) v-ECLEs that were discovered spectroscopically by their strong coronal line emission and were later associated with TDEs; 
and (b) coronal-line TDEs (CL-TDEs), which were discovered as transients in wide-field photometric surveys and for which subsequent spectroscopic observations revealed the presence of strong coronal lines.

\begin{figure*}
    \centering
    \includegraphics[width=\linewidth]{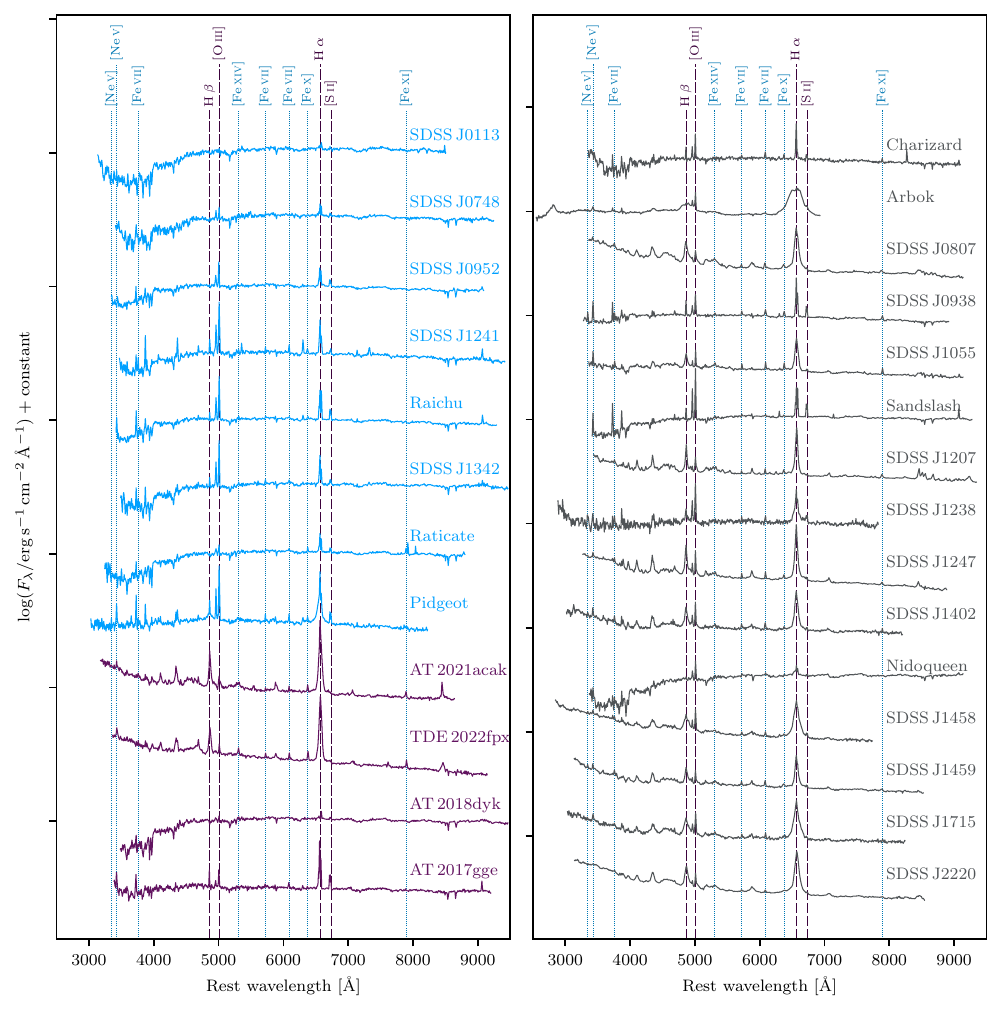}
    \caption{DESI spectra of the ECLEs in our sample.
    The left-hand panel shows the spectra of the v-ECLEs (cyan) and the CL-TDEs (purple);
    the right-hand plot shows the spectra of the nv-ECLEs (grey).
    The spectra have been smoothed for clarity.
    The locations of key emission lines are indicated by vertical dotted and dashed lines for the coronal and low-ionisation lines, respectively.}
    \label{fig:desi_spectra}
\end{figure*}

An overview of the sample is given below in Table~\ref{tab:objects}.
\textcolor{black}{In Appendix~\ref{sec:appendix} we describe the discovery, observations, and classification of each source in turn.}
For each source we have obtained available optical spectra from public archives and other private data from direct correspondence with the authors of previous studies.
We have limited our selection to spectra with sufficient spectral resolution ($R\gtrsim1000$) and signal-to-noise (S/N) ratios to enable robust measurements of the relatively weak and narrow spectral features of interest.
We have additionally included recent spectra from DESI, where available; these are shown in Fig.~\ref{fig:desi_spectra}.
Example optical spectra of SDSS\,J0748 are shown in Fig.~\ref{fig:sdssj0748_spectra} in which we highlight the evolution of key emission lines.
The spectroscopic observations are summarised in Tables~\ref{tab:spectra_1} and \ref{tab:spectra_2}.

\begin{table*}
    \centering
    \caption{Summary of the ECLEs in this study}
    \begin{tabular}{llcccl}
    \hline
         Host galaxy or              &  Short name & RA           & Dec          & $z$    & Selected  \\
         transient name              &             & J2000 [deg.] & J2000 [deg.] &        & reference(s) \\
    \hline
         \multicolumn{6}{c}{Variable sources: v-ECLEs} \\
         SDSS\,J011306.68$+$093712.2 & SDSS\,J0113 &  18.2779     &  $+$9.6201   & 0.1537 & \cite{Callow25} \\
         SDSS\,J074820.67$+$471214.3 & SDSS\,J0748 & 117.0861     & $+$47.2040   & 0.0616 & \cite{Wang11}; \cite{Clark24} \\
         SDSS\,J095209.56$+$214313.3 & SDSS\,J0952 & 148.0399     & $+$21.7204   & 0.0795 & \cite{Komossa08,Komossa09}; \cite{Clark24} \\
         SDSS\,J124134.26$+$442639.2 & SDSS\,J1241 & 190.3927     & $+$44.4442   & 0.0419 & \cite{Wang12}; \cite{Clark24} \\
         DESI\,J204.3990$+$03.8775   & Raichu      & 204.3990     &  $+$3.8776   & 0.0566 & \cite{Clark26} \\
         SDSS\,J134244.42$+$053056.1 & SDSS\,J1342 & 205.6851     &  $+$5.5160   & 0.0366 & \cite{Wang12}; \cite{Clark24} \\
         SDSS\,J135001.49$+$291609.7 & SDSS\,J1350 & 207.5062     & $+$29.2694   & 0.0777 & \cite{Wang12}; \cite{Clark24} \\
         DESI\,J216.5891$+$00.1436   & Raticate    & 216.5892     &  $+$0.1437   & 0.1149 & \cite{Clark26} \\
         DESI\,J243.3798$+$55.7528   & Pidgeot     & 243.3799     & $+$55.7528   & 0.1931 & \cite{Clark26} \\
         \hline
         \multicolumn{6}{c}{Variable sources: CL-TDEs} \\
         TDE\,2022upj                 &             &   5.987      & $-$14.4233   & 0.054  & \cite{Fremling22}; \cite{Newsome24} \\
         AT\,2018gn                 &             &  26.6766     & $+$32.5076   & 0.0375 & \cite{Wang24} \\
         AT\,2021dms                 &             &  50.3503     & $-$11.1460   & 0.0311 & \cite{Forster21}; \cite{Hinkle24a} \\
         TDE\,2019qiz                 &             &  71.6579     & $-$10.2264   & 0.0151 & \cite{Nicholl20}; \cite{Hung21}; \cite{Short23} \\
         AT\,2021acak                &             & 158.7000     & $+$15.4896   & 0.136  & \cite{Li23} \\ 
         TDE\,2022fpx                 &             & 232.7654     & $+$53.4054   & 0.073  & \cite{Perez-Fournon22}; \cite{Koljonen24} \\
         AT\,2018dyk                 &             & 233.2834     & $+$44.5356   & 0.0367 & \cite{Fremling18}; \cite{Clark25} \\
         SDSS\,J162034.99$+$240726.5 & AT\,2017gge & 245.1458     & $+$24.1240   & 0.0665 & \cite{Onori22}; \cite{Hinkle24a} \\
         TDE\,2021qth                 &             & 302.9122     & $-$21.1602   & 0.0805 & \cite{Yao23} \\
         AT\,2018bcb*                &             & 340.9286     & $-$16.9856   & 0.1192 & \cite{Neustadt20}; \cite{Hinkle24a} \\
         \hline
        \multicolumn{6}{c}{Non-variable sources: nv-ECLEs} \\
        DESI\,J027.9525$+$04.1951   & Charizard   &  27.95254 & $+$4.19518 & 0.079   & \cite{Clark26} \\
        DESI\,J115.5730$+$39.4366   & Arbok       & 115.57303 & $+$39.43668 & 0.416   & \cite{Clark26} \\
        SDSS\,J080727.31$+$140537.0 & SDSS\,J0807 & 121.86382 & $+$14.09364 & 0.0739 & \cite{Callow24} \\ 
        SDSS\,J093801.64$+$135317.0 & SDSS\,J0938 & 144.50682 &  $+$13.88807 & 0.1006 & \cite{Wang12} \\
        SDSS\,J105526.43$+$563713.3 & SDSS\,J1055 & 163.86007 & $+$56.62031 & 0.07396 & \cite{Wang12} \\
        DESI\,J172.6675$+$50.6179   & Sandslash   & 172.66753 & $+$50.61793 & 0.058   & \cite{Clark26} \\
        SDSS\,J120719.81$+$241155.8 & SDSS\,J1207 & 181.83254 & $+$24.19886 & 0.05027 & \cite{Callow24} \\
        SDSS\,J123829.58$+$185237.5 & SDSS\,J1238 & 189.62329 & $+$18.8771 & 0.25328 & \cite{Callow24} \\
        SDSS\,J124726.37$+$070525.0 & SDSS\,J1247 & 191.85988 & $+$7.0903 & 0.1043 & \cite{Callow24} \\
        SDSS\,J140204.75$+$293946.8 & SDSS\,J1402 & 210.51982 & $+$29.66302 & 0.19637 & \cite{Callow24} \\
        DESI\,J212.9556$+$52.7676   & Nidoqueen   & 212.95561 & $+$52.76762 & 0.074   & \cite{Clark26} \\
        SDSS\,J145849.72$+$191033.5 & SDSS\,J1458 & 224.70719 & $+$19.17598 & 0.26813 & \cite{Callow24} \\
        SDSS\,J145926.06$+$404538.5 & SDSS\,J1459 & 224.85862 & $+$40.76071 & 0.15109 & \cite{Callow24} \\
        SDSS\,J171504.28$+$564715.8 & SDSS\,J1715 & 258.76787 & $+$56.78773 & 0.19051 & \cite{Callow24} \\
        SDSS\,J222055.73$-$075317.8 & SDSS\,J2220 & 335.23221 & $-$7.88829 & 0.14862 & \cite{Callow24} \\
    \hline
    \end{tabular}
    \parbox{17cm}{\footnotesize
    \textit{Notes:}
    *We have no useful spectra of AT\,2018bcb, so it is not included in subsequent analyses.}
    \label{tab:objects}
\end{table*}

\begin{table*}
    \centering
    \caption{Summary of the optical spectra of variable ECLEs analysed in this study}
    \begin{tabular}{lcllllc}
    \hline
         Short name   & $E(B-V)$ & Observatory/ & Obs.\ date & MJD & Pre/Post &  $\mathrm{FWHM_{inst}}$ \\
                      & [mag]    & Spectrograph                         & [UTC]      &     &          & [km\,s$^{-1}$] \\
    \hline
    \multicolumn{7}{c}{Variable sources: v-ECLEs} \\
         SDSS\,J0113  & 0.0470   & SDSS/BOSS        & 2012-09-19 & 56189 & Post &  72    \\
                      &          & Gemini/GMOS      & 2023-12-24 & 60302 & Post &   230 (blue), 140 (red)    \\
                      &          & Mayall/DESI      & 2025-01-21 & 60687 & Post &   35    \\
         SDSS\,J0748  & 0.0588   & SDSS/SDSS        & 2004-02-20 & 53055 & Post &  72   \\
                      &          & MMT/BCS     & 2011-12-26 & 55921 & Post &  210 \\    
                      &          & Mayall/DESI      & 2021-12-05 & 59918 & Post &  35 \\                      
         SDSS\,J0952  & 0.0236   & SDSS/SDSS        & 2005-12-30 & 53734 & Post &  72 \\
                      &          & MMT/BCS     & 2011-12-26 & 55921 & Post & 210 \\
                      &          & Mayall/DESI      & 2021-11-18 & 59536 & Post & 35 \\
         SDSS\,J1241  & 0.0144   & SDSS/SDSS        & 2004-02-27 & 53062 & Post &  72 \\
                      &          & MMT/BCS     & 2011-12-26 & 55921 & Post &  210 \\
                      &          & Mayall/DESI      & 2025-03-04 & 60738 & Post &  35 \\
         Raichu       & 0.0222   & Mayall/DESI      & 2021-04-02 & 59306 & Post &  35 \\
         SDSS\,J1342  & 0.0218   & SDSS/SDSS        & 2002-04-09 & 52373 & Post &  72 \\
                      &          & MMT/BCS     & 2011-12-26 & 55921 & Post &  210 \\
                      &          & Mayall/DESI      & 2021-03-06 & 59279 & Post &  35 \\                    
         SDSS\,J1350  & 0.0131   & SDSS/SDSS        & 2006-03-23 & 53848 & Post &  72 \\
                      &          & MMT/BCS     & 2011-12-26 & 55921 & Post &  210 \\
         Raticate     & 0.0279   & Mayall/DESI      & 2021-05-13 & 59347 & Post &  35 \\              
         Pidgeot      & 0.0072   & Mayall/DESI      & 2021-04-06 & 59310 & Post &  35 \\
    \hline
    \multicolumn{7}{c}{Variable sources: CL-TDEs} \\
         TDE\,2022upj  & 0.0257   & SOAR/GHTS        & 2022-11-29 & 59912 & Post & 150 \\  
                      &          & SOAR/GHTS        & 2022-12-04 & 59917 & Post &  150 \\
                      &          & SOAR/GHTS        & 2024-01-04 & 60313 & Post &   150 \\
                      &          & SOAR/GHTS        & 2024-01-08 & 60317 & Post &  150 \\
         AT\,2018gn  & 0.0354   & FLWO/FAST        & 2018-01-15 & 58133 & Post &  540 \\
                      &          & FLWO/FAST        & 2018-02-11 & 58160 & Post &  540 \\
                      &          & Palomar/DBSP     & 2022-06-29 & 59759 & Post &  300 \\
                      &          & Palomar/DBSP     & 2022-09-02 & 59824 & Post &  300 \\
                      &          & Palomar/DBSP     & 2023-09-09 & 60196 & Post &  300 \\
         AT\,2021dms  & 0.0676   & Magellan/LDSS    & 2021-08-13 & 59439 & Post &  350 \\
         TDE\,2019qiz  & 0.0939   & Keck/LRIS        & 2019-09-25 & 58751 & Post &   300   \\
                      &          & VLT/X-shooter    & 2019-09-28 & 58754 & Post &  35 (VIS)   \\
                      &          & VLT/X-shooter    & 2019-10-10 & 58766 & Post &  35 (VIS)   \\
                      &          & VLT/X-shooter    & 2019-11-13 & 58800 & Post &  35 (VIS)   \\
                      &          & VLT/X-shooter    & 2019-12-13 & 58830 & Post &   35 (VIS)   \\
                      &          & VLT/X-shooter    & 2020-12-15 & 59198 & Post &   35 (VIS)   \\
                      &          & VLT/X-shooter    & 2022-01-26 & 59605 & Post &    35 (VIS)   \\
         AT\,2021acak & 0.0320   & Lijiang/YFOSC      & 2022-04-23 & 59692 & Post &  170 \\ 
                      &          & Lijiang/YFOSC      & 2022-04-23 & 59692 & Post &  170 \\ 
                      &          & Lijiang/YFOSC       & 2022-04-24 & 59693 & Post &  170 \\ 
                      &          & Lijiang/YFOSC       & 2022-04-25 & 59694 & Post &  170 \\ 
                      &          & Lijiang/YFOSC       & 2022-06-03 & 59733 & Post &  170 \\ 
                      &          & Mayall/DESI      & 2023-04-15 & 60049 & Post &  35 \\
         TDE\,2022fpx  & 0.0144   & Mayall/DESI      & 2023-05-29 & 60093 & Post &  35 \\
         AT\,2018dyk  & 0.0164   & SDSS/SDSS        & 2002-07-11 & 52466 & Pre  &  72   \\
                      &          & SDSS/MaNGA       & 2017-01-06 & 57759 & Pre  &  150   \\ 
                      &          & Keck/LRIS        & 2018-08-08 & 58338 & Post &  63   \\ 
                      &          & Mayall/DESI      & 2023-09-05 & 60192 & Post &  35   \\ 
         AT\,2017gge  & 0.0606   & SDSS/SDSS        & 2004-09-09 & 53226 & Pre  &  72 \\
                      &          & Gemini/GMOS      & 2018-09-15 & 58376 & Post &  178 \\ 
                      &          & Gemini/GMOS      & 2019-02-26 & 58540 & Post &  178 \\ 
                      &          & VLT/X-shooter    & 2022-03-28 & 59666 & Post &  35 (VIS) \\
                      &          & Mayall/DESI      & 2022-05-13 & 59712 & Post &  35 \\
         TDE\,2021qth  & 0.0924   & Palomar/DBSP     & 2021-08-04 & 59430 & Post &  250 (blue), 307 (red)   \\
                      &          & Keck/LRIS        & 2022-05-26 & 59725 & Post &  407 (blue), 297 (red)   \\ 
    \hline
    \end{tabular}
    \parbox{12.5cm}{\footnotesize
    \textit{Notes:} 
    $E(B-V)$ gives the colour excess value used to correct the spectra for Galactic reddening. 
    Pre/Post indicates whether the spectrum was recorded before or after the transient outburst.
    Instrumental spectral broadening (expressed as $\mathrm{FWHM_{inst}}$ in km\,s$^{-1}$) is either taken from the literature or derived from the reported spectral resolution, $R$, assuming $\Delta v = c/R$.}
    \label{tab:spectra_1}
\end{table*}

\begin{table*}
    \centering
    \caption{Summary of the optical spectra of non-variable ECLEs analysed in this study}
    \begin{tabular}{lclllc}
    \hline
         Short name   & $E(B-V)$ & Observatory/ & Obs.\ date & MJD  &  $\mathrm{FWHM_{inst}}$ \\
                      & [mag]    & Spectrograph                         & [UTC]      &      &  [km\,s$^{-1}$] \\
    \hline
    \multicolumn{6}{c}{Non-variable sources: nv-ECLEs} \\
         Charizard    & 0.0319   & Mayall/DESI      & 2021-02-06 & 59251 & 35 \\
                      &          & Gemini/GMOS  & 2021-05-12 & 59436 & 270 (blue), 140 (red) \\ 
         Arbok        & 0.0484   & SDSS/SDSS    & 2010-10-11 & 55480 &  72 \\
                      &          & Mayall/DESI  & 2021-03-09 & 59282 &  35 \\
         SDSS\,J0807  & 0.0327   & SDSS/SDSS    & 2005-11-08 & 53682 & 72 \\
                      &          & Mayall/DESI  & 2024-12-13 & 60657 & 35 \\
         SDSS\,J0938  & 0.0383   & SDSS/SDSS    & 2006-12-23 & 54092 & 72 \\
                      &          & MMT/BCS & 2011-12-26 & 55921 & 210 \\
                      &          & Mayall/DESI  & 2023-01-24 & 59968 & 35 \\
         SDSS\,J1055  & 0.0075   & SDSS/SDSS    & 2002-04-09 & 52373 & 72 \\
                      &          & MMT/BCS & 2011-12-26 & 55921 & 210 \\
                      &          & Mayall/DESI  & 2024-02-19 & 60359 & 35 \\
         Sandslash    & 0.0104   & SDSS/SDSS    & 2002-04-01 & 52365 & 72 \\
                      &          & Mayall/DESI  & 2021-03-24 & 59297 & 35 \\
         SDSS\,J1207  & 0.0224   & SDSS/SDSS    & 2008-01-19 & 54484 & 72 \\
                      &          & Mayall/DESI  & 2025-02-08 & 60714 & 35 \\
         SDSS\,J1238  & 0.0228   & SDSS/SDSS    & 2008-01-16 & 54481 & 72 \\
                      &          & Mayall/DESI  & 2023-12-28 & 60306 & 35 \\
         SDSS\,J1247  & 0.0228   & SDSS/SDSS    & 2006-05-21 & 53876 & 72 \\
                      &          & Mayall/DESI  & 2023-03-17 & 60020 & 35 \\
         SDSS\,J1402  & 0.0109   & SDSS/SDSS    & 2007-03-19 & 54178 & 72 \\
                      &          & Mayall/DESI  & 2023-04-24 & 60058 & 35 \\
         Nidoqueen    & 0.0084   & SDSS/SDSS    & 2002-07-13 & 52468 & 72 \\
                      &          & Mayall/DESI  & 2021-05-05 & 59339 & 35 \\
         SDSS\,J1458  & 0.0294   & SDSS/SDSS    & 2008-03-14 & 54539 & 72 \\
                      &          & Mayall/DESI  & 2021-06-13 & 59378 & 35 \\
         SDSS\,J1459  & 0.0122   & SDSS/SDSS    & 2004-05-21 & 53146 & 72 \\
                      &          & Mayall/DESI  & 2022-04-14 & 59683 & 35 \\
         SDSS\,J1715  & 0.0189   & SDSS/SDSS    & 2000-09-01 & 51788 & 72 \\
                      &          & Mayall/DESI  & 2025-04-22 & 60787 & 35 \\
         SDSS\,J2220  & 0.0610   & SDSS/SDSS    & 2001-10-21 & 52203 & 72 \\
                      &          & Mayall/DESI  & 2022-09-24 & 59846 & 35 \\
    \hline
    \end{tabular}
    \vspace{2mm}
    \parbox{10.5cm}{\footnotesize
    \textit{Notes:}
    $E(B-V)$ gives the colour excess value used to correct the spectra for Galactic reddening. 
    Instrumental spectral broadening (expressed as $\mathrm{FWHM_{inst}}$ in km\,s$^{-1}$) is either taken from the literature or derived from the reported spectral resolution, $R$, assuming $\Delta v = c/R$.}
    \label{tab:spectra_2}
\end{table*}

\section{Methods}\label{sec:methods}
\subsection{Emission line measurements}
Using the extinction curve of \cite{CCM89} we first correct each spectrum for Galactic reddening using the $E(B-V)$ colour excess given in Tables~\ref{tab:spectra_1} and \ref{tab:spectra_2}, taken from \cite{SF11}, then transform the spectrum to the host galaxy rest frame using the redshift quoted in Table~\ref{tab:objects}. 
For each spectrum we then estimate and subtract the global continuum by fitting a cubic spline to the fluxes in several emission line free continuum windows spanning the optical band.
Each emission line of interest is then fit in velocity space within a window wide enough to measure its profile above the neighbouring local continuum (typically fitting windows of a few thousand~km\,s$^{-1}$).
Emission lines are fit with a single Gaussian profile, with the exception of the permitted recombination lines H\,$\upalpha~\lambda6564$, H\,$\upbeta~\lambda4865$, and He\,\textsc{ii}~$\lambda4686$ which are fit with both broad and narrow components where necessary.
In addition to the Gaussian emission line profiles, a linear local continuum is also included in the fit, to account for residual continuum flux not subtracted as part of the global continuum.
Blended and closely-spaced lines are fitted together, and certain lines have linked parameters: for example, the [O\,\textsc{iii}]~$\lambda\lambda4960,5007$, [N\,\textsc{ii}]~$\lambda\lambda6549,6585$, and [S\,\textsc{ii}]~$\lambda\lambda6718,6732$ doublet lines are constrained to have the same full widths at half maximum (FWHMs) and velocity offsets. 
We fix the flux ratios of [O\,\textsc{iii}]~$\lambda\lambda4960,5007$ and [N\,\textsc{ii}]~$\lambda\lambda6549,6585$ to 1:3 (\citealt{Storey00}; \citealt{Dimitrijevic07}; \citealt{Bon25}).

Before translating each measured emission line width into a virial gas radius, as described below, we first correct the measurements for instrumental broadening by subtracting the instrumental width, $\mathrm{FWHM_{inst}}$, from the measured width in quadrature.
The instrumental widths have either been taken from the relevant study describing the original observations or calculated from the resolving power quoted in the instrument documentation.\footnote{We discuss this approximate correction further in Section~\ref{sec:disc-obs_limitations}.} $\mathrm{FWHM_{inst}}$ for each spectrum is given in Tables~\ref{tab:spectra_1} and \ref{tab:spectra_2}. 
This correction allows meaningful comparisons of gas distances derived from spectra obtained with instruments of differing resolving power.
The emission line widths and luminosities are given in Table~\ref{tab:fwhm_table}; a portion is shown here and the complete table is available in the online supplementary material.

\begin{table*}
\centering
\caption{Line widths FWHM (km s$^{-1}$) and luminosities $\log(L/\mathrm{erg\,s^{-1}})$ for prominent emission lines.}
\label{tab:fwhm_logL_example}
\begin{tabular}{lllcccccc}
\hline
\multirow{2}{*}{Source} & \multirow{2}{*}{Spectrum} & \multirow{2}{*}{MJD} & \multicolumn{2}{c}{[O\,\textsc{iii}]\,$\lambda5007$} & \multicolumn{2}{c}{[Fe\,\textsc{vii}]\,$\lambda5720$} & \multicolumn{2}{c}{[Fe\,\textsc{vii}]\,$\lambda6087$} \\
\cmidrule(lr){4-5}\cmidrule(lr){6-7}\cmidrule(lr){8-9}
 & & & FWHM & $\log(L)$ & FWHM & $\log(L)$ & FWHM & $\log(L)$ \\
\hline
SDSS\,J0113 & SDSS & 56189 & $508\pm50$ & $40.43\pm0.06$ & $534\pm82$ & $40.15\pm0.09$ & $850\pm120$ & $40.29\pm0.08$ \\
 & DESI & 60687 & \ldots & \ldots & \ldots & \ldots & \ldots & \ldots \\
SDSS\,J0748 & SDSS & 53055 & $78\pm19$ & $39.00\pm0.14$ & \ldots & \ldots & \ldots & \ldots \\
 & MMT & 55921 & $294\pm16$ & $39.82\pm0.03$ & \ldots & \ldots & \ldots & \ldots \\
 & DESI & 59918 & $143\pm11$ & $39.67\pm0.04$ & \ldots & \ldots & \ldots & \ldots \\
SDSS\,J0952 & SDSS & 53734 & $339\pm20$ & $40.40\pm0.03$ & $469\pm33$ & $40.23\pm0.04$ & $301\pm21$ & $40.20\pm0.04$ \\
 & MMT & 55921 & $257.8\pm4.9$ & $40.56\pm0.01$ & $139\pm22$ & $39.27\pm0.09$ & $222\pm32$ & $39.35\pm0.08$ \\
 & DESI & 59536 & $294.4\pm7.4$ & $40.51\pm0.01$ & \ldots & \ldots & \ldots & \ldots \\
SDSS\,J1241 & SDSS & 53062 & $203.7\pm3.8$ & $39.97\pm0.01$ & $175\pm19$ & $39.14\pm0.06$ & $194.1\pm9.2$ & $39.37\pm0.03$ \\
 & MMT & 55921 & \ldots & \ldots & \ldots & \ldots & \ldots & \ldots \\
 & DESI & 60738 & $183.2\pm3.3$ & $40.00\pm0.01$ & $360\pm60$ & $38.76\pm0.09$ & $237\pm23$ & $38.88\pm0.05$ \\
Raichu & DESI & 59306 & $273.2\pm2.5$ & $40.79\pm0.01$ & $346\pm56$ & $39.20\pm0.09$ & $246\pm53$ & $38.97\pm0.12$ \\
\hline
\end{tabular}
\parbox{14.5cm}{\footnotesize
\textit{Notes:}
We give only upper limits on FWHM for emission lines which were significantly detected but had measured widths below the assumed instrumental FWHM.
A portion of the table is shown here for illustrative purposes; a machine-readable version of the full table is available in the online supplementary material.}
\label{tab:fwhm_table}
\end{table*}

\subsection{Black hole masses}\label{sec:mass}
Once corrected for instrumental broadening, our emission line FWHMs can be translated into radial distances, under the assumption that they represent the line-of-sight velocity dispersions of line-emitting gas in orbit around the SMBH.
The characteristic radius $r$ of the line-emitting gas is related to the FWHM of the emission line profile as $r=GM_\mathrm{BH}/\mathrm{FWHM}^2$ where $G$ is the gravitational constant and the FWHM is measured in velocity space.
To apply this relation, we need to know the BH mass, $M_\mathrm{BH}$.

We found $M_\mathrm{BH}$ estimates for approximately one third of our sample in the literature, 
which were obtained using a variety of methods. 
For many of these, masses were derived from the relation between $M_\mathrm{BH}$ and the bulge stellar velocity dispersion $\sigma_\star$ \citep{Ferrarese00}.
Stellar velocity dispersions are calculated as part of the SDSS spectroscopic pipeline for galaxy spectra (i.e.\ non-stellar and non-QSO spectra).
Other groups have also provided catalogs of SDSS galaxy properties, including $\sigma_\star$ estimates.
For sources with SDSS optical spectra, we search the catalogs of galaxy properties from both the Portsmouth Group \citep{Thomas13} and the MPA-JHU Group (\citealt{Kauffmann03,Brinchmann04,Tremonti04}). 
For sources with DESI spectra, we attempted to measure $\sigma_\star$ using the DESI spectral synthesis and emission line fitting code \textsc{FastSpecFit} \citep{Moustakas23}; 
this was successful for the DESI spectrum of Raichu, but the velocity dispersion fit failed in most cases.

\textcolor{black}{For several of the CL-TDEs in our sample, BH mass estimates were derived by previous studies (listed in Table~\ref{tab:masses}) using \mosfit\ \citep{Guillochon18, Mockler19}, which fits the TDE lightcurves assuming reprocessed accretion disc emission, and \tdemass\ \citep{Ryu20}, which estimates the BH mass from the peak luminosity and temperature of the UV/optical emission.}

For the non-variable, AGN-linked objects, a handful of BH masses estimated using the single-epoch virial method (via the $R$-$L$ relation, e.g., \citealt{Bentz09}) are available in value-added catalogs of SDSS quasars.
\cite{Rakshit20} performed a spectroscopic analysis of all SDSS QSO spectra up to and including DR14, from which they calculated single-epoch virial BH masses from the measured broad line widths and optical luminosities.
Four of our ECLEs (all non-variable, AGN-linked sources) appear within their catalogue, from which we take the fiducial virial mass.
All four have BH masses in excess of $10^8~\mathrm{M}_\odot$, the maximum mass at which a main sequence star would be tidally disrupted by a non-spinning (Schwarzchild) BH \citep{Hills75,Evans89}.
For other nv-ECLEs with broad lines, we estimate masses using the single-epoch virial relation of \cite{Kynoch23}, employing both the width and luminosity of broad \ha, as measured from the available spectra.
By using the luminosity of the emission line itself in the mass estimation (rather than the nearby 5100~\si\angstrom\ continuum luminosity, as is the more common practice) we avoid the complication of separating the nuclear and host galaxy continua.
The BH masses derived using this relation assume a virial factor $f=1$, consistent with the calibration of \cite{Mejia-Restrepo16} on which it is based.

Finally, we are able to source BH mass estimates for the majority of sources in our sample.
No estimates were found or made for AT\,2021dms, Raticate, or Charizard.  
As described in Section~\ref{sec:spec-at2021dms}, AT\,2021dms lacks sufficient emission line measurements to make a map of the gas;
for Raticate and Charizard we adopt a BH mass of $10^{7}$~M$_\odot$ to make the maps presented in Figs.~\ref{fig:v-ECLE_maps} and \ref{fig:nonvar_ecle_maps} but exclude the distance measurements from our subsequent analyses.

A compilation of the mass estimates and their origins is presented in Table~\ref{tab:masses}.
For sources with multiple mass estimates, we have adopted the geometric mean mass of those given in the table to make the maps, showing the virial distance inferred for each detected emission line, ordered by their ionisation potential (given in Table~\ref{tab:emlines}).
The assumed mass for each map is indicated in Figs.~\ref{fig:v-ECLE_maps}, \ref{fig:cl-tde_maps}, and \ref{fig:nonvar_ecle_maps}.

\begin{table*}
    \centering
    \caption{Black hole mass estimates for the objects in this study}
\resizebox{\textwidth}{!}{\begin{tabular}{lcll}
    \hline
         Short name   & $\log(M_\mathrm{BH}/\mathrm{M}_\odot)$ & Basis measurement                                           & Reference \\
    \hline
    \multicolumn{4}{c}{Variable sources: v-ECLEs} \\
         SDSS\,J0113  &  7.69 & $\sigma_*=155$~km\,s$^{-1}$ [BOSS DR12] \protect\citep{Thomas13}; $M_\mathrm{BH}$--$\sigma_*$ relation of \cite{Ferrarese05}  & This work  \\
         SDSS\,J0748  &  7.25 & $\sigma_*=126$~km\,s$^{-1}$ [SDSS DR8] \protect\citep{Thomas13}; $M_\mathrm{BH}$--$\sigma_*$ relation of \cite{Ferrarese05} & This work  \\
                      &  6.27 & $\sigma_*=79$~km\,s$^{-1}$ [SDSS] \protect\citep{Blanton05}; $M_\mathrm{BH}$--$\sigma_*$ relation of \cite{Ferrarese05} & This work \\
         SDSS\,J0952  &  6.85 & $\sigma_*=95$~km\,s$^{-1}$ [SDSS]; $M_\mathrm{BH}$--$\sigma_*$ relation of \cite{Ferrarese05} & \cite{Komossa08} \\
         SDSS\,J1241  &  6.42 & $\sigma_*=85$~km\,s$^{-1}$ [SDSS] \protect\citep{Blanton05} & This work  \\
         Raichu       &  7.28 & $\sigma_*=128$~km\,s$^{-1}$ [DESI]; $M_\mathrm{BH}$--$\sigma_*$ relation of \cite{Ferrarese05} & This work  \\
         SDSS\,J1342  &  6.06 & $\sigma_*=72$~km\,s$^{-1}\,^{\dagger}$ [SDSS DR8] \protect\citep{Thomas13}; $M_\mathrm{BH}$--$\sigma_*$ relation of \cite{Ferrarese05} & This work  \\
                      &  6.18 & $\sigma_*=76$~km\,s$^{-1}$ [SDSS DR8] \protect\citep{Blanton05}; $M_\mathrm{BH}$--$\sigma_*$ relation of \cite{Ferrarese05} & This work  \\
         SDSS\,J1350  &  7.00 & FWHM and luminosity of broad \ha\ [SDSS]; $M_\mathrm{BH}$ relation of \cite{Kynoch23} & This work  \\
         Raticate     &  (7.00) & No measurement found &  \\
         Pidgeot      &  7.32 & FWHM and luminosity of broad \ha\ [DESI 2021/04/06]; $M_\mathrm{BH}$ relation of \cite{Kynoch23} & This work  \\
                      &  7.35 & FWHM and luminosity of broad \ha\ [DESI 2021/04/10]; $M_\mathrm{BH}$ relation of \cite{Kynoch23} & This work  \\
    \hline
    \multicolumn{4}{c}{Variable sources: CL-TDEs} \\
         TDE\,2022upj  &  5.8  & \mosfit\ analysis of $g$-band lightcurve                                                     & \cite{Newsome24} \\
                      &  6.2  & \tdemass\ analysis of $g$-band lightcurve                                                    & \cite{Newsome24} \\
                      &  5.9  & Stellar mass from \bagpipes; $M_\mathrm{BH}$--$M_\mathrm{bulge}$ relation of \cite{Scott13}  & \cite{Newsome24} \\
                      &  6.0  & Stellar mass from \bagpipes; $M_\mathrm{BH}$--$M_\mathrm{bulge}$ relation of \cite{Greene20} & \cite{Newsome24} \\

         AT\,2018gn  &  6.77 & \mosfit\ analysis of multiband lightcurves & \cite{Wang24} \\
                      &  6.65 & $\sigma_*=76$~km\,s$^{-1}$ [Palomar]; $M_\mathrm{BH}$--$\sigma_*$ relation of \cite{Kormendy13} & \cite{Wang24} \\
         AT\,2021dms  &  (7.00) & No measurement found &  \\
         TDE\,2019qiz  &  5.89 & \mosfit\ analysis of multiband lightcurves & \cite{Nicholl20} \\
                      &  6.14 & \mosfit\ analysis of multiband lightcurves & \cite{Hung21} \\
                      &  5.82 & $\sigma_*=72$~km\,s$^{-1}$ [X-shooter]; $M_\mathrm{BH}$--$\sigma_*$ relation of \cite{McConnell13} & \cite{Short23} \\
                      &  6.24 & $\sigma_*=72$~km\,s$^{-1}$ [X-shooter]; $M_\mathrm{BH}$--$\sigma_*$ relation of \cite{Gultekin09}  & \cite{Short23} \\
                      &  6.54 & $\sigma_*=72$~km\,s$^{-1}$ [X-shooter]; $M_\mathrm{BH}$--$\sigma_*$ relation of \cite{Kormendy13}  & \cite{Short23} \\
         AT\,2021acak &  7.31 & FWHM and luminosity of broad \ha\ [DESI]; $M_\mathrm{BH}$ relation of \cite{Kynoch23} & This work \\ 
         TDE\,2022fpx  &  7.46 & \mosfit\ analysis of multiband lightcurves  & \cite{Koljonen24} \\ 
                      &  6.71 & \tdemass\ analysis of multiband lightcurves & \cite{Koljonen24} \\ 
                      &  6.80 & TDE peak luminosity relation with $M_\mathrm{BH}$ \citep{Mummery24} & \cite{Koljonen24} \\ 
                      &  7.19 & TDE radiated energy relation with $M_\mathrm{BH}$ \citep{Mummery24} & \cite{Koljonen24} \\
                      &  6.62 & Galaxy mass from SED fit; $M_\mathrm{BH}$--$M_\mathrm{gal}$ relation of \cite{Greene20} & \cite{Koljonen24} \\
         AT\,2018dyk  &  7.56 & $\sigma_*=123$~km\,s$^{-1}$ [SDSS/MaNGA]; $M_\mathrm{BH}$--$\sigma_*$ relation of \cite{Kormendy13} & \cite{Clark25} \\

         AT\,2017gge  &  6.55 & $\sigma_*=97$~km\,s$^{-1}$ [X-shooter]; $M_\mathrm{BH}$--$\sigma_*$ relation of \cite{McConnell13} & \cite{Onori22} \\
                      &  6.80 & $\sigma_*=97$~km\,s$^{-1}$ [X-shooter]; $M_\mathrm{BH}$--$\sigma_*$ relation of \cite{Gultekin09} & \cite{Onori22} \\
         TDE\,2021qth  &  5.95 & Galaxy mass from SED fit; $M_\mathrm{BH}$--$M_\mathrm{gal}$ relation of \cite{Yao23} & \cite{Yao23} \\
         AT\,2018bcb  &  7.6  & V-band bulge luminosity; $M_\mathrm{BH}$--$L_V$ relation of \cite{McConnell13} & \cite{Neustadt20} \\
    \hline
    \multicolumn{4}{c}{Non-variable sources: nv-ECLEs} \\
        Charizard    &  (7.00) & No measurement found &  \\
        Arbok        &  9.56 & Single-epoch virial mass estimate from SDSS spectrum & \cite{Rakshit20}  \\
        SDSS\,J0807  &  7.69 & FWHM and luminosity of broad \ha\ [SDSS]; $M_\mathrm{BH}$ relation of \cite{Kynoch23} & This work \\
                     &  7.64 & FWHM and luminosity of broad \ha\ [DESI]; $M_\mathrm{BH}$ relation of \cite{Kynoch23} & This work \\
        SDSS\,J0938  &  6.93 & $\sigma_*=108$~km\,s$^{-1}$ [SDSS DR8] \protect\citep{Thomas13}; $M_\mathrm{BH}$--$\sigma_*$ relation of \cite{Ferrarese05} & This work \\
                     &  7.26 & $\sigma_*=127$~km\,s$^{-1}$ [SDSS] \protect\citep{Blanton05}; $M_\mathrm{BH}$--$\sigma_*$ relation of \cite{Ferrarese05} & This work \\
        SDSS\,J1055  &  7.12 & FWHM and luminosity of broad \ha\ [SDSS]; $M_\mathrm{BH}$ relation of \cite{Kynoch23} & This work \\
                     &  7.03 & FWHM and luminosity of broad \ha\ [DESI]; $M_\mathrm{BH}$ relation of \cite{Kynoch23} & This work \\
        Sandslash    &  6.73 & $\sigma_*=99$~km\,s$^{-1}$ [SDSS DR8] \protect\citep{Thomas13}; $M_\mathrm{BH}$--$\sigma_*$ relation of \cite{Ferrarese05} & This work  \\
                     &  7.07 & $\sigma_*=116$~km\,s$^{-1}$ [SDSS] \protect\citep{Blanton05}; $M_\mathrm{BH}$--$\sigma_*$ relation of \cite{Ferrarese05} & This work  \\
        SDSS\,J1207  &  7.06 & FWHM and luminosity of broad \ha\ [SDSS]; $M_\mathrm{BH}$ relation of \cite{Kynoch23} & This work \\
        SDSS\,J1238  &  7.66 & FWHM and luminosity of broad \ha\ [SDSS]; $M_\mathrm{BH}$ relation of \cite{Kynoch23} & This work \\
                     &  7.57 & FWHM and luminosity of broad \ha\ [DESI]; $M_\mathrm{BH}$ relation of \cite{Kynoch23} & This work \\
        SDSS\,J1247  &  7.27 & FWHM and luminosity of broad \ha\ [SDSS]; $M_\mathrm{BH}$ relation of \cite{Kynoch23} & This work  \\
                     &  7.33 & FWHM and luminosity of broad \ha\ [DESI]; $M_\mathrm{BH}$ relation of \cite{Kynoch23} & This work  \\
        SDSS\,J1402  &  8.03 & Single-epoch virial mass estimate from SDSS spectrum & \cite{Rakshit20} \\
        Nidoqueen    &  7.47 & $\sigma_*=140$~km\,s$^{-1}$ [SDSS DR8] \protect\citep{Thomas13}; $M_\mathrm{BH}$--$\sigma_*$ relation of \cite{Ferrarese05} & This work  \\
                     &  7.39 & $\sigma_*=135$~km\,s$^{-1}$ [SDSS] \protect\citep{Blanton05}; $M_\mathrm{BH}$--$\sigma_*$ relation of \cite{Ferrarese05} & This work  \\
        SDSS\,J1458  &  8.14 & Single-epoch virial mass estimate from SDSS spectrum & \cite{Rakshit20} \\
        SDSS\,J1459  &  7.26 & FWHM and luminosity of broad \ha\ [SDSS]; $M_\mathrm{BH}$ relation of \cite{Kynoch23} & This work  \\
                     &  7.28 & FWHM and luminosity of broad \ha\ [DESI]; $M_\mathrm{BH}$ relation of \cite{Kynoch23} & This work  \\
        SDSS\,J1715  &  7.86 & FWHM and luminosity of broad \ha\ [SDSS]; $M_\mathrm{BH}$ relation of \cite{Kynoch23} & This work  \\
                     &  7.89 & FWHM and luminosity of broad \ha\ [DESI]; $M_\mathrm{BH}$ relation of \cite{Kynoch23} & This work  \\
        SDSS\,J2220  &  8.22 & Single-epoch virial mass estimate from SDSS spectrum & \cite{Rakshit20} \\
    \hline
    \end{tabular}}
    \parbox{\textwidth}{Parenthesised values indicate sources for which no BH mass estimate could be found in the literature or derived from the available data; for these we have adopted the fiducial value $\log(M_\mathrm{BH}/\mathrm{M}_\odot)=7.00$.}
    \label{tab:masses}
\end{table*}

\section{Spectroscopic analysis}\label{sec:spectro}
Having measured the emission line widths, we translate these to virial radii of the emission line gas using the BH masses measured above in order to construct our emission line maps.
Maps of the v-ECLEs, CL-TDEs, and nv-ECLEs are presented in Figs.~\ref{fig:v-ECLE_maps}, \ref{fig:cl-tde_maps}, and \ref{fig:nonvar_ecle_maps}, respectively.
Below we describe the detected lines, measurements and spectroscopic evolution over multiple epochs for each source in our sample.
We summarise these spectroscopic analyses in Table~\ref{tab:specsummary}.
Throughout this section, the emission line luminosities are given as $\log(L_\mathrm{line})$ where $L_\mathrm{line}$ has units of erg\,s$^{-1}$. 

\begin{table}
    \centering
    \caption{Properties of optical emission lines}
    \begin{tabular}{lccc}
    \hline
    Line                & $\lambda$      & Critical density   & Ionisation potential \\
                        & [\si\angstrom] & [cm$^{-3}$]        & [eV] \\
    \hline
    {[Ne\,\textsc{v}]}  & 3345.821       & $1.14\times10^{7}$ &  97.11 \\
    {[Ne\,\textsc{v}]}  & 3425.881       & $1.90\times10^{7}$ &  97.11 \\  
    \fevii\             & 3758.920       & $4.02\times10^{7}$ &  99.00 \\   
    \fev\               & 3839.270       & $1.00\times10^{8}$ &  54.80 \\ 
    \fev\               & 3891.280       & $1.61\times10^{8}$ &  54.80 \\
    \heii\              & 4687.02~       &                    &  54.42 \\
    \hb\                & 4862.68~       &                    &  13.7 \\
    \fevii\             & 4893.370       & $3.09\times10^{6}$ &  99.00 \\
    \oiii\              & 4960.295       & \textcolor{black}{$7.0\times10^{5}$}  &  54.94 \\
    \oiii\              & 5008.240       & \textcolor{black}{$7.0\times10^{5}$}  &  54.94 \\ 
    \fevi\              & 5145.750       & $2.29\times10^{7}$ &  75.00 \\
    \fevii\             & 5158.890       & $3.44\times10^{6}$ &  99.00 \\
    \fevi\              & 5176.040       & $3.29\times10^{7}$ &  75.00 \\
    \fevii\             & 5276.380       & $2.98\times10^{6}$ &  99.00 \\
    \fexiv\             & 5302.860       & $3.99\times10^{8}$ & 361.00 \\
    \cav\               & 5309.110       & $6.63\times10^{7}$ &  67.10 \\
    \fevi\              & 5335.180       & $6.32\times10^{6}$ &  75.00 \\
    \arx\               & 5533.265       & $2.36\times10^{9}$ & 422.60 \\
    \fevii\             & 5720.700       & $3.72\times10^{7}$ &  99.00 \\ 
    \hei\               & 5877.29~       &                    &  24.6 \\
    \fevii\             & 6087.000       & $4.46\times10^{7}$ &  99.00 \\
    \fex\               & 6374.510       & $4.45\times10^{8}$ & 235.04 \\
    \nii\               & 6549.85~       & \textcolor{black}{$7.9\times10^{4}$}  &  14.53 \\
    \ha\                & 6564.61~       &                    &  13.7 \\
    \nii\               & 6585.28~       & \textcolor{black}{$7.9\times10^{4}$}  &  14.53 \\
    \sii\               & 6718.29        & \textcolor{black}{$1.6\times10^{3}$}  &  23.34 \\
    \sii\               & 6732.67        & \textcolor{black}{$4.0\times10^{3}$}  &  23.34 \\
    \sxii\              & 7611.000       & $7.09\times10^{9}$ & 504.78 \\
    \fexi\              & 7891.800       & $6.39\times10^{8}$ & 262.10 \\
    \hline
    \end{tabular}
    \label{tab:emlines}
\end{table}

\subsection{Variable ECLEs}
\subsubsection{SDSS\,J0113}
Several strong coronal lines are evident in the 2012 SDSS/BOSS spectrum, including \nev~$\lambda3425$, \fevii~$\lambda3758, 5720, 6087$, and \fexi~$\lambda7891$.
The narrow low-ionisation \oiii~$\lambda\lambda4960, 5007$ lines are also present.
There is a weak broad component to H\,$\upalpha$ with $\mathrm{FWHM}=1948\pm291$~km\,s$^{-1}$.
By the time of the 2021 Gemini spectrum, the coronal lines and broad H\,$\upalpha$ had faded although narrow H\,$\upalpha$ and \nii~$\lambda\lambda6549, 6585$ were still clearly present, although telluric absorption in this part of the spectrum means we are unable to reliably measure these lines.
As noted by \cite{Callow25}, \oiii\ emission is much weaker in the Gemini spectrum and we are unable to obtain a robust measurement of its profile.
All emission lines in the new, 2025 DESI spectrum are very weak (if present at all), although it is just possible to measure the profiles of narrow H\,$\upalpha$ and  \nii~$\lambda\lambda6549, 6585$.
We see from the SDSS spectrum that the coronal line profiles are broader than \oiii~$\lambda5007$, which in turn is broader than narrow \ha, implying an ionisation stratification of the emission line gas.

\subsubsection{SDSS\,J0748}
The 2004 SDSS spectrum clearly shows prominent high- and low-ionisation emission lines. 
The coronal lines \fex~$\lambda6374$ and \fexi~$\lambda7891$ are strong features with broader profiles ($\mathrm{FWHM}\approx300$--600~km\,s$^{-1}$) than the low-ionisation lines ($\mathrm{FWHM}\approx100$--200~km\,s$^{-1}$), implying an ionisation stratification.
The very high-ionisation line [Ar\,\textsc{xiv}]~$\lambda4414$ (ionisation potential 685.5~eV) is also possibly weakly detected with $\mathrm{FWHM}=447\pm105$~km\,s$^{-1}$. 
In contrast, the lower-ionisation \fevii\ lines are not present.
A very shallow, broad component to H\,$\upalpha$ is perhaps detected in the SDSS spectrum, but this feature is not present in the later spectra. 
In the 2011 MMT spectrum, all of the coronal lines have disappeared, whereas the \oiii\ lines are much stronger.
The low-ionisation lines are still present in the 2021 DESI spectrum and have narrower profiles; no coronal lines appear or reappear.

\subsubsection{SDSS\,J0952}
As described in Section~\ref{sec:sample:sdssj0952}, the coronal lines faded progressively between 2005 and 2021.
The 2005 SDSS spectrum shows \fevii, \fex, \fexi, and \fexiv\ coronal lines, as well as broad Balmer and \heii\ lines and the narrow \oiii\ 
doublet; weak \sxii~$\lambda7611$ is possibly also detected.

Ionisation stratification of the emission line gas is evident;
the coronal lines appear between $\approx0.05$--0.3~pc, low-ionisation narrow lines at $\gtrsim0.3$~pc, and the short-lived broad lines lie at $<0.03$~pc.
\fevii~$\lambda5720$ appears much narrower in the MMT spectrum compared with SDSS, but otherwise there is little evolution in linewidth across the three spectra.

\subsubsection{SDSS\,J1241}
As described in Section~\ref{sec:sample:sdssj1241}, coronal lines were present in the 2004 SDSS and 2011 MMT spectra but were not detected in the 2021 Kast spectrum.
The 2004 SDSS spectrum shows \fevii, \fex, and \fexi\ coronal lines; unfortunately the red part of the 2011 MMT spectrum was not recorded with the slit centred on the nucleus, so it is not possible to check the longer-wavelength lines at that epoch.

We present here for the first time the recent (2025 March) DESI spectrum of SDSS\,J1241.
\fevii~$\lambda6087$ is robustly detected in this spectrum (luminosity $\log[L]=38.88\pm0.05$) and weak \fevii~$\lambda3758$ and 5720 also appear to be present (luminosities $\log[L]=38.81\pm0.13$ and $\log[L]=38.76\pm0.09$, respectively).
The presence of \fex~$\lambda6374$ is highly uncertain; its luminosity is $\log[L]=38.39\pm0.20$.
\fexi~$\lambda7891$, which was significantly detected in the SDSS spectrum, is not seen in the DESI spectrum.

Narrow Balmer lines and \oiii\ emission are clearly seen in the SDSS, Kast, and DESI spectra.
Again restating our caution in comparing emission line luminosities measured from spectra recorded at different observatories,
we note that \oiii~$\lambda5007$ has a slightly greater luminosity ($\log[L]=40.00\pm0.01$) in DESI compared with SDSS ($\log[L]=39.97\pm0.01$), although the Balmer line luminosities and Balmer decrement have reduced substantially.
Narrow \ha\ weakens from $\log[L]=40.124\pm0.006$ to $39.67\pm0.02$ and \hb\ from $\log[L]=39.31\pm0.07$ to $39.10\pm0.13$.

Linewidths measured in the SDSS spectrum are mostly very similar, with the narrow Balmer lines, \heii, \oiii, and the coronal lines \fevii~$\lambda3758$, 5720, 6087, \fex~$\lambda6374$, and \fexi~$\lambda7891$ all having $\mathrm{FWHM}\approx200$~km\,s$^{-1}$.
There appears to be greater variation in linewidths in the DESI spectrum: \sii\ (which was very weak in the SDSS spectrum) is clearly very narrow with $\mathrm{FWHM}=90\pm9$~km\,s$^{-1}$; \oiii~$\lambda5007$ is broader with $\mathrm{FWHM}=183\pm3$~km\,s$^{-1}$, and the strongest coronal line, \fevii~$\lambda6087$, is broader still with $\mathrm{FWHM}=237\pm23$~km\,s$^{-1}$.
Narrow \ha\ has a consistent width between the SDSS and DESI spectra, as do narrow \hb\ and \heii\ (although these have much greater uncertainties).
\oiii\ is slightly narrower in the DESI spectrum compared with SDSS ($\mathrm{FWHM}=204\pm4$~km\,s$^{-1}$).
It is unclear from these measurements whether the emission line gas is stratified in ionisation state.

It appears that coronal line emission in SDSS\,J1241 has persisted but weakened over two decades of observations.
This was not obvious from the 2021 Kast spectrum (in which coronal lines were only marginally detected, if at all) but the new, high-quality DESI spectrum confirms that \fevii\ emission is present at a lower level than previously observed.

\subsubsection{Raichu (DESI\,J204.3990$+$03.8775)}
The 2021 DESI discovery spectrum of Raichu shows the coronal lines \nev~$\lambda3425$, \fevii~$\lambda5720$, 6087, \fex~$\lambda6374$, and possibly weak \fexi~$\lambda7891$ (\citealt{Clark26} noted that the latter is contaminated by skyline emission).
Low-ionisation lines include narrow Balmer and \heii, \sii, \oii, and strong \oiii.

The virial radii calculated for the coronal lines suggest they originate on scales $\lesssim1$~pc with the low-ionisation lines having greater virial radii. 
However, given the uncertainties on the coronal line FWHMs, the evidence of stratification of the emission line gas is tentative. 
The two strongest coronal lines (\nev~$\lambda3425$ and \fevii~$\lambda6087$) have widths consistent with \oiii~$\lambda5007$; only weak \fex~$\lambda6374$ has a width that is significantly broader than \oiii~$\lambda5007$.

\subsubsection{SDSS\,J1342}
As described in Section~\ref{sec:sample:sdssj1342}, the highest-ionisation coronal lines faded between 2002 and 2011, \fevii\ appeared, and \oiii\ emission strengthened substantially.
\fevii\ emission was still evident in the DESI spectrum of 2021.
The \oiii\ doublet is even stronger than it appeared in 2011: the luminosity of \oiii~$\lambda5007$ is $\log(L)=40.377\pm0.008$ in the DESI spectrum, compared with 
$39.80\pm0.02$ in the MMT spectrum and only $39.49\pm0.03$ in the SDSS spectrum.

In the SDSS spectrum, the strong \fex\ and \fexi\ coronal lines are broader than the \oiii\ and Balmer lines which implies a stratification in ionisation.
The situation is less clear in the later spectra: in the MMT spectrum, the widths of \fevii~$\lambda5720$ and 6087 are consistent with each other and both are narrower than \oiii~$\lambda5007$, but in the DESI spectrum the width of \fevii~$\lambda5720$ is consistent with that of \oiii~$\lambda5007$ (within the large uncertainty on the former).
Both narrow \ha\ and \sii\ emission lines become progressively narrower (the latter is possibly unresolved in the low-resolution MMT spectrum).

\subsubsection{SDSS\,J1350}
As described in Section~\ref{sec:sample:sdssj1350}, the high-ionisation coronal lines faded and \fevii\ appeared between the 2006 SDSS and 2011 MMT spectra of SDSS\,J1350, with broad H and He lines also disappearing.
\fevii~$\lambda5720$ has a measured width $\mathrm{FWHM}=211\pm34$~km\,s$^{-1}$ so is not resolved by the MMT spectrograph.
Strong \oiii\ and weak \sii\ lines appear in both spectra, with \oiii\ having a greater luminosity in the latter.
\fexi~$\lambda7891$ was also a strong feature in the SDSS spectrum, but this wavelength was not covered by MMT.
\cite{Clark24} noted the possible presence of a low-S/N \fexi~$\lambda7891$ feature in the 2021 NTT spectrum, 
although we do not make use of that spectrum here because of its low resolution.
The galaxy has not yet been observed by DESI although it is a target (ID 39628465351164568), 
so a future spectrum may clarify whether this feature is still present.

Unlike many other sources, in SDSS\,J1350 the coronal lines are slightly narrower than low-ionisation lines such as those of \oiii\ and \sii\ (although the differences are not great, considering the scale of the uncertainties).
For example, in the SDSS spectrum \fex~$\lambda6374$ and \fexi~$\lambda7891$ have consistent widths $\mathrm{FWHM}=234\pm27$ and $237\pm30$~km\,s$^{-1}$, respectively, while \oiii~$\lambda5007$ has width $\mathrm{FWHM}=292\pm29$~km\,s$^{-1}$.
In the MMT spectrum \fevii~$\lambda6087$ is also found to be slightly narrower than \oiii~$\lambda5007$.
The circumnuclear gas therefore does not show the ionisation stratification typical of ECLEs, with coronal line gas apparently located at larger radii than the low-ionisation line emitting gas.

\subsubsection{Raticate (DESI\,J216.5891$+$00.1436)}
Raticate is another of the \cite{Clark26} DESI EDR v-ECLEs.
It was identified by strong \fexi~$\lambda7891$ emission in addition to the detection of \fex~$\lambda6374$.
Their widths are consistent at $\mathrm{FWHM}=291\pm35$ and $254\pm29$~km\,s$^{-1}$, respectively.
These lines are only marginally broader than \oiii~$\lambda5007$ ($\mathrm{FWHM}=210\pm14$~km\,s$^{-1}$).
Weak \fevii~$\lambda6087$ is also detected ($\log[L]=39.1\pm0.1$) and \fevii~$\lambda5720$ marginally so ($\log[L]=39.2\pm0.2$).
There is no clear ionisation stratification of the emission line gas.

\subsubsection{Pidgeot (DESI\,J243.3798$+$55.7528)}
The 2021 DESI spectrum of Pidgeot contains strong \oiii\ emission and Balmer lines with shallow broad components (broad \ha\ $\mathrm{FWHM}\approx2800$~km\,s$^{-1}$).
Coronal line emission includes \nev\ and \fevii\ coronal lines. 
\fex~$\lambda6374$ is possibly seen, but it falls in a region of overlap between detectors that is also affected by telluric absorption and other contamination (\citealt{Clark26}).
We do not see strong evidence of stratification of the gas from the few lines measured: all narrow lines are emitted on scales of $\sim2$~pc.


\begin{figure*}
\centering
\setlength{\tabcolsep}{4pt} 
\renewcommand{\arraystretch}{1.0} 

\begin{tabular}{@{}cc@{}}
\includegraphics[width=0.45\textwidth]{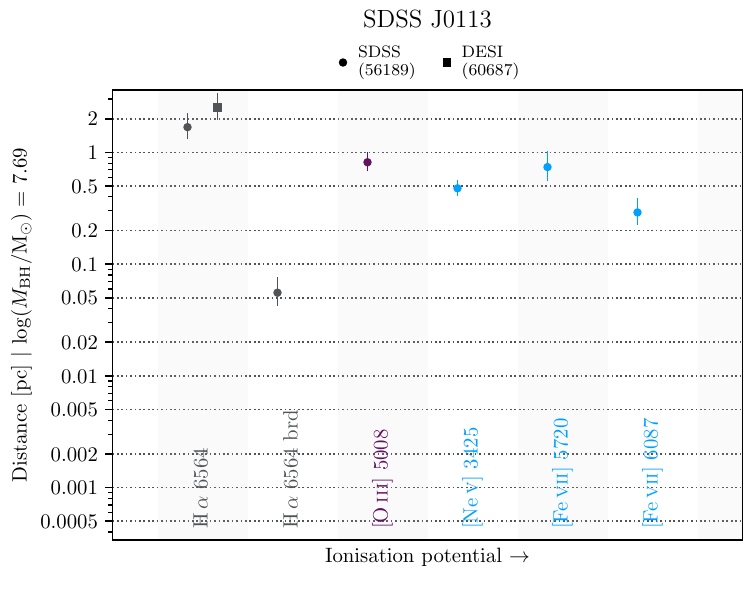} & 
\includegraphics[width=0.45\textwidth]{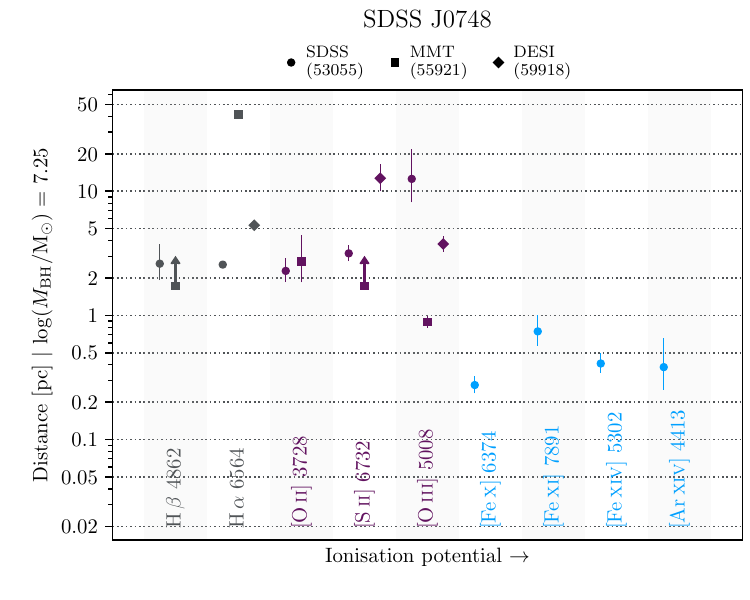} \\ 
\includegraphics[width=0.45\textwidth]{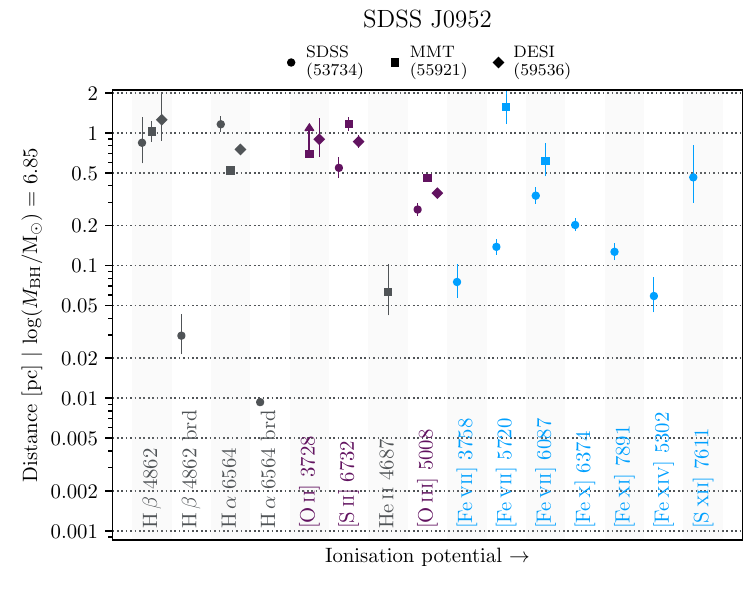} &  
\includegraphics[width=0.45\textwidth]{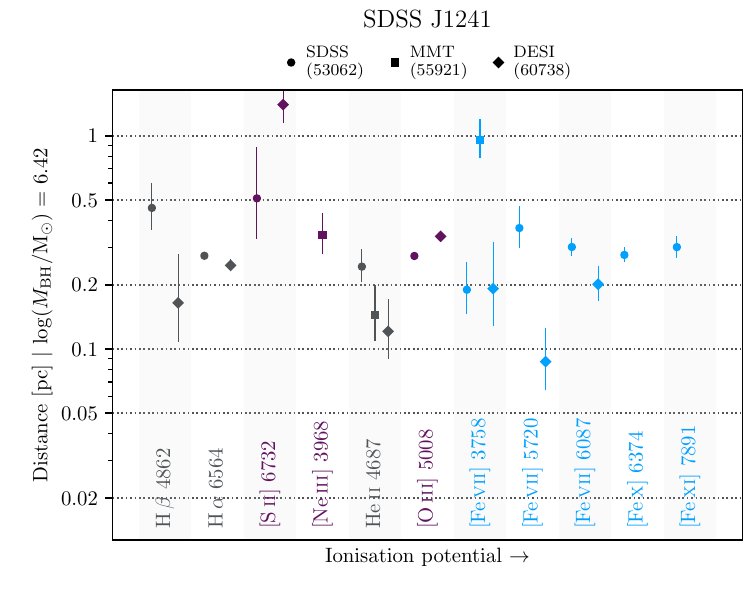} \\  
\includegraphics[width=0.45\textwidth]{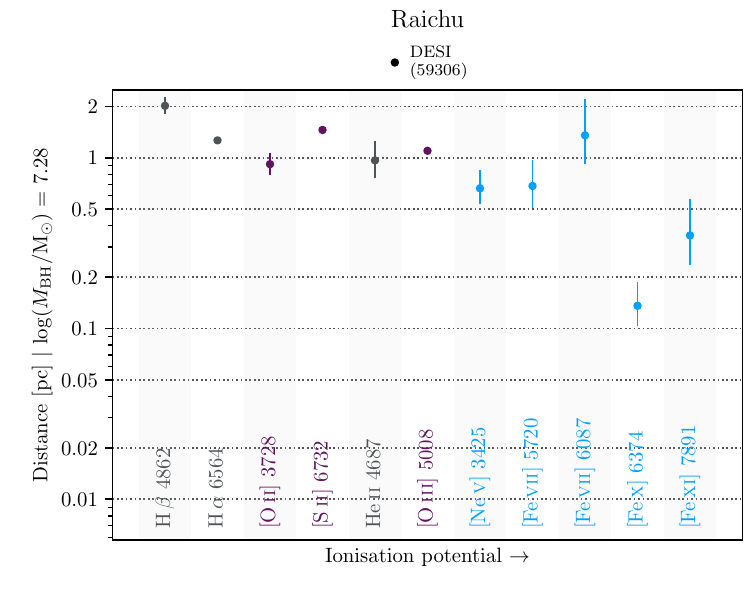} & 
\includegraphics[width=0.45\textwidth]{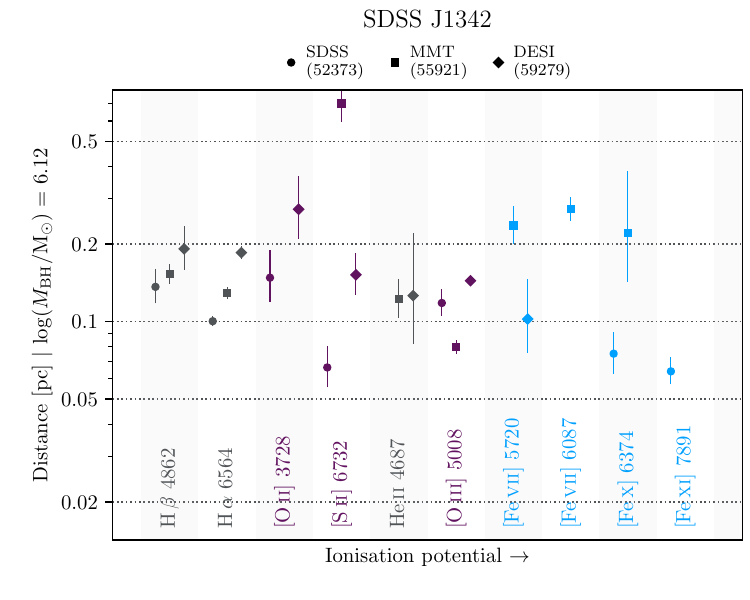} \\  
\end{tabular}

\caption{Emission line maps for v-ECLEs.
         We show the FWHM-inferred distances of emission line gas versus ionisation potential for each emission line. 
         Transitions are ordered by increasing ionisation potential along the x-axis. 
         Permitted transitions are coloured grey, low-ionisation forbidden lines purple, and coronal lines cyan.
         Where multiple epochs are available, measurements are grouped by line and plotted left to right in chronological order, with the observatory distinguished by marker type as indicated in the legend.
         The assumed BH mass ($M_\mathrm{BH}$) used in the distance calculations is given in the y-axis label.}
\label{fig:v-ECLE_maps}
\end{figure*}

\begin{figure*}
\addtocounter{figure}{-1} 
\centering
\setlength{\tabcolsep}{4pt}
\renewcommand{\arraystretch}{1.0}

\begin{tabular}{@{}cc@{}}
\includegraphics[width=0.45\textwidth]{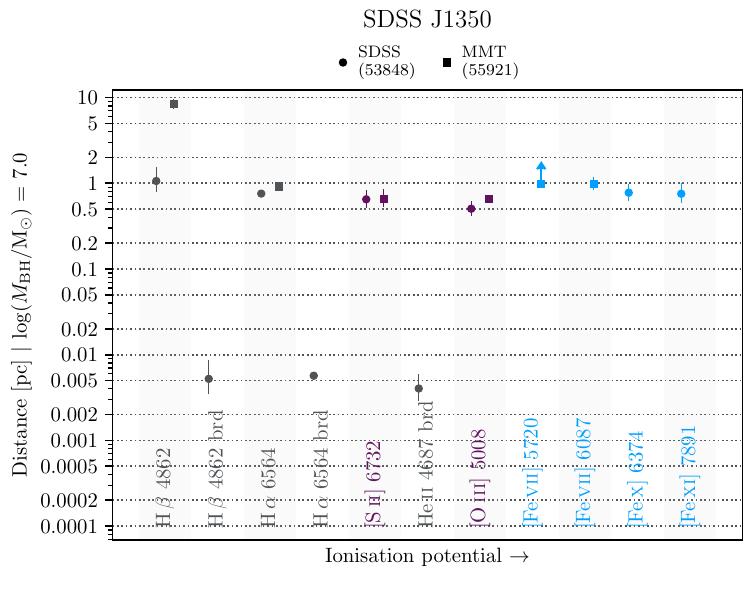} & 
\includegraphics[width=0.45\textwidth]{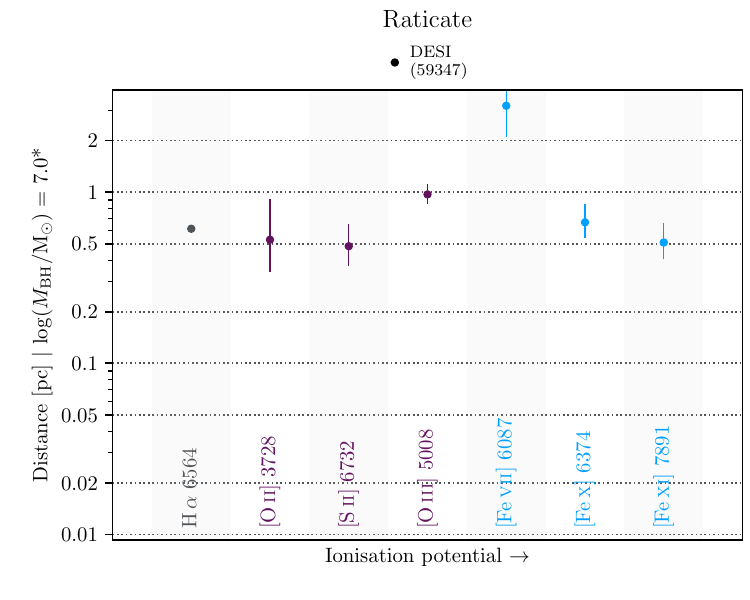} \\ 
\includegraphics[width=0.45\textwidth]{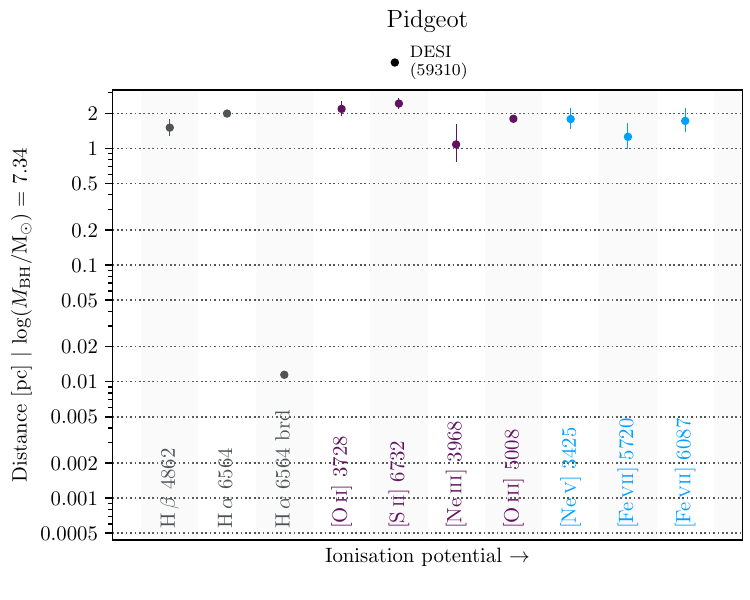} & 

\end{tabular}

\caption{Emission line maps for v-ECLEs (continued).}
\end{figure*}


\subsection{CL-TDEs}
\subsubsection{TDE\,2022upj}
An emission line map for TDE\,2022upj was previously presented by \cite{Newsome24}.
We do not have access to their spectra for a detailed analysis of the spectroscopic evolution of the source,
but \cite{Newsome24} provide median FWHM measurements for key emission lines allowing us to produce a distance versus ionisation potential map for this source for comparison with the rest of our sample.
For our plot, we assume a mass $\log(M_\mathrm{BH}/\mathrm{M}_\odot)=5.98$, the logarithmic mean value of those determined by \cite{Newsome24} (see Table~\ref{tab:masses}).
As seen in Fig.~\ref{fig:cl-tde_maps}, \cite{Newsome24} found that the \fevii, \fex, and \fexiv\ coronal lines were all broader than \oiii~$\lambda5007$.
They noted that whilst the coronal lines were present during the optical peak of the transient, \oiii\ emission only appeared later ($\approx400$~d after the peak).
The width of the narrow Balmer lines was intermediate between \fevii\ and \fex\ (which were narrower) and \fexiv\ (which was broader).
Additionally, they reported very broad \heii\ emission in all spectra.
This initially strong feature faded over the $\approx1$~yr of observations following the peak emission.
Its mean width was found to be $14,700\pm3400$~km\,s$^{-1}$, corresponding to a virial distance of $\approx2\times10^{-5}$~pc.

\subsubsection{AT\,2018gn (ASASSN-18ap)}
We have assumed instrumental broadening $\mathrm{FWHM_{inst}}=540$~km\,s$^{-1}$ for the two FLWO spectra, calculated assuming a resolution of 7.2~\si\angstrom\ for the FAST 300 grating (\citealt{Mink21}; \citealt{Wang24}).
This means that none of the narrow lines present are resolved, although we are still able to measure the width of broad \ha.
Consistent with the analysis performed by \cite{Wang24}, we observe a broad component to \ha\ that decreases in both flux and width as time progresses.
Its width in the earliest spectrum is $3671\pm359$~km\,s$^{-1}$ and in the latest is just $1784\pm72$~km\,s$^{-1}$.
The luminosity of this component peaks at $\log(L)=41.09\pm0.03$ in the 2018 February spectrum, and has reduced by about two thirds to $40.53\pm0.03$ in the last Palomar spectrum.

We already detect the lowest-ionisation coronal line \nev~$\lambda3425$ in the second FLWO spectrum of 2018 February 11 (1~d before the optical peak), 
although it was not detected in the January 15 spectrum and no other coronal lines are detected until the 2022 Palomar spectra.
We estimate its luminosity at $\log(L)=40.80\pm0.05$.
In the first Palomar spectrum taken 52 months later the line is still present although with a much reduced luminosity of $40.30\pm0.03$ and $\mathrm{FWHM}=411\pm20$~km\,s$^{-1}$.
The bluer \nev~$\lambda3345$ line is also covered by the Palomar spectrum and has a similar strength and width ($\log[L]=39.88\pm0.08$ and $\mathrm{FWHM}=513\pm71$~km\,s$^{-1}$).
The strengths of the lower-ionisation lines \oii~$\lambda\lambda3726,3728$ and \sii~$\lambda\lambda6718,6732$ are very similar between the last FLWO and first Palomar spectra, whereas the doublet \oiii~~$\lambda\lambda4960,5007$ is much stronger at the time of the later observation.

The coronal line profiles are similar in strength and shape over the 14 months covered by the three Palomar spectra.
The \nev\ lines become progressively broader: \nev~$\lambda3345$ increases from $\mathrm{FWHM}=416\pm58$ to $596\pm50$~km\,s$^{-1}$ between the 2022 June and 2023 September spectra and \nev~$\lambda3425$ increases from $\mathrm{FWHM}=281\pm14$ to $459\pm18$~km\,s$^{-1}$ in the same period.
Widths of the other forbidden emission lines remain broadly consistent.

\fex~$\lambda6374$ is seen in the Palomar spectra, although we obtain a good measurement of its profile only in the second, in which it is both weaker and broader than the other coronal lines.
Unlike \cite{Wang24} we do not see convincing \fexiv\ emission in the Palomar spectra.
\fexi~$\lambda7891$ is located in a spectral region affected by telluric absorption. 
A very narrow (apparently unresolved, $\mathrm{FWHM}=157\pm21$~km\,s$^{-1}$) feature appears at the expected location of \fexi~$\lambda7891$ in the 2022 June Palomar spectrum, but it does not appear in the later spectra.
The luminosity of the feature is $\log(L)=39.55\pm0.08$, less than half that in the \fevii\ lines: $39.84\pm0.03$ and $40.00\pm0.03$ for \fevii~$\lambda5720$ and 6087, respectively.
If we consider this a genuine detection, then the brief appearance of this feature is consistent with a trend seen in other sources in which the highest-ionisation lines fade first, whilst lower-ionisation lines remain or appear later (e.g., SDSS\,J0748, SDSS\,J0952, SDSS\,J1342).

From the Palomar spectra, we see that the coronal lines appear to be produced at a distance intermediate between the broad emission line gas and the low-ionisation, narrow line gas.
The widths of the coronal lines are greater than those of \oiii\ and \sii\ which, in turn, are broader than the narrow Balmer lines.

\subsubsection{AT\,2021dms}\label{sec:spec-at2021dms}
The low-resolution 2021 Magellan spectrum is the only available optical spectrum of the source.
In it, the low-ionisation lines (e.g., H, He, and \oiii) are not detected, nor are the \fevii\ coronal lines.
Both the very high-ionisation \fex~$\lambda6374$ and \fexi~$\lambda7891$ lines are present in the spectrum and have similar widths of $\mathrm{FWHM}=651\pm7$ and $598\pm14$~km\,s$^{-1}$ (corrected for instrumental broadening).
Assuming a BH mass of $10^7$~M$_\odot$, these correspond to distances 0.10 and 0.12~pc, respectively.
\fexiv~$\lambda5302$ is also clearly present and its profile appears somewhat broader than both \fex\ and \fexi, although we do not obtain a reliable measurement.
Given the scarcity of measurements, we do not produce an emission line map for this source.

\subsubsection{TDE\,2019qiz}
Detailed spectroscopic analyses of TDE\,2019qiz have been presented by \cite{Nicholl20}, \cite{Hung21}, and \cite{Short23}; 
here we describe our own measurements, focusing on the emission line profiles relevant to our distance estimates.
The early spectra show very broad ($\mathrm{FWHM}\sim15{,}000$~km\,s$^{-1}$), shallow, asymmetric H and He lines; an intermediate-width component narrows progressively throughout the 2019 spectra, approaching $\mathrm{FWHM}\approx2000$~km\,s$^{-1}$.
In the 2020--22 spectra the very broad components have disappeared and an intermediate-width component remains in \ha.
Very weak \sii\ lines can be discerned in the 2019 September Keck spectrum, but \oii\ is not clearly detected.  
Both doublets are present in the 2019 September X-shooter spectrum taken a few days later.
The \oiii\ doublet is not seen in the Keck spectrum and these are very weak features initially, but are strong in the 2020 December X-shooter spectrum and remain strong in 2022.

Flux excesses around the locations of optical coronal lines are visible in some of the early spectra, however they are only convincingly detected in the later spectra (as previously noted by \citealt{Short23}).
In the 2020 December X-shooter spectrum we see the very high ionisation lines \fexi~$\lambda7891$ and \fex~$\lambda6374$ and possibly weak \fexiv~$\lambda5302$ and \fevii~$\lambda5720$ and 6087; unlike \cite{Short23}, we do not find measureable \nev\ lines.
All of these lines, including \nev~$\lambda\lambda3345, 3425$, are clearly present in the 2022 January spectrum, with the exception of \fevii~$\lambda6087$.

We see that the widths of the \oii, \oiii, and \sii\ doublets become progressively narrower with time.
Between MJD 58754 and 59605 the FWHM of \sii\ decreases from $129\pm5$ to $93\pm8$~km\,s$^{-1}$.
The broad and narrow components of \ha\ also decrease in width.

The coronal lines are generally broader than the low-ionisation forbidden lines, except for the \nev\ lines that we measure in the 2022 spectrum, which are considerably narrower.
Similar to \cite{Short23}, we are able to approximately identify three emission line regions: low-ionisation forbidden lines arise on scales $\gtrsim0.2$~pc; the coronal lines are emitted at $\sim0.01$--0.2~pc; and broad lines at $\lesssim0.005$~pc.  

\subsubsection{AT\,2021acak}
Detailed spectral modelling of the Lijiang spectra was performed by \cite{Li23}.
We confirm their main findings, including the presence of the high-ionisation coronal lines \fexiv~$\lambda5302$ and \fex~$\lambda6375$ having widths of several hundred km\,s$^{-1}$ and the absence of \fevii\ and \oiii\ lines.
\cite{Li23} reported H and He lines with both broad ($\mathrm{FWHM}\approx2000$--3000~km\,s$^{-1}$) and narrow ($\mathrm{FWHM}\approx1000$~km\,s$^{-1}$) components in the Lijiang spectra.\footnote{Their fits of the broad \heii\ component were highly uncertain because of the limited S/N and possible blend with Fe\,\textsc{ii} emission.}
Our simpler fitting routine often struggles to separate the broad and narrow Balmer line components, and to convincingly measure broad \heii, so we do not report the values here.
In the later DESI spectrum, the broad and narrow Balmer lines are separable; we find narrow components with $\mathrm{FWHM}\approx200$~km\,s$^{-1}$ and broad components $\mathrm{FWHM}\approx1500$~km\,s$^{-1}$, narrower than were observed earlier.

\fex~$\lambda6375$ is narrower in the DESI spectrum than was found previously ($\mathrm{FWHM}=427\pm50$~km\,s$^{-1}$).
\fexiv~$\lambda5302$ has weakened substantially, and we are unable to obtain a reliable measurement of its width.
\fexi~$\lambda7891$ is clearly detected, also with a narrower profile than observed in the Lijiang spectra ($\mathrm{FWHM}=223\pm29$~km\,s$^{-1}$).
The \fevii~$\lambda3758$, 5720, and 6087 coronal lines newly appear in the DESI spectrum.  
They have widths in the range 141--287~km\,s$^{-1}$, similar to \fexi~$\lambda7891$ and the narrow Balmer lines.  
Additionally, \oiii\ emission is clearly detected in this new DESI spectrum ($\mathrm{FWHM}=266\pm34$~km\,s$^{-1}$).

\subsubsection{TDE\,2022fpx}
As described in Section~\ref{sec:sample:at2022fpx}, early-time spectra presented by \cite{Koljonen24} showed strong coronal line emission; 
however, given their low spectral resolution ($R\approx350$--360) we do not use them in this analysis.
Here we present measurements from the new, high-quality DESI spectrum of the source, taken on 2023 May 29 (310~d after the optical peak).
We do not signficantly detect [S\,\textsc{xii}]~$\lambda7613$ in the DESI spectrum, although \fex~$\lambda6374$, \fexi~$\lambda7891$, and \fexiv~$\lambda5302$ are all present, along with strong \fevii~$\lambda5720$ and 6087 emission.
The latter was discernible in the NOT spectrum (see fig.~1 of \citealt{Koljonen24}) although no claim was made of its detection.
\oiii\ emission was very weak in the earlier spectra presented by \cite{Koljonen24} and appears somewhat stronger in the 2023 DESI spectrum;
\sii\ is also now significantly detected, although very weak.

We measure broad components to the Balmer lines and \heii\ with $\mathrm{FWHM}\approx1500$~km\,s$^{-1}$; \heii\ is considerably narrower than the $2500\pm100$~km\,s$^{-1}$ reported by \cite{Koljonen24}.
The narrow components of these lines have similar widths $\mathrm{FWHM}\approx170$--270~km\,s$^{-1}$, narrower than the weak \oiii~$\lambda5007$ line ($\mathrm{FWHM}=449\pm40$~km\,s$^{-1}$).  
Given the relatively large uncertainties, the widths of the coronal lines are consistent with that of \oiii~$\lambda5007$, meaning there is no strong evidence for ionisation stratification of the emission line gas.
The very weak \sii\ doublet appears to be much narrower than all other lines measured, with $\mathrm{FWHM}=65\pm11$~km\,s$^{-1}$.

\subsubsection{AT\,2018dyk}
Emission line maps for AT\,2018dyk were previously presented by \cite{Clark25}.
The pre-flare SDSS and MaNGA\footnote{The MaNGA spectrum used here is extracted from the spaxel centred on the nucleus of the galaxy. At $z=0.036$, the hypotenuse of a spaxel ($0.707^{\prime\prime}$) corresponds to 0.515~kpc.} spectra (from 2002 and 2017, respectively) show very weak, narrow \ha\ emission as well as \oii, \nii, \sii, and weak \oiii.
The linewidths are all similar ($\mathrm{FWHM}\approx200$--300~km\,s$^{-1}$) and consistent between the two spectra.

Strong Fe coronal lines, including \fevii~$\lambda3758$, 5720, 6087, \fex~$\lambda6374$, \fexi~$\lambda7891$, and \fexiv~$\lambda5302$ are seen in the Keck spectrum from 2018 August, taken 19~d after the optical peak.
\hb\ and \heii\ emission is also newly seen in the Keck spectrum.
Their profiles have $\mathrm{FWHM}\approx1000$~km\,s$^{-1}$, much broader than narrow \ha\ in the earlier spectra ($\mathrm{FWHM}\approx300$~km\,s$^{-1}$), and we are unable to distinguish separate broad and narrow components.
The coronal lines are all substantially broader than the low-ionisation forbidden lines, with the highest-ionisation \fex--\fexiv\ lines being the broadest of all.

As noted by \cite{Clark25}, the 2023 DESI spectrum closely resembles the pre-outburst SDSS spectrum, with a similar continuum shape and the absence of either coronal lines or broad emission lines.
Whilst the Balmer, \heii, \nii\ and \sii\ lines all return to their pre-outburst strengths, \oiii\ emission remains much stronger than it was in 2002 and 2017.
Linewidths generally remain consistent between the spectra
(narrow \ha\ is anomalously broad in the Keck spectrum, likely because the relatively narrow broad component has confounded our fit).

\subsubsection{AT\,2017gge}
As described in Section~\ref{sec:sample:at2017gge}, coronal lines appeared late ($>200$~d) following the optical peak.
Here we analyse a subset of the spectra presented by \cite{Onori22}, together with an archival pre-outburst SDSS spectrum.
The 2004 SDSS spectrum shows narrow Balmer, \nii, \sii, \oii, and weak \oiii\ emission lines, all with widths $\mathrm{FWHM}\approx200$~km\,s$^{-1}$, providing a pre-outburst baseline.
In the first post-outburst (phase 408~d) Gemini/GMOS spectrum we see the coronal lines \fevii~$\lambda5720,6087$, \fex~$\lambda6374$, and \fexiv~$\lambda5302$ (although we do not obtain a reliable measurement of the profile of the latter).
Shallow broad components to the H and He lines are also visible.
These same lines are visible in the next Gemini/GMOS spectrum taken 164~d later.
The luminosity of \oiii~$\lambda5007$ increases progressively throughout these observations, from $\log(L)=39.3\pm0.1$ through $39.76\pm0.02$ to $40.24\pm0.02$.

The longer wavelength coverage of X-shooter allows the \nev~$\lambda\lambda3345, 3425$ and \fexi~$\lambda7891$ lines to be detected in the 2022 March spectrum (phase 1698~d).
\fexiv~$\lambda5302$ has disappeared by this time whilst the \fevii\ and \fex\ lines remain, and the broad lines are much weaker.
In the later 2022 May DESI spectrum the \fevii\ lines are still present, but \fex\ and \fexi\ are not (\nev~$\lambda3425$ is not robustly detected at the blue end of the spectrum). 
\cite{Onori22} reported the narrowing over time of the broad Balmer lines, which we see in the spectra studied here.
After correcting for instrumental broadening, we find that the \fevii\ and \fex\ coronal lines, \oiii\ and \sii\ forbidden lines, and narrow H and He lines are all broader in the 2018--19 Gemini spectra than in the 2022 X-shooter and DESI spectra.
Unusually, we see that the coronal line profiles tend to be narrower than both \oiii\ and \sii, so there is no strong evidence of the usual decrease of inferred gas distance with ionisation potential.

\subsubsection{TDE\,2021qth}
The early-time Palomar spectrum of TDE\,2021qth (taken 27~d after the transient optical peak) contains narrow Balmer lines and broad \ha.
\oiii\ and \sii\ emission are not significantly detected, and no coronal lines are seen.
In contrast, strong \oiii\ and coronal emission lines are seen in the late-time Keck spectrum taken at a phase of 300~d.
\fex~$\lambda6374$ is broader than \oiii~$\lambda5007$, although the other coronal lines \nev~$\lambda3425$, \fevii~$\lambda5720$, 6087, and \fexi~$\lambda7891$ have widths that are consistent with it within the uncertainties.
Evidence for gas stratification by ionisation potential is tentative at best in this source.


\begin{figure*}
\centering
\setlength{\tabcolsep}{4pt} %
\renewcommand{\arraystretch}{1.0} %

\begin{tabular}{@{}cc@{}}
\includegraphics[width=0.45\textwidth]{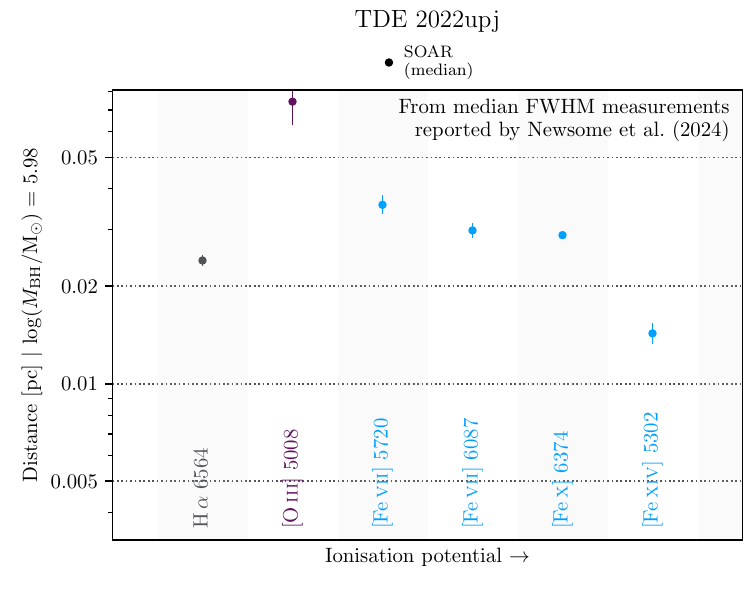} & 
\includegraphics[width=0.45\textwidth]{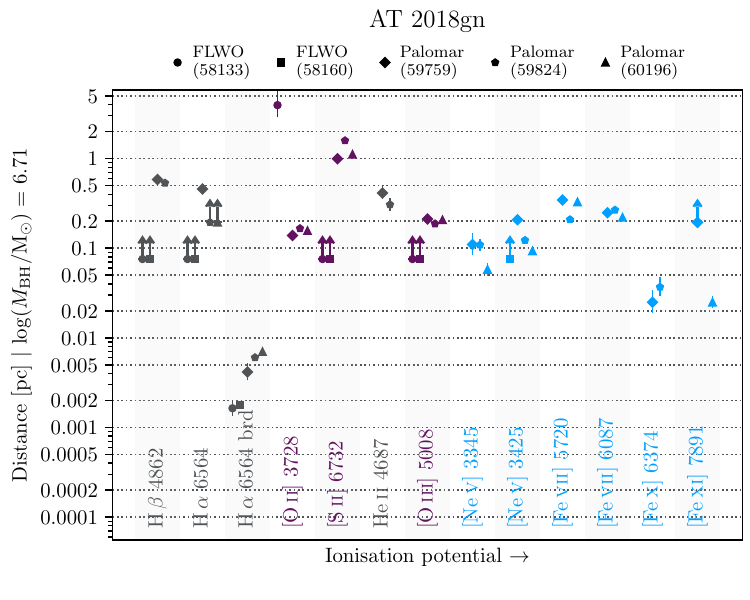} \\ 
\includegraphics[width=0.45\textwidth]{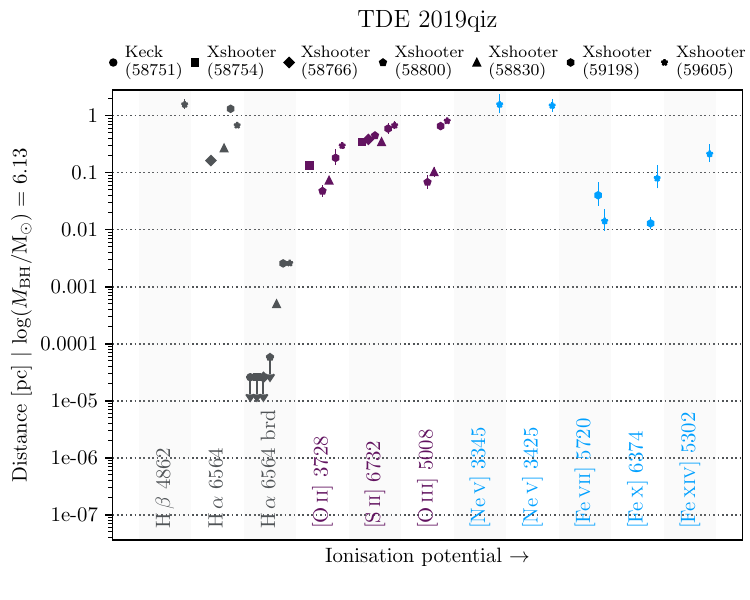} & 
\includegraphics[width=0.45\textwidth]{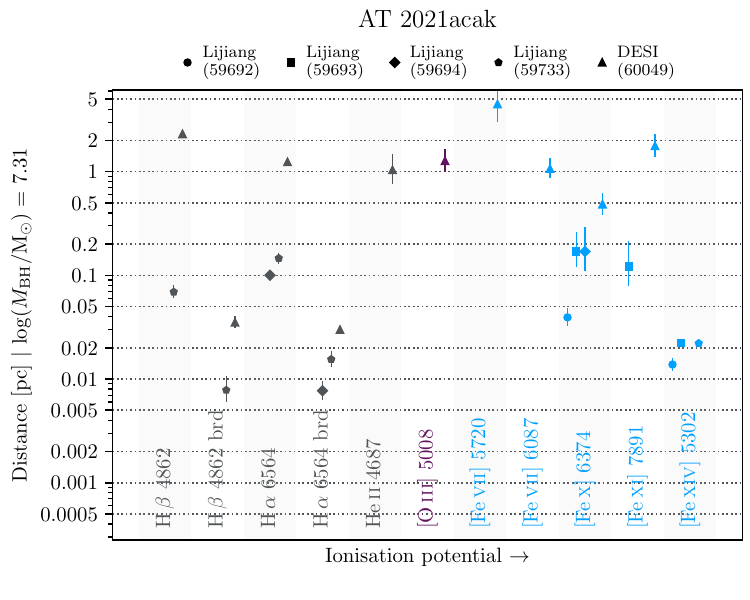} \\ 
\includegraphics[width=0.45\textwidth]{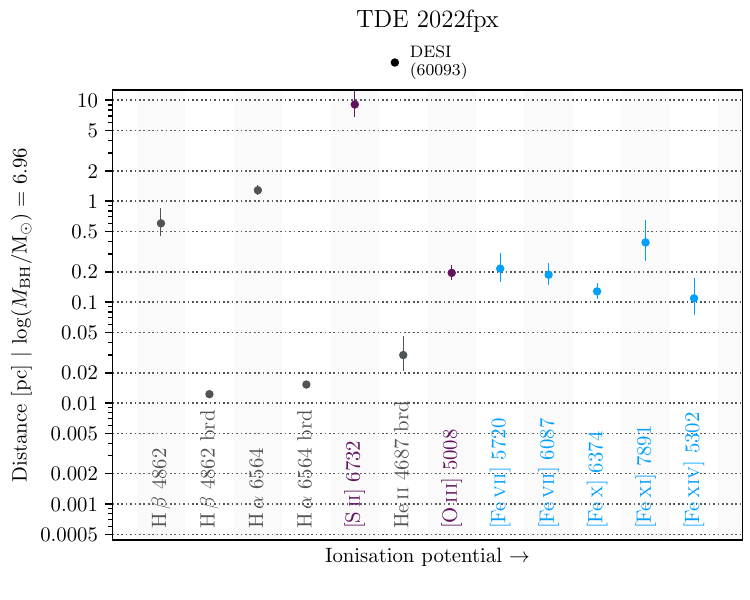} & 
\includegraphics[width=0.45\textwidth]{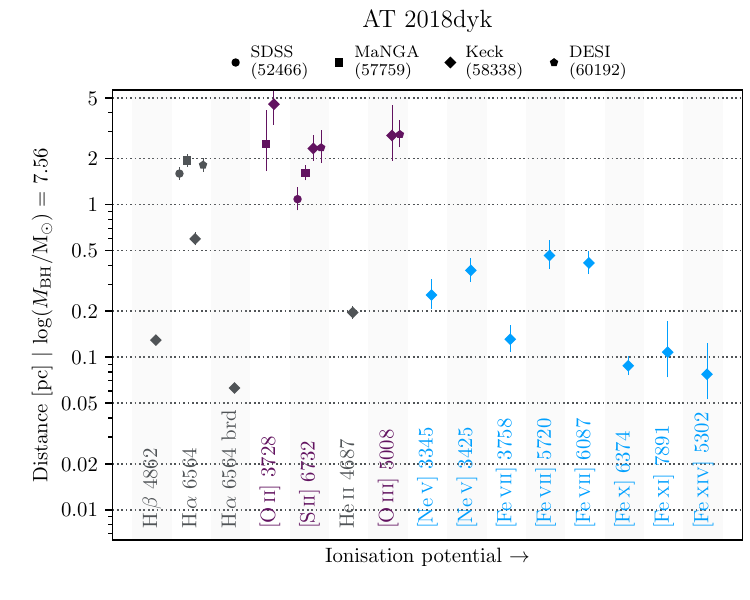} \\  
\end{tabular}

\caption{Emission line maps for CL-TDEs.
         We show the FWHM-inferred distances of emission line gas versus ionisation potential for each emission line. 
         Transitions are ordered by increasing ionisation potential along the x-axis. 
         Permitted transitions are coloured grey, low-ionisation forbidden lines purple, and coronal lines cyan.
         Where multiple epochs are available, measurements are grouped by line and plotted left to right in chronological order, with the observatory distinguished by marker type as indicated in the legend.
         The assumed BH mass ($M_\mathrm{BH}$) used in the distance calculations is given in the y-axis label.}
\label{fig:cl-tde_maps}
\end{figure*}

\begin{figure*}
\addtocounter{figure}{-1} %
\centering
\setlength{\tabcolsep}{4pt}
\renewcommand{\arraystretch}{1.0}

\begin{tabular}{@{}cc@{}}
\includegraphics[width=0.45\textwidth]{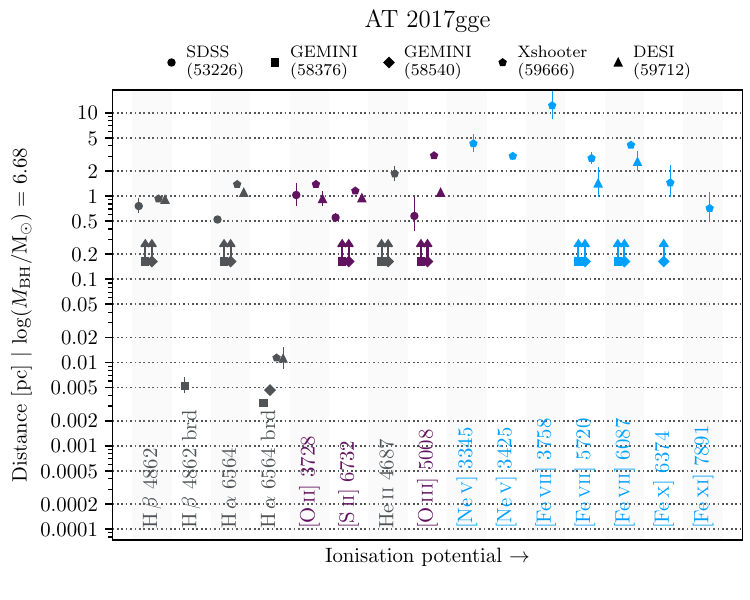} &  
\includegraphics[width=0.45\textwidth]{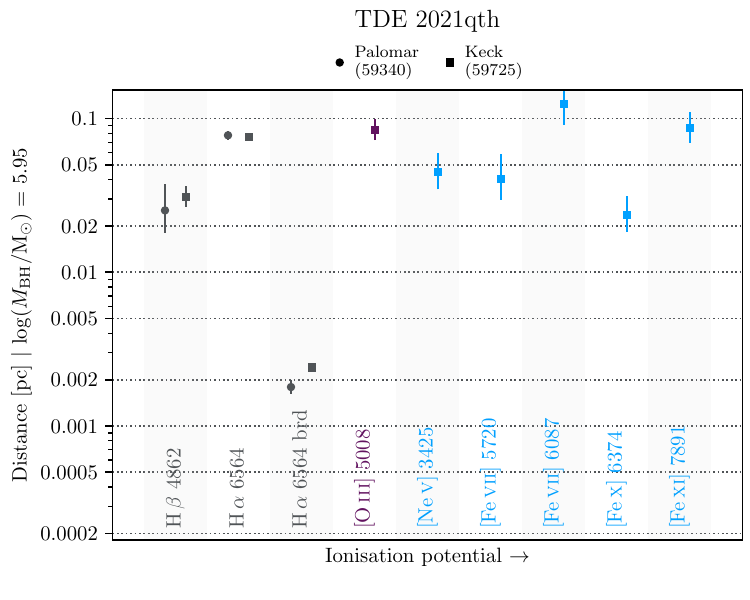} \\ 
\end{tabular}

\caption{Emission line maps for CL-TDEs (continued).}
\end{figure*}


\subsection{Non-variable ECLEs}
\subsubsection{Charizard (DESI\,J027.9525$+$04.1951)}
Narrow Balmer, \oii, \oiii, and \sii\ lines are seen in the DESI discovery spectrum of Charizard in addition to the coronal lines \fevii~$\lambda5720$, 6087, \fex~$\lambda6374$, and \fexi~$\lambda7891$.
The same lines are visible in the follow-up Gemini/GMOS spectrum taken a few months later 
(except for \fex~$\lambda6374$, which \citealt{Clark26} note was subject to contamination in the later spectrum).
There is little change in the emission line profiles over these three months and all of the lines have similar widths $\mathrm{FWHM}\approx100$--200~km\,s$^{-1}$.
We do not therefore find clear evidence of ionisation stratification of the emission line gas.

\subsubsection{Arbok (DESI\,J115.5730$+$39.4366)}
The 2021 DESI discovery spectrum of Arbok is very similar to the archival SDSS spectrum taken in 2010.
The most striking features in the optical spectra are the very strong and broad ($\mathrm{FWHM}\gtrsim10,000$~km\,s$^{-1}$) Balmer lines;
narrow \oiii\ emission is also strong.
Weak \fevii~$\lambda5720$ and 6087 coronal lines are seen, although only \fevii~$\lambda6087$ is detected with high signficance (in the DESI spectrum).
However, \nev~$\lambda3425$ emission is strong and is significantly detected in both spectra, in which its width is consistently $\mathrm{FWHM}\approx900$~km\,s$^{-1}$.
Given the uncertainties, \nev\ is marginally broader than \oiii~$\lambda5007$ ($\mathrm{FWHM}\approx570$~km\,s$^{-1}$), meaning no strong evidence of ionisation stratification is found.

\subsubsection{SDSS\,J0807}
SDSS\,J0807 exhibits coronal lines across a wide range of ionisation potentials in both its 2005 SDSS and 2024 DESI spectra.
There is little change in line widths over the 19~yr between observations, although the broad Balmer lines appear both stronger and narrower in the DESI spectrum.
The coronal lines appear marginally stronger in 2024, whereas the \oiii\ and \sii\ line profiles are very similar in both strength and width in both epochs.
We measure narrow Balmer line components with $\mathrm{FWHM}\approx600$--800~km\,s$^{-1}$, wider than the \oiii\ and \sii\ lines ($\mathrm{FWHM}\approx300$--500~km\,s$^{-1}$), however the broad and narrow components are not clearly distinguished in the observed profiles.
There is a clear stratification of the ionised gas running from \sii\ (ionisation potential of $23$~eV) to \fexiv\ (ionisation potential of $361$~eV).
The coronal line emission arises from scales of $\sim0.1$~pc, intermediate between the broad lines ($\sim0.01$~pc) and low-ionisation narrow lines ($\sim1$~pc).

\subsubsection{SDSS\,J0938}
Similar to SDSS\,J0807, emission from the coronal lines \fevii, \fex, \fexi, and \fexiv\ can be seen in both the 2006 SDSS, 2011 MMT, and 2023 DESI spectra of SDSS\,J0938.
Additionally, the \nev\ lines are accessible and detected in the DESI spectrum.
No broad emission lines are detected in the spectra.  
A broad trend of increasing line width with ionisation potential can be seen; the coronal line profiles have widths of several hundred km\,s$^{-1}$ and are broader than all of the low-ionisation lines (with FWHMs $\approx300$~km\,s$^{-1}$), except \heii~$\lambda4687$ which has a similar width to the coronal lines in all three spectra.
The widths of the lines remain similar across the observations.
Some of the low-ionisation lines are slightly narrower in 2023, but the widths of the coronal lines are broadly consistent within their larger uncertainties.

\subsubsection{SDSS\,J1055}
SDSS\,J1055 also shows \fevii, \fex, and \fexi\ coronal lines, several low-ionisation emission lines, and broad Balmer lines in its 2002 SDSS, 2011 MMT, and 2024 DESI spectra.
Weak \fexiv~$\lambda5302$ emission appears in these spectra, although a reliable measurement is only obtained from the MMT spectrum. 
There is little evolution of the source across the three epochs.
The low-ionisation lines and \fevii\ coronal lines appear narrower in the later spectra, whereas \fex~$\lambda6374$ and \fexi~$\lambda7891$ have consistent widths between SDSS and DESI.
Similar to many other ECLEs, the emission line gas is stratified, with coronal lines being emitted at $\lesssim1$~pc scales, intermediate between the broad lines ($\sim0.01$~pc) and low-ionisation narrow lines ($\gtrsim1$~pc). 

\subsubsection{Sandslash (DESI\,J172.6675$+$50.6179)}
Similar to Arbok, Sandslash was identified by weak \fevii\ emission in its 2021 DESI spectrum, which can also be seen in the archival SDSS spectrum of 2002.
The two spectra are very similar overall, with narrow H and He lines and strong \oiii\ emission seen in both.
We detect \fevii\ lines significantly only in the DESI spectrum, although they appear to be present in the earlier observation as well.
The bluer wavelength coverage of DESI also picks up \nev~$\lambda3425$ with high significance;
its width is consistent with that of \oiii~$\lambda5007$ with both lines having $\mathrm{FWHM}\approx240$~km\,s$^{-1}$, similarly to the \sii\ and narrow H and He lines.
\fex~$\lambda6374$ appears very broad in the SDSS spectrum, although the profile is irregular (the line is not significantly detected in the DESI spectrum).
We do not see a clear ionisation stratification of the emission line gas.

\subsubsection{SDSS\,J1207}
The strengths and shapes of the emission line profiles are remarkably similar between the 2008 SDSS spectrum and the 2025 DESI spectrum. 
The coronal lines all have widths of the order $\sim500$~km\,s$^{-1}$, generally a little narrower than \oiii\ and significantly narrower than \sii\ ($\mathrm{FWHM}\approx100$--200~km\,s$^{-1}$).
In addition to shallow, broad ($\mathrm{FWHM}\approx2500$~km\,s$^{-1}$) components, the Balmer lines also appear to have `narrow' components with greater widths than the other low-ionisation lines ($\mathrm{FWHM}\sim1000$~km\,s$^{-1}$).  
The \nii\ doublet is not clearly seen in the \ha\ profile, and the broad and narrow components are not clearly distinguished, meaning the decomposition and profile measurements are questionable.
There is no strong evidence of ionisation stratification of the emission line gas in this source.

\subsubsection{SDSS\,J1238}
SDSS\,J1238 is one of the most distant sources in our sample, so its spectra have relatively low S/N, making the detection and measurement of emission lines more challenging.
SDSS\,J1238 exhibits both broad and narrow permitted lines in addition to prominent \oiii\ and \fevii\ lines.
The broad Balmer lines appear weaker in the DESI spectrum, whereas the profiles of the \oii, \oiii, and \fevii~$\lambda5720$ and 6087 lines are very similar in strength and shape.
There is no clear ionisation stratification of the emission line gas.
The two strongest coronal lines, \fevii~$\lambda5720$ and 6087, are detected in both spectra and have widths $\approx400$~km\,s$^{-1}$, a little narrower than \oiii\ ($\mathrm{FWHM}\approx650$~km\,s$^{-1}$) and the narrow Balmer line components.
The \sii\ doublet is detected in both spectra, although we only obtain a reliable measurement of its width in the DESI spectrum: $\mathrm{FWHM}=313\pm52$~km\,s$^{-1}$. 
\nev~$\lambda3345$ and \fevii~$\lambda3758$ are possibly seen weakly in the SDSS spectrum and are broader ($\mathrm{FWHM}\approx900$~km\,s$^{-1}$, although with large uncertainties).

\subsubsection{SDSS\,J1247}
SDSS\,J1247 has a rich emission line spectrum showing both broad and narrow H and He lines, low-ionisation forbidden lines and coronal lines of \fevii, \fex, and weak \fexi. 
\nev\ is also present in the 2023 DESI spectrum, which covers shorter wavelengths.
Weak \arxiv~$\lambda4414$ is detected in the DESI spectrum (its luminosity is $\log[L]=40.0\pm0.1$), but only tentatively in the 2006 SDSS spectrum ($\log[L]=39.9\pm0.2$).
Broad \heii\ emission is approximately 50~per cent stronger in the DESI spectrum, but otherwise the line shapes and profiles are very similar between the two epochs.

The strong coronal lines (\nev~$\lambda3425$, \fevii~$3758$, 5720 and 6087) all have similar widths and are broader than the low-ionisation \sii\ and \oiii\ lines.
However, the narrow Balmer and \heii\ lines, in addition to \neiii~$\lambda3968$, have line widths equal to or broader than the coronal lines. 
Unlike sources where a clear trend between line width and ionisation potential is observed, the circumnuclear gas in SDSS J1247 does not therefore show a well-defined stratification.

\subsubsection{SDSS\,J1402}
The 2007 SDSS spectrum of SDSS\,J1402 shows both broad and narrow H and He emission lines and strong \fevii\ coronal lines, as noted by \cite{Callow24}.
In addition to these, we also find strong \nev\ emission lines. 
Weak \fex~$\lambda6374$ ($\log[L]=40.6\pm0.1$) is seen in the 2007 SDSS spectrum but not the 2023 DESI spectrum where, if present, it is lost in noise.
The low-ionisation lines \oii, \neiii, and \sii\ are very weak (consequently the line widths have large uncertainties), although the \oiii\ emission is strong.
The coronal lines all have widths $\mathrm{FWHM}\approx500$~km\,s$^{-1}$, broader than \oiii~$\lambda5007$ ($\mathrm{FWHM}\approx400$~km\,s$^{-1}$) and the other low-ionisation forbidden lines.
We determine that the emission line gas is stratified by ionisiation potential.
The \hb\ profile is fairly noisy in the DESI spectrum and our fit includes an anomalously narrow component, possibly attempting to fit a noise spike.
Overall, the emission line profiles are very similar in the two spectra.

\subsubsection{Nidoqueen (DESI\,J212.9556$+$52.7676)}
Similar to Arbok and Sandslash, an archival SDSS spectrum of Nidoqueen was found for comparison with its 2021 DESI discovery spectrum.
Again, the two spectra are found to be very similar in spite of the nearly two decades between observations.
Both show clear \oiii\ and \ha\ emission lines, the latter having a broad component with FWHM of a few thousand km\,s$^{-1}$.
As noted by \cite{Clark26}, the \fevii\ coronal lines are weak (we detect only \fevii~$\lambda6087$ in the SDSS spectrum), but strong \nev~$\lambda3425$ is detected in the DESI spectrum.
We do not find evidence for ionisation stratification of the emission line gas; in the DESI spectrum all narrow-line widths are consistent with only \fevii~$\lambda6087$ being marginally narrower than the others.

\subsubsection{SDSS\,J1458}
The 2008 SDSS and 2021 DESI spectra of SDSS\,J1458 both show strong \fevii~$\lambda3758$ and 6087 coronal lines and Balmer lines with both broad and narrow components.
\heii\ emission is also present, but it is not strong enough for us to measure its profile.
Weak \fevii~$\lambda5720$ ($\log[L]=40.5\pm0.1$) is detected in the DESI spectrum, but not in the SDSS spectrum.
Weak \nev~$\lambda3425$ is detected in both spectra and very weak \fex~$\lambda6374$ ($\log[L]=40.7\pm0.2$) is marginally detected in the SDSS spectrum, but not in DESI.\footnote{\cite{Callow24} noted that \fex\ was coincident with an optical skyline.}
There is no obvious ionisation stratification of the gas in SDSS\,J1458; the widths of all narrow lines are similar in both spectra, except for the \sii\ doublet which is significantly narrower than the others.
Broad \hb\ appears to be narrower in the DESI spectrum, but the \ha\ and \nii\ blend looks very similar in shape and strength.
The emission line profiles are very similar in the two spectra.

\subsubsection{SDSS\,J1459}
The 2004 SDSS and 2022 DESI spectra of SDSS\,J1459 both contain the \nev~$\lambda3425$ and \fevii~$\lambda3758$, 5720, and 6087 coronal lines in addition to Balmer lines with both broad and narrow components.
Weak \fex~$\lambda6374$ and \fexi~$\lambda7891$ are possibly seen in the DESI spectrum but not in the SDSS spectrum.

Narrow features in the residuals indicate that our procedure has not fit the narrow \ha\ and \hb\ components well in the DESI spectrum, and these may be narrower than implied by our results.
In the SDSS spectrum, narrow \ha\ is found to have $\mathrm{FWHM}=198\pm9$~km\,s$^{-1}$.
Generally, the emission line gas is stratified, with the coronal lines all being broader than the low-ionisation lines.
The strong \nev\ and \fevii\ coronal lines have similar widths $\mathrm{FWHM}\approx500$--600~km\,s$^{-1}$;
if genuine, the higher-ionisation \fex~$\lambda6374$ and \fexi~$\lambda7891$ lines are much broader with $\mathrm{FWHM}\approx1500$~km\,s$^{-1}$.
The broad permitted emission lines appear slightly stronger in the DESI spectrum, otherwise the line profiles are very similar.

\subsubsection{SDSS\,J1715}
SDSS\,J1715 was identified by \cite{Callow24} on the basis of strong \fevii~$\lambda6087$ emission in its 2000 SDSS spectrum, in addition to prominent \oiii\ emission and broad Balmer lines.
Its recent (2025) DESI spectrum, presented here for the first time, is very similar and shows those same features.  
We confidently detect \fevii~$\lambda5720$ in both spectra, although \fevii~$\lambda3758$ emission is uncertain.

The narrow Balmer line components are more distinct from the broad components in the SDSS spectrum than in the DESI spectrum, in which the inflection is less clear.
Our fit procedure erroneously fits an unresolved line to a noise spike at the location of \heii\ in the SDSS spectrum.  
Very weak \heii\ ($\log[L]=40.5\pm0.2$) is marginally detected in the DESI spectrum.

Given the weakness of the low-ionisation narrow lines and the scarcity of strong coronal lines, it is difficult to ascertain a convincing trend of line width with ionisation potential.
However, the strong coronal line \fevii~$\lambda6087$ has widths $615\pm103$ and $559\pm83$~km\,s$^{-1}$ in the SDSS and DESI spectra, respectively;
these are broader than the widths measured for \oiii~$\lambda5007$ ($457\pm18$ and $438\pm16$~km\,s$^{-1}$).
Given the scarcity of the measurements, it is unclear whether ionisation stratification of the emission line gas is present in this source.

\subsubsection{SDSS\,J2220}
We confirm the appearance of \fevii\ coronal lines in the optical spectra of SDSS\,J2220 reported by \cite{Callow24}.
\fevii~$\lambda6087$ is strong and clearly detected in both 2001 SDSS and 2022 DESI spectra.
The other \fevii\ lines are weak: \fevii~$\lambda5720$ is only marginally detected in the SDSS spectrum
and \fevii~$\lambda3758$ is only marginally detected in both spectra.
\fexi~$\lambda7891$ is significantly detected at the red end of the DESI spectrum only.
As noted by \cite{Callow24}, the \oiii~$\lambda5007$ line increases slightly in luminosity: here we measure $\log(L)=41.20\pm0.04$ in SDSS and $41.27\pm0.02$ in DESI.
In contrast, \fevii~$\lambda6087$ fades from $\log(L)=40.75\pm0.05$ to $40.51\pm0.06$.
The widths of the lines remain consistent between the two spectra, with \fevii~$\lambda6087$ being broader than \oiii\ and the weak \sii\ doublet, implying ionisation stratification of the emission line gas.

As with some of our other sources, the broad and narrow Balmer lines are not well-separated.
A very narrow emission line in the \hb\ profile (probably emission from a nearby star falling within the SDSS / DESI fibres) has confused the fit in the SDSS spectrum.


\begin{figure*}
\centering
\setlength{\tabcolsep}{4pt} %
\renewcommand{\arraystretch}{1.0} %

\begin{tabular}{@{}cc@{}}
\includegraphics[width=0.45\textwidth]{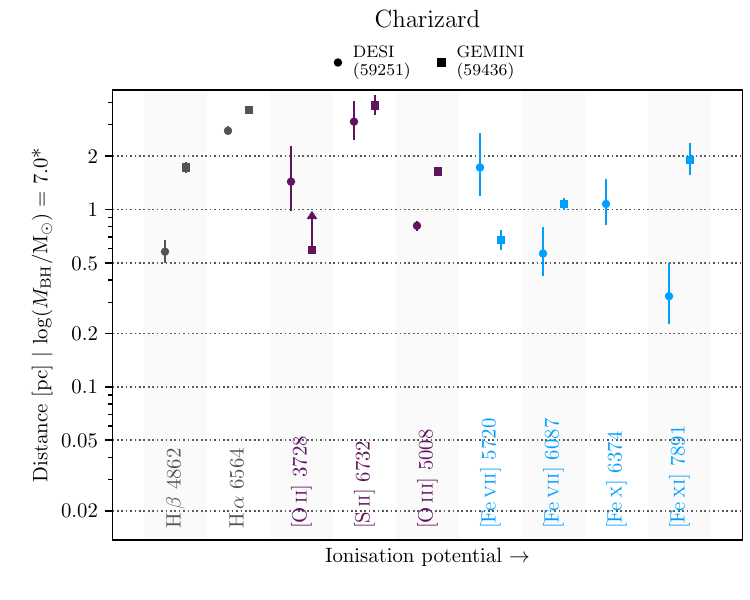} &  
\includegraphics[width=0.45\textwidth]{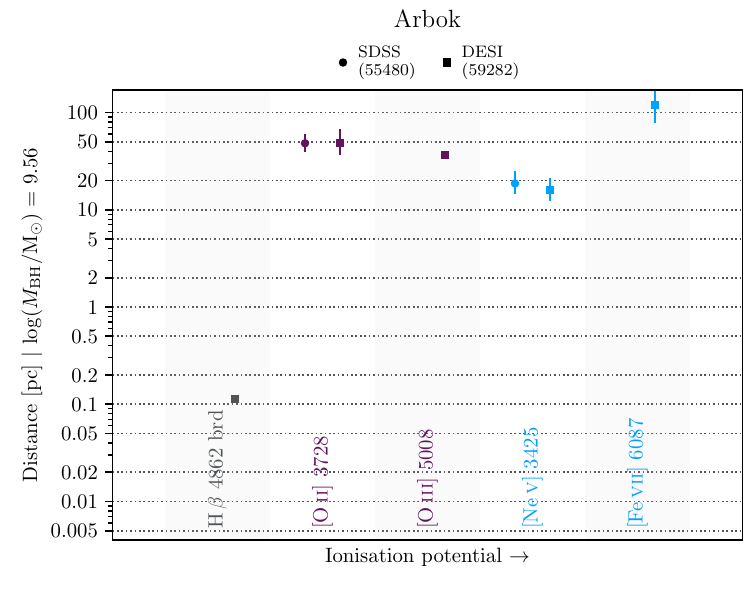} \\ 
\includegraphics[width=0.45\textwidth]{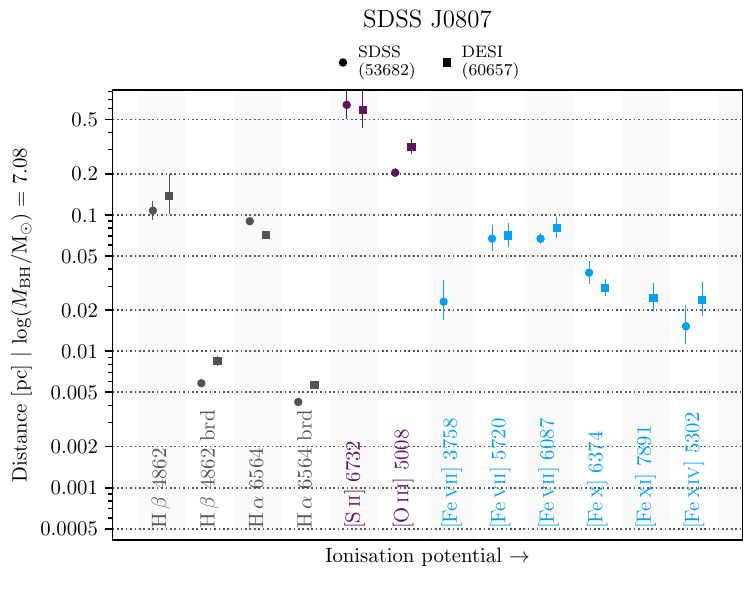} &  
\includegraphics[width=0.45\textwidth]{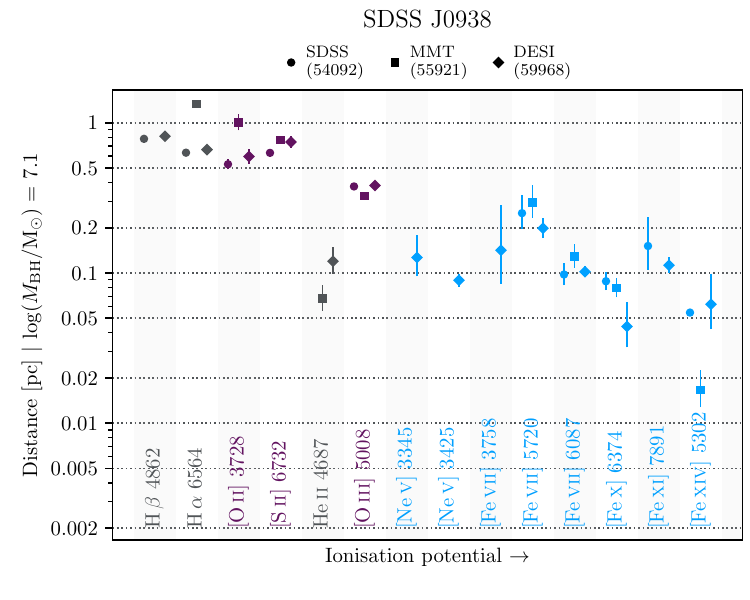} \\ 
\includegraphics[width=0.45\textwidth]{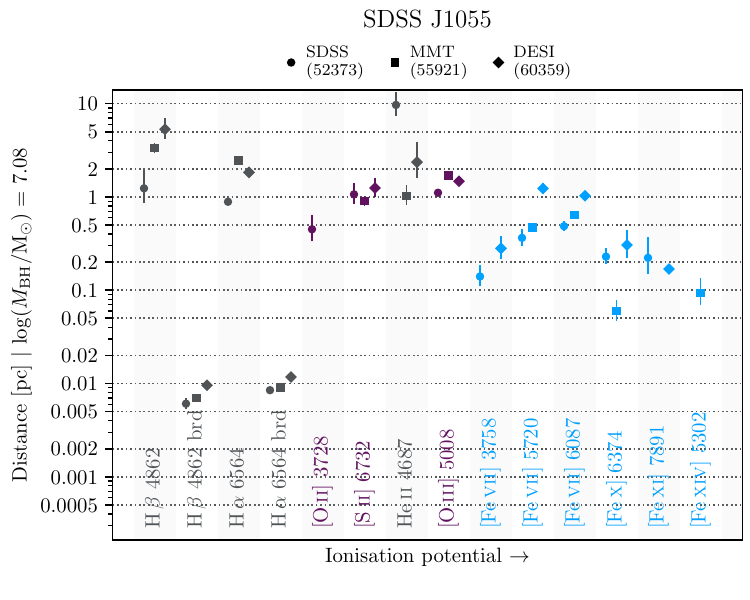} &  
\includegraphics[width=0.45\textwidth]{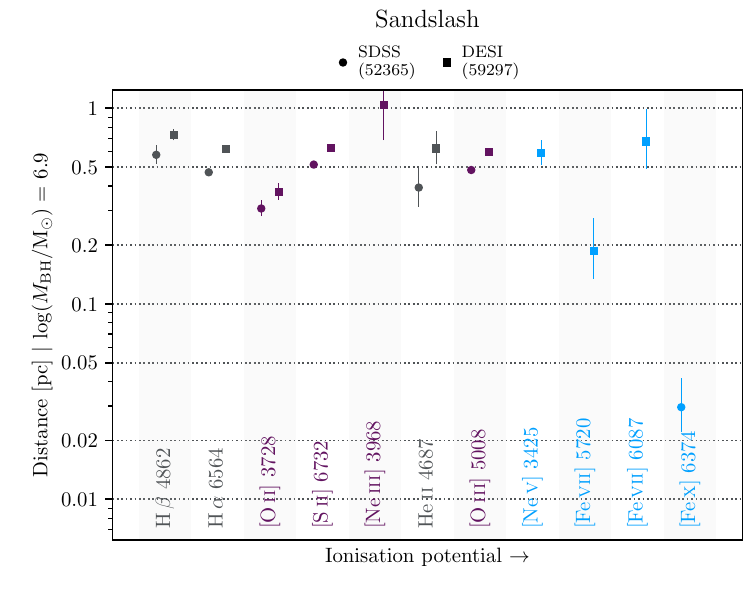} \\ 

\end{tabular}

\caption{Emission line maps for nv-ECLEs.
         We show the FWHM-inferred distances of emission line gas versus ionisation potential for each emission line. 
         Transitions are ordered by increasing ionisation potential along the x-axis. 
         Permitted transitions are coloured grey, low-ionisation forbidden lines purple, and coronal lines cyan.
         Where multiple epochs are available, measurements are grouped by line and plotted left to right in chronological order, with the observatory distinguished by marker type as indicated in the legend.
         The assumed BH mass ($M_\mathrm{BH}$) used in the calculation of the distances is given in the y-axis label.}
\label{fig:nonvar_ecle_maps}
\end{figure*}

\begin{figure*}
\addtocounter{figure}{-1} %
\centering
\setlength{\tabcolsep}{4pt}
\renewcommand{\arraystretch}{1.0}

\begin{tabular}{@{}cc@{}}
\includegraphics[width=0.45\textwidth]{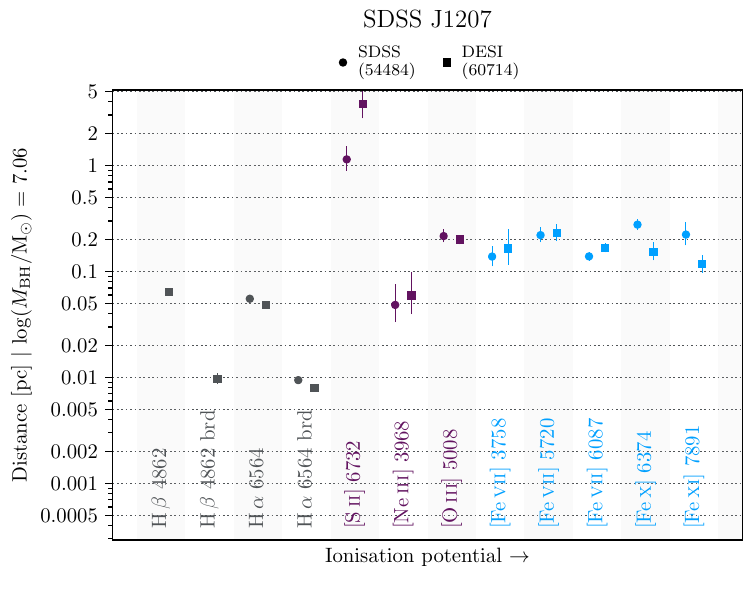} &  
\includegraphics[width=0.45\textwidth]{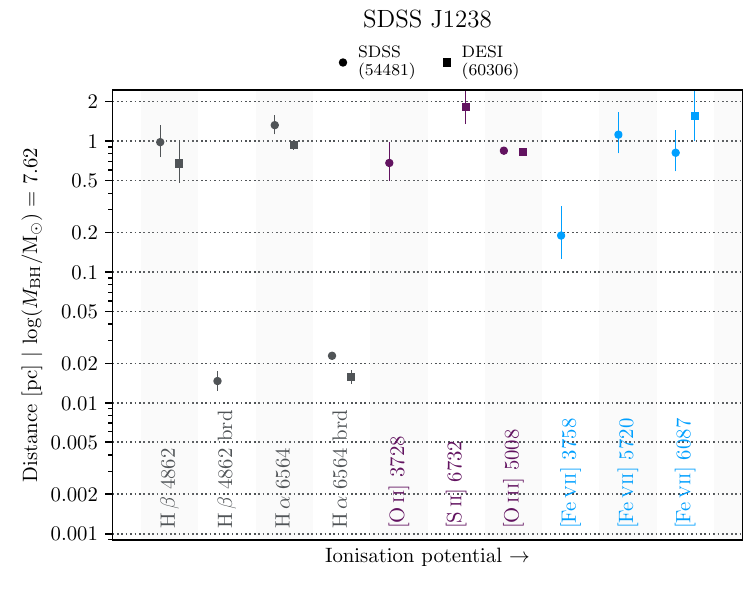} \\ 
\includegraphics[width=0.45\textwidth]{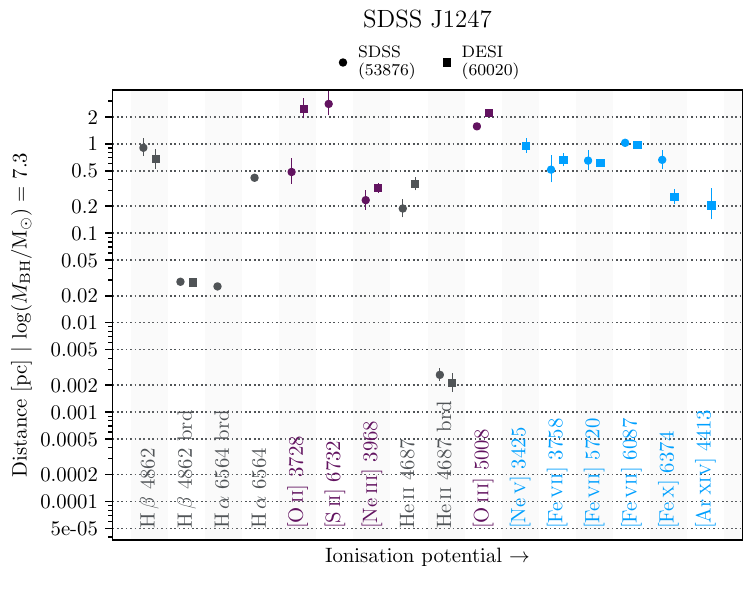} &  
\includegraphics[width=0.45\textwidth]{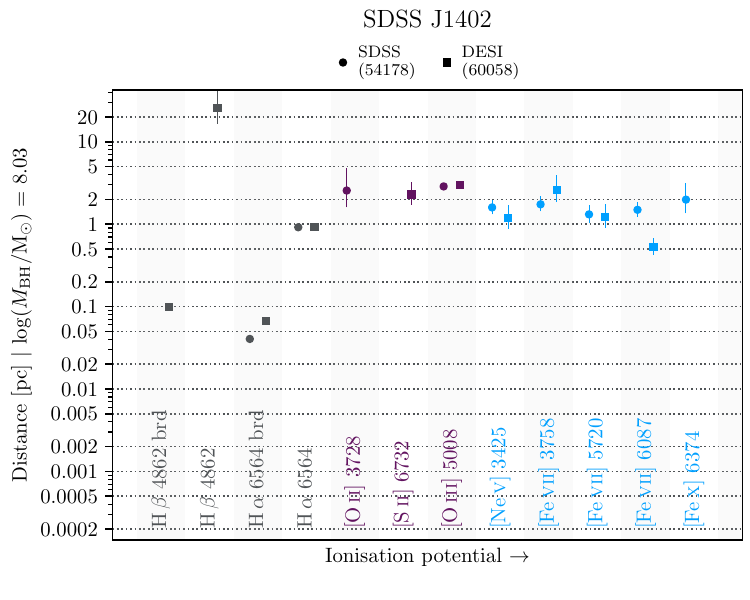} \\ 
\includegraphics[width=0.45\textwidth]{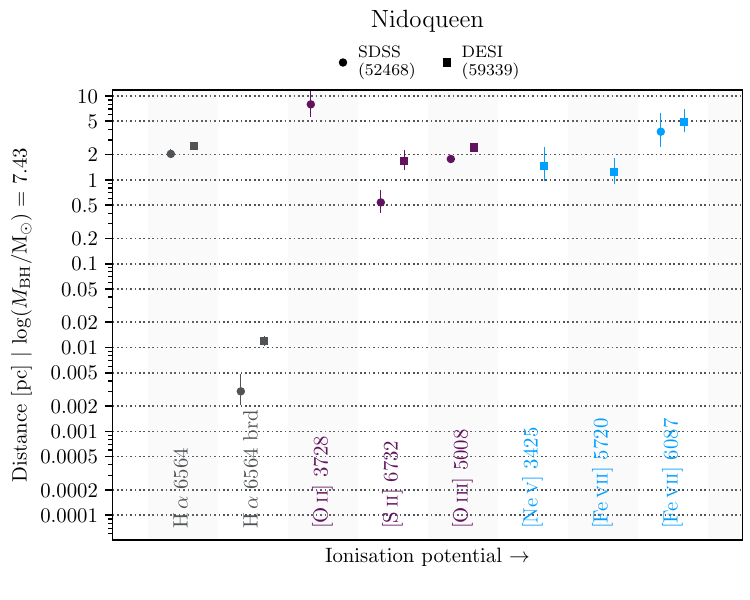} &  
\includegraphics[width=0.45\textwidth]{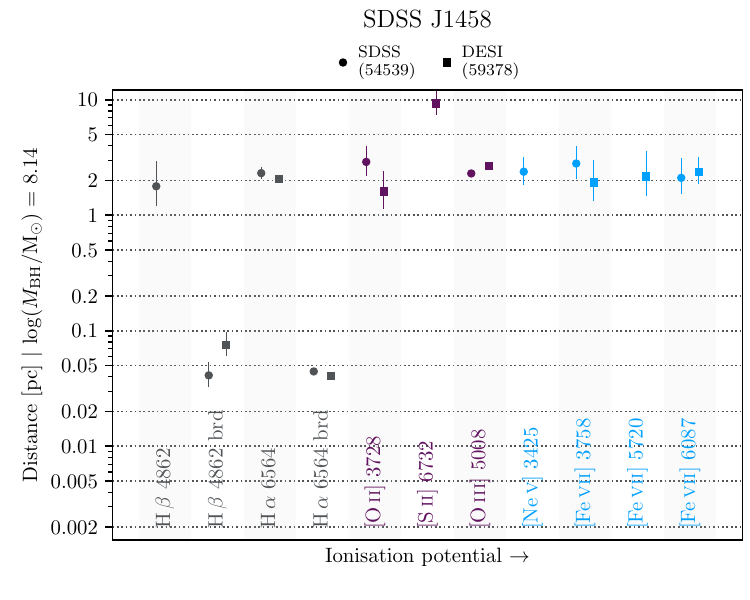} \\ 

\end{tabular}

\caption{Emission line maps for nv-ECLEs (continued).}
\end{figure*}

\begin{figure*}
\addtocounter{figure}{-1} %
\centering
\setlength{\tabcolsep}{4pt}
\renewcommand{\arraystretch}{1.0}

\begin{tabular}{@{}cc@{}}
\includegraphics[width=0.45\textwidth]{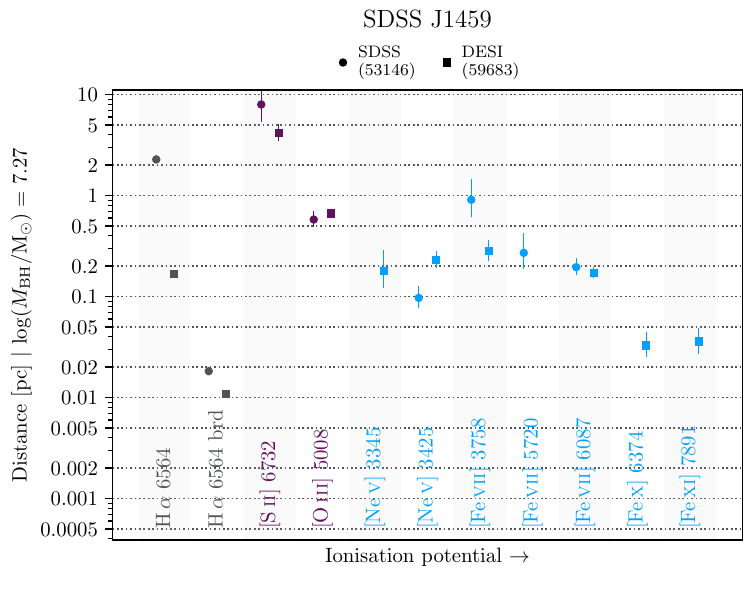} &  
\includegraphics[width=0.45\textwidth]{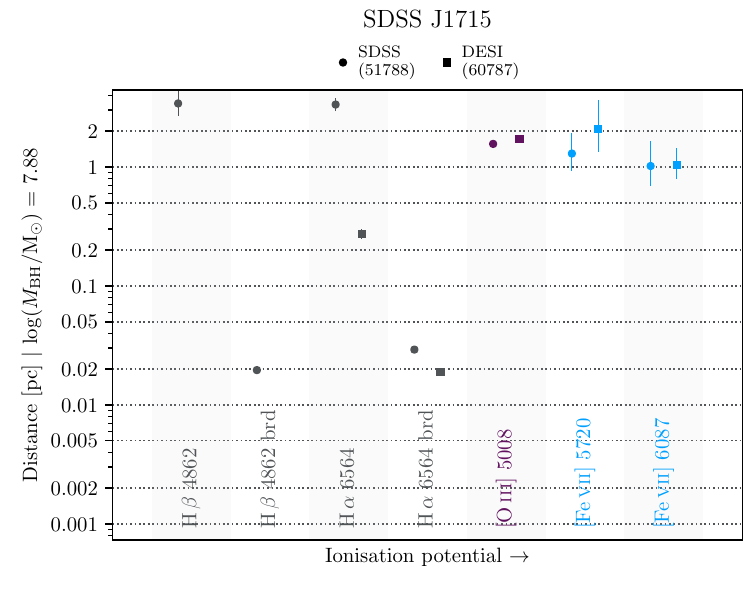} \\ 
\includegraphics[width=0.45\textwidth]{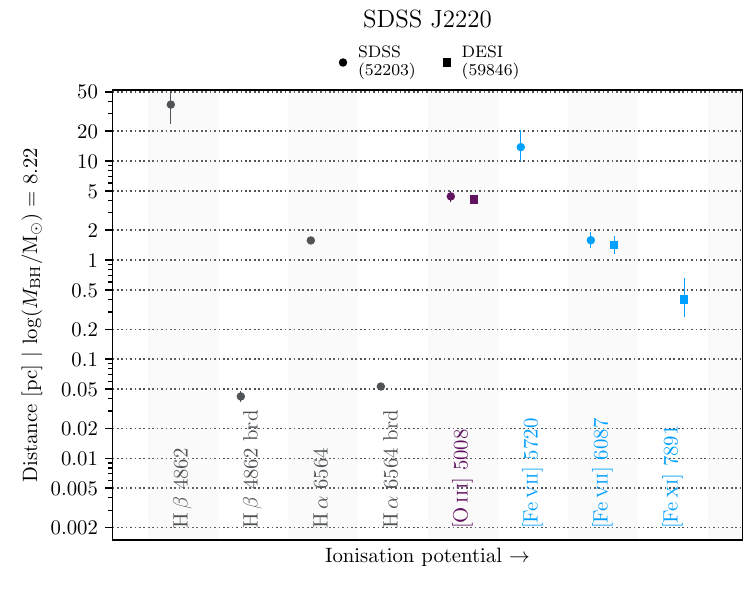} &  \\ 
\end{tabular}

\caption{Emission line maps for nv-ECLEs (continued).}
\end{figure*}


\begin{table*}
    \centering
    \caption{Summary of spectroscopic analysis}
    \begin{tabular}{llp{1.3in}p{1.8in}p{1.8in}}
    \hline
         Source & Stratification & Coronal lines detected & Coronal line evolution & Low-ionisation line evolution \\
    \hline
        \multicolumn{5}{c}{Variable sources: v-ECLEs} \\

        SDSS\,J0113  & Yes     & \nev, \fevii, \fexi\               & Disappear between 2012 \& 2023 & Broad \ha\ disappears, \oiii\ weakens \\ 
        SDSS\,J0748  & Yes     & \fex, \fexi, \fexiv, \arxiv\       & Disappear between 2004 and 2011 & \oiii\ strengthens between 2004 and 2011, still strong in 2025; Balmer lines weaken \\
        SDSS\,J0952  & Yes     & \fevii, \fex\ \fexi, \fexiv, \sxii & Strong in 2005; only weak \fevii\ remains by 2011 & Broad \ha\ disappears and \oii\ strengthens between 2005 and 2011 \\
        SDSS\,J1241  & Unclear & \fevii, \fex, \fexi\                & Coronal lines weaker but still present $\approx20$~yr post-discovery & Balmer lines weaken and \oiii\ strengthens between 2004 and 2025 \\
        Raichu       & Unclear & \fevii, \fex, \fexi\                & Insufficient data & Insufficient data \\
        SDSS\,J1342  & Unclear & \fevii, \fex, \fexi\                & \fex\ and \fexi\ disappear as \fevii\ appears & \oiii\ strengthens between 2002-21 \\
        SDSS\,J1350  & No      & \fevii, \fex, \fexi\                & \fex\ and \fexi\ disappear as \fevii\ appears & Broad lines fade and \oiii\ strengthens between 2006-11 \\
        Raticate     & No      & \fevii, \fex, \fexi\                & Insufficient data & Insufficient data \\
        Pidgeot      & No      & \nev, \fevii\                       & Insufficient data & Insufficient data \\
    \hline    
        \multicolumn{5}{c}{Variable sources: CL-TDEs} \\
        TDE\,2022upj  & Yes     & \fevii, \fex, \fexiv, \sxii\       & \fex\ and \fexiv\ strong in early spectra and weaken; \fevii\ lines weak in early spectra and stronger $\approx400$~d post-peak & \oiii\ appears 400~d post-peak; broad \heii\ fades  \\
        AT\,2018gn  & Yes     & \nev, \fevii, \fex, (\fexi\ ?)     & Most lines appear $\approx4$~yr post-peak, little change seen in following 14~months & \oiii\ strengthens, broad \ha\ weakens and narrows \\
        AT\,2021dms  & Unknown & \fex, \fexi, \fexiv\               & Insufficient data & Insufficient data \\
        TDE\,2019qiz  & Yes     & \nev, \fevii, \fex, \fexi, \fexiv\ & Appear $\approx400$~d post, peak then weaken & Broad \ha\ brightens then fades twice between 2019-22, \oiii\ strengthens then stable over 2020-22 \\
        AT\,2021acak & Unclear & \fevii, \fex, \fexi, \fexiv\ & \fevii\ lines newly appear in 2023 DESI spectrum; \fexiv\ weakens & Broad \ha\ persists between 2022-23; \oiii\ not seen in 2002 spectra, detected in DESI spectrum \\
        TDE\,2022fpx  & No      & \fevii, \fex, \fexi, \fexiv, \sxii\ & \fex--\fexiv, \sxii\ appear near peak; \fevii\ lines detected 310~d post-peak & Broad Balmer and \heii\ lines still present and \oiii\ stronger 310~d post-peak \\
        AT\,2018dyk  & Yes     & \nev, \fevii, \fex, \fexi, \fexiv\  & Disappear between 2018--23 & \oiii\ strengthens between 2018-23 \\
        AT\,2017gge  & No      & \nev, \fevii, \fex, \fexi, \fexiv\  & \fexiv\ disappears by 2022 May; only \fevii\ remains in 2022 March & Broad lines fade and narrow; \oiii\ strengthens \\
        TDE\,2021qth  & No      & \nev, \fevii, \fex, \fexi\          & Seen at $+300$~d but not at $+28$~d & Narrow Balmer lines and broad \ha\ seen at $+28$~d; \oiii\ seen at $+300$~d \\
    \hline
        \multicolumn{5}{c}{Non-variable sources: nv-ECLEs} \\
        Charizard   & No      & \fevii, \fex, \fexi\               & \multicolumn{2}{l}{Little change in coronal or low-ionisation lines between 2021 February and May} \\
        Arbok       & No      & \nev, \fevii\                      & \multicolumn{2}{l}{Little change in coronal or low-ionisation lines between 2010 and 2021} \\
        SDSS\,J0807 & Yes     & \fevii, \fex, \fexi, \fexiv\       & \multicolumn{2}{l}{Little change in coronal or low-ionisation lines between 2005 and 2024} \\
        SDSS\,J0938 & Yes     & \nev, \fevii, \fex, \fexi, \fexiv\ & \multicolumn{2}{l}{Little change in coronal or low-ionisation lines between 2006, 2011, 2023} \\
        SDSS\,J1055 & Yes     & \fevii, \fex, \fexi, \fexiv\       & \multicolumn{2}{l}{Little change in coronal or low-ionisation lines between 2002, 2011, 2024} \\
        Sandslash   & Unclear & \nev, \fevii, \fex\                & \multicolumn{2}{l}{Little change in coronal or low-ionisation lines between 2002 and 2021} \\
        SDSS\,J1207 & Unclear & \fevii, \fex, \fexi\               & \multicolumn{2}{l}{Little change in coronal or low-ionisation lines between 2008 and 2025} \\
        SDSS\,J1238 & No      & \fevii\                            & Similar between 2008 and 2023 & Broad lines shallower in 2023 \\
        SDSS\,J1247 & Yes     & \nev, \fevii, \fex, \arxiv\        & \multicolumn{2}{l}{Little change in coronal or low-ionisation lines between 2006 and 2023} \\
        SDSS\,J1402 & Yes     & \nev, \fevii, \fex\                & \multicolumn{2}{l}{Little change in coronal or low-ionisation lines between 2007 and 2023} \\
        Nidoqueen   & No      & \nev, \fevii\                      & \multicolumn{2}{l}{Little change in coronal or low-ionisation lines between 2002 and 2021} \\
        SDSS\,J1458 & No      & \nev, \fevii\                      & \multicolumn{2}{l}{Little change in coronal or low-ionisation lines between 2008 and 2021} \\
        SDSS\,J1459 & Yes     & \nev, \fevii, \fex, \fexi\         & Similar between 2004 and 2022 & Broad lines stronger in 2022 \\
        SDSS\,J1715 & Unclear & \fevii\                            & Similar between 2004 and 2025 & Broad lines weaker in 2025 \\
        SDSS\,J2220 & Yes     & \fevii, \fexi\                     & Similar between 2001 and 2022 & Broad lines narrower in 2022 \\
    \hline
    \end{tabular}
    \label{tab:specsummary}
\end{table*}

\subsection{Trends in estimated gas distances}\label{sec:dist-trends}
\begin{figure*}
    \centering
    \includegraphics[width=7in]{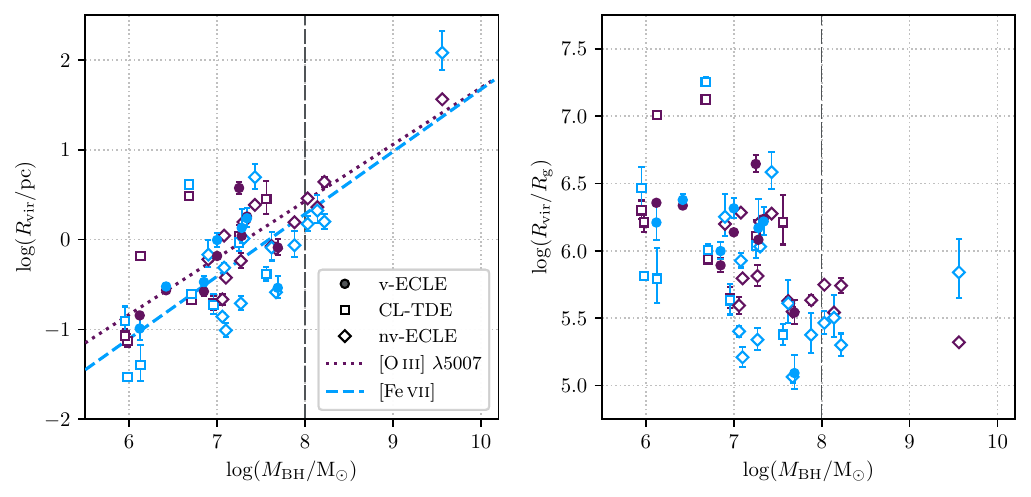}
    \caption{Representative emission line gas distances, inferred from the FWHM of \oiii~$\lambda5007$ (purple) and an \fevii\ coronal line (cyan) for sources in our sample.  
    On the left we show the absolute distances in pc and on the right mass-scaled distances in gravitational radii.
    The different source classifications are represented by different markers, as indicated in the legend.
    Best-fitting linear relations between $\log(\mathrm{Distance}$) and $\log(M_\mathrm{BH})$ for \oiii\ and \fevii\ are shown by purple dotted and cyan dashed lines, respectively.
    The dashed grey vertical line indicates the Hills mass ($10^8~\mathrm{M}_\odot$), above which we do not expect to observe TDEs.}
    \label{fig:dists}
\end{figure*}

Having calculated virial radii $R_\mathrm{vir}$ for the emission lines observed in each source in the sample, we now consider the sample as a whole.
We compare between low-ionisation and coronal line gas radii using distances inferred from \oiii~$\lambda5007$ and \fevii, respectively, since these are amongst the most frequently observed and strongest emission lines in our spectra.
For each source we select robust measurements of the \oiii~$\lambda5007$ and \fevii\ virial radii\footnote{We use \fevii~$\lambda5720$ for TDE\,2019qiz and SDSS\,J1342 and \fevii~$\lambda6087$ for the remaining sources.} and plot these as a function of BH mass in Fig.~\ref{fig:dists} 
(we exclude from this analysis AT\,2021dms, Raticate and Charizard, for which we have no mass estimates). 
Data points for \oiii\ and \fevii\ are coloured purple and cyan, respectively, and the source classifcation is indicated by the marker shape: v-ECLE (circles), CL-TDEs (squares) and nv-ECLEs (diamonds).
For each line, we see that there is a correlation between the absolute gas distance (in pc) and BH mass.
Performing a simple, unweighted, linear regression of the form $\log(R_\mathrm{vir}/\mathrm{pc})=\alpha\log(M_\mathrm{BH}/\mathrm{M_\odot})+\beta$, we find slopes $\alpha_\mathrm{[O{\textsc{iii}}]}=0.63\pm0.08$ and $\alpha_\mathrm{[Fe{\textsc{vii}}]}=0.71\pm0.12$.
Since the measurement uncertainties on the virial radii are small compared to the scatter in the relation, the simple regression yields a chi-squared per degree of freedom of $\chi^{2}_\nu\sim100$-200, indicating that significant intrinsic scatter is present beyond the measurement errors.
We therefore adopt a maximum likelihood approach that simultaneously fits for the slope, intercept, and an intrinsic scatter term $\varepsilon$ added in quadrature with the measurement uncertainties (e.g., \citealt{Tremaine02}).
From these fits, we determine $\alpha_\mathrm{[O{\textsc{iii}}]}=0.63\pm0.08$, $\beta_\mathrm{[O{\textsc{iii}}]}=-4.6\pm0.6$, $\varepsilon_\mathrm{[O{\textsc{iii}}]}=0.32\pm0.04$~dex, $\alpha_\mathrm{[Fe{\textsc{vii}}]}=0.69\pm0.12$, $\beta_\mathrm{[Fe{\textsc{vii}}]}=-5.3\pm0.8$, and $\varepsilon_\mathrm{[Fe{\textsc{vii}}]}=0.44\pm0.06$~dex.
The fit statistics are $\chi^{2}_\nu=1.07$ and 1.06 for \oiii\ and \fevii, respectively, indicating good fits once intrinsic scatter is accounted for.
Each slope is inconsistent with zero at $\gtrsim5\sigma$, confirming that the distance--mass correlation is statistically significant for both emission lines.

In the right-hand panel of Fig.~\ref{fig:dists} we show the relative (BH mass-scaled) distances in gravitational radii, $R_\mathrm{g}$.
Here, an inverse relationship between distance and mass is seen, with emission lines in more massive systems originating at relatively smaller radii.
We address these relations in the following section.

We note that all variable sources have BH masses lower than the Hills mass ($10^8~\mathrm{M}_\odot$), above which we do not expect to observe TDEs \citep{Hills75}.
This provides another line of evidence for associating variable ECLEs with TDEs.

\section{Discussion}\label{sec:disc}
\subsection{Evolution of emission line spectra}
As one would expect, there is a marked difference of behaviour between variable and non-variable ECLEs in terms of the evolution of their emission line spectra.
Of the 14 nv-ECLEs for which we have assessed multiple spectra, none show any major changes on timescales spanning months to two decades between observations.
Where noticeable changes between spectra are present, these are mostly in the widths and strengths of the broad emission lines, which is expected behaviour given the typical $\sim$light-day scales of the BLR and its response to accretion disc continuum variations (e.g., \citealt{Wang20}).
Additionally, the same emission lines (including coronal lines) persist across the observed timescales, in contrast to the majority of v-ECLEs.

The v-ECLEs do generally exhibit spectroscopic changes over months--decades.
One trend we observed is the early appearance and disappearance of the highest-ionisation lines (e.g., \fex--\fexiv) and the later appearance and fading of \fevii\ lines.
We saw this clearly in five of the v-ECLEs (both v-ECLEs and CL-TDEs): SDSS\,J0952, SDSS\,J1342, SDSS\,J1350, TDE\,2022fpx, and AT\,2017gge, and it has also been reported for TDE\,2022upj \citep{Newsome24}.
However, because of the uneven spectroscopic sampling of sources, in addition to the unknown phases of v-ECLE spectra relative to their peaks, we are unable to determine whether similar behaviour does or does not occur in other systems.
In sources such as AT\,2021dms we have observed only the highest-ionisation coronal lines \fex-\fexiv, whereas in Pidgeot we have observed only the lowest-ionisation coronal lines \nev\ and \fevii.

Another very common observation was the strengthening of \oiii\ emission over time, which we saw in 11 v-ECLEs studied here and has again been reported for TDE\,2022upj \citep{Newsome24} as well as Pidgeot \citep{Clark26}.
\cite{Clark26} reported no significant change in the \oiii\ luminosity of Raichu between the 2021 DESI spectrum and the 2025 Keck spectrum.
We are unable to assess changes in \oiii\ luminosity in AT\,2021dms because we know of only one optical spectrum of the source, taken $\approx6$~months after its transient report, that exhibits no \oiii\ emission; follow-up observations of this source may show if \oiii\ has subsequently appeared.

A \textit{weakening} of \oiii\ emission over 11 and 4~y has been reported for SDSS\,J0113 \citep{Callow25} and Raticate \citep{Clark26}, respectively; these objects may have been observed at a later phase of their evolution and are returning to quiescence.

The commonly observed sequence, in which the highest-ionisation coronal lines fade before the appearance or strengthening of lower-ionisation species such as \fevii\ and \oiii, is broadly consistent with the time-dependent photoionisation of circumnuclear gas following a TDE flare. 
Similar coronal line variability has been reported in previous studies of objects in this sample, and has been interpreted as the response of pre-existing gas to the transient ionising continuum (e.g., \citealt{Yang13}; \citealt{Short23}; \citealt{Newsome24}). 
In this picture the evolution may reflect recombination in gas at roughly fixed radius as the ionising continuum fades and softens, the outward propagation of an ionisation echo through a radially stratified medium, or a combination of both.

Among the v-ECLEs, only SDSS\,J1241 and SDSS\,J1342 show evidence of coronal line emission over a baseline of decades.
In the majority of v-ECLEs the coronal lines completely disappear after several years.
SDSS\,J1342 exhibited evolution of its coronal lines, with high-ionisation \fex, \fexi, and \fexiv\ lines appearing in its 2002 SDSS spectrum (but no \fevii\ lines), fading by the time of the 2011 MMT spectrum, in which \fevii\ lines appear \citep{Yang13}.
We confirm here the observation of \cite{Clark24} that \fevii\ emission was still present in the 2021 DESI spectrum of SDSS\,J1342.

SDSS\,J1241 is unusual and has characteristics of both variable and nv-ECLEs.  
As mentioned previously, \cite{Yang13} originally classified the source as a nv-ECLE because \fevii\ emission was still present in a long-term follow up spectrum.
\cite{Clark24} then reclassified it as variable having found no significant coronal line emission in a later follow-up spectrum 
and noting the long-term, monotonic dimming of its MIR luminosity.
Here, in the new (2025) DESI spectrum, we significantly detect \fevii~$\lambda5720$ and make tentative detections of \fevii~$\lambda3758$ and 6087.
The source has some emission line characteristics in common with v-ECLEs mentioned above: the highest-ionisation coronal lines disappear first (\fexi~$\lambda7891$ was strong in the 2004 SDSS spectrum but is not detected in either the 2021 Kast or 2025 DESI spectra) whilst \fevii\ lines remain and \oiii\ emission becomes stronger with time \citep{Yang13}.
The MIR evolution of SDSS\,J1241 is also intermediate between variable and nv-ECLEs.
Whilst showing a progressive decline in its MIR fluxes over $\approx12$~yr of \textit{WISE} observations,
the fading of SDSS\,J1241 was much less pronounced than that of the other four v-ECLEs in that sample (SDSS\,J0748, SDSS\,J0952, SDSS\,J1342, and SDSS\,J1350, which were all more similar to each other than to SDSS\,J1241).\footnote{See fig.~5 of \cite{Clark24}.}
SDSS\,J1241 initially had a $\mathrm{W1}-\mathrm{W2}$ colour of $0.63\pm0.01$~mag (well within the \citealt{Stern12} $\mathrm{W1}-\mathrm{W2}<0.8$~mag `non-AGN' region) and decreasing monotonically thereafter.
Again, this behaviour was more similar to the v-ECLEs that also had declining MIR colour indices than to the nv-ECLEs that maintained a steady AGN-like colour.  
Although, as noted by \cite{Clark24}, the SDSS\,J1241 host galaxy contribution to the W1 band is poorly-constrained, which impacts the assessment of both its nuclear flux and colour evolution.

Given the sparse spectroscopic sampling of these sources it is unclear whether the later detections of coronal lines reflect sustained emission over decades or a re-emergence of high-ionisation emission lines.
Quasi-periodic eruptions (QPEs) have been reported in two v-ECLEs, TDE\,2019qiz \citep{Nicholl24} and TDE\,2022upj \citep{Chakraborty25}, suggesting that the bright ionising continuum can recur.
Furthermore, \cite{Chakraborty25} noted that two of the three known optical TDE QPE hosts are ECLEs and that this is unlikely to be just a coincidence, rather this points to a physical link between the ECLE and QPE phenomena.
However, higher-cadence optical and MIR photometry of SDSS\,J1241 and SDSS\,J1342 do not reveal any strong rebrightening episodes \citep{Clark24}.

A related question is whether any of the v-ECLEs occurred in galaxies that already hosted an active nucleus. 
Pre-existing AGN activity can complicate TDE identification, since both phenomena produce broad permitted lines and blue UV/optical continua. 
In this sample, AT\,2018dyk was initially classified as a changing-look LINER \citep{Frederick19} before its TDE nature was established \citep{Huang23, Clark25}.
\cite{Li23} have proposed that AT\,2021acak was a TDE ocurring in an AGN host galaxy.
Additionally, AT\,2018bcb shares characteristics with both TDE-like and AGN-like nuclear variability \citep{Neustadt20}.
Multiwavelength follow-up could help to disentangle potential TDE/AGN emission in these objects.

\subsection{Location and stratification of the emission line gas}
\subsubsection{Location}
The few previous studies that mapped the circumnuclear gas of CL-TDEs found that the coronal line region was located on $<\mathrm{pc}$ scales, intermediate between the low-ionisation NLR and the BLR (\citealt{Short23}; \citealt{Newsome24}; \citealt{Clark25}).
Furthermore, \cite{Clark25} suggested that the higher BH mass of AT\,2018dyk compared with that of TDE\,2022upj may explain why the coronal line region of the former was larger than the latter.
They found the ratio of virial radii, 0.9--3.6, was much lower than the ratio of the masses, 3.6--14.5 (with the exact value depending on which mass estimate was adopted for TDE\,2022upj), but much closer to the ratio of the square roots of the masses ($\approx1.9
$--3.8). 
This is consistent with the expectation from photoionisation equilibrium that $r\propto L^{0.5}$ (\citealt{Bentz09}; \citealt{Wu25}); if luminosity scales roughly linearly with BH mass then $r\propto M_\mathrm{BH}^{0.5}$.
We test this scaling by taking our calculated distances of \oiii~$\lambda5007$ and a strong coronal line (typically \fevii~$\lambda6087$) for each source and plotting them as a function of the BH mass in Fig.~\ref{fig:dists}.
We see that there is a trend of increasing gas distance with BH mass.
Linear regression gives the slopes between $\log{R}$ and $\log{M_\mathrm{BH}}$ as $0.63\pm0.08$ and $0.69\pm0.12$ for \oiii\ and \fevii, respectively (Section~\ref{sec:dist-trends}).
Both the \oiii\ and \fevii\ radius-mass relations are shallower than 1:1, with the \oiii\ slope strongly inconsistent with unity.
Whilst the fitted slopes are somewhat steeper than the simplistic $M_\mathrm{BH}^{0.5}$ expectation, the differences are only marginal ($2\sigma$ for \oiii\ and $1.8\sigma$ for \fevii).
A larger sample size and higher quality spectroscopic data would enable a more thorough investigation of the trends. 
Overall, it appears plausible that the locations of the emission line regions are determined by photoionisation.
The alternative scenario, in which the high-ionisation species are produced by collisional ionisation in shocks rather than by the central radiation field (e.g., \citealt{Contini01}), would not predict a simple dependence of emission-line radius on BH mass, since the ionisation state is set by the shock velocity and post-shock temperature rather than by the luminosity of the central source. 
The observed correlation therefore supports photoionisation as the dominant ionisation mechanism in our sample.

\subsubsection{Stratification}
We calculated distance-mass correlations for \oiii\ and \fevii\ separately, and found that across the whole sample \fevii\ coronal lines had lower virial radii than \oiii\ (as seen in the systematic offset between trend lines in Fig.~\ref{fig:dists}).
In the individual source maps presented in Figs.~\ref{fig:v-ECLE_maps}, \ref{fig:cl-tde_maps}, and \ref{fig:nonvar_ecle_maps} we frequently observed ionisation stratification in the circumnuclear gas, with higher-ionisation species exhibiting broader line widths and therefore smaller inferred radii.
Of the 33 objects in our sample AT\,2021dms lacks sufficient measurements to assess stratification, 
and a further seven remain inconclusive owing to large uncertainties, limited line coverage, or scatter in the inferred radii. 
Among the 25 objects for which a clear classification is possible, 14 (56 per cent) show evidence of stratified gas. 
The incidence of stratification is comparable in the two subsamples (7/14 variable versus 7/11 non-variable objects).
The 11 objects that show no evidence for stratification are similarly divided between variable (six) and non-variable (five) systems. 
We therefore find no indication that stratification preferentially occurs in either TDE-linked (variable) or AGN-linked (non-variable) ECLEs.
There is no evidence that stratification depends on BH mass in our sample: the mean $\log(M_\mathrm{BH}/\mathrm{M_\odot})=7.20$ for stratified ECLEs and 7.41 for the non-stratified sources.
Nor do we find a significant dependence of stratification on the source redshift.
The lack of a clear difference in the location and structure of gas between variable and non-variable ECLEs suggests that the circumnuclear environments of the two categories are similar
and that in TDEs the gas is only temporarily illuminated, whereas in AGNs the illumination is much longer-term.

\subsection{Limitations of the present study and future work}
\subsubsection{Methodological limitations}
Our estimates of the gas distances assume that the observed emission line widths trace virial motions in the gravitational potential of the central SMBH.
Under this assumption the measured FWHM can be translated into characteristic radii for the emitting gas.
However, particularly in the case of TDEs (in the aftermath of the accretion episode), the circumnuclear environment may be dynamically disturbed and the observed line widths could include contributions from non-virial motion such as outflows, inflows, or turbulence.
In AGNs, coronal line emission has commonly been associated with outflowing ionised gas \textcolor{black}{(e.g., \citealt{Rodriguez-Ardila06}; \citealt{MullerSanchez11}; \citealt{May18}; \citealt{Kynoch22})}.
The observed line widths may also depend on the geometry and orientation of the emitting gas relative to the line of sight.
Considering these limitations, the inferred radii should be interpreted as approximate characteristic scales rather than precise orbital distances. 
Even if non-gravitational motions and geometric effects influence the line widths, systematic trends between different ionic species and between objects can still provide useful information on the relative distribution of the emitting gas.
Furthermore, this approach has precedent in previous studies of CL-TDEs (\citealt{Short23}; \citealt{Newsome24}; \citealt{Clark24}) and our aim here was to apply the method more widely to a sample of ECLEs, allowing comparative mapping of the circumnuclear gas distributions. 

BH mass estimates are required to translate the measured emission line widths into virial gas distances. 
For approximately one third of the sample, published mass estimates are available in the literature, derived using a variety of methods including stellar velocity dispersion measurements, TDE light-curve modelling, host-galaxy scaling relations, and single-epoch virial estimators based on broad emission lines. 
In several cases multiple estimates are reported for the same source, from which we adopt average values. 
For objects lacking published measurements we estimate masses using the most readily available method permitted by the data, and as a result the BH masses used in this work are not derived using a uniform methodology.
However, where multiple independent estimates exist they typically agree to within factors of $\sim5$–$7$ (see, e.g., TDE\,2022upj, TDE\,2019qiz, and TDE\,2022fpx in Table~\ref{tab:masses}), comparable to the intrinsic uncertainties of many commonly used SMBH mass estimators. 
Given that our sample spans $\sim4$ dex in estimated BH mass, these uncertainties are unlikely to dominate the overall trends presented here.
Since distance $r\propto M_\mathrm{BH}/\mathrm{FWHM}^2$, the uncertainties in mass propagate linearly into the estimated virial radii, but these remain small compared with the range of distances explored and are unlikely to dominate the global trends presented here.

\subsubsection{Instrumental / observational limitations}\label{sec:disc-obs_limitations}
The spectroscopic data used in this work are heterogeneous, having been obtained with a variety of telescopes and instruments having different observing configurations, wavelength coverage, spectral resolutions, and extraction apertures. 
As a result, the observations sample different physical regions of the host galaxies and may include varying contributions from nuclear and host-galaxy light. 
In addition, many spectra (particularly follow-up observations of the CL-TDEs) were reduced by different groups using independent pipelines and may therefore have less uniform flux calibration or wavelength solutions than large survey data products. 
We have made best efforts to mitigate these effects by correcting emission line widths for instrumental broadening using the reported spectral resolutions, although these corrections are approximate.

We noticed that spectra of nv-ECLEs obtained from large surveys such as SDSS and DESI show very similar emission line profiles and relative strengths over timescales of many years, suggesting that systematic differences between instruments do not strongly affect the measurements used here.

A further limitation is the uneven temporal sampling of the ECLE population. 
While the CL-TDEs often have relatively dense spectroscopic follow-up during their outburst evolution, 
the v-ECLEs (identified retrospectively) have only a small number of spectra obtained at unknown phases relative to the flare, 
and the nv-ECLEs typically have only two or three spectra separated by decades. 
Consequently, the absence of particular emission lines or evolutionary trends in some objects will reflect limited temporal coverage rather than intrinsic differences between sources.

Finally, the currently known ECLE population remains relatively small. 
Our sample contains only a dozen v-ECLEs and nv-ECLEs with sufficient spectroscopic data for analysis, limiting the statistical power of comparisons between subclasses.
The limitations described above reflect the small number of currently known ECLEs and the largely serendipitous nature of their discovery.
Future systematic searches within massive, multiplexed, spectroscopic surveys offer a path toward overcoming these limitations. 
Our team is conducting a search for ECLE candidates within nightly DESI observations and recent DESI data releases, 
building on the \textsc{sleipnir} detection pipeline that was applied to the DESI EDR and yielded three of the v-ECLEs included here \citep{Clark26}. 
Expanding the ECLE sample will improve the statistical power of population studies, while identifying TDE-linked ECLEs shortly after outburst will enable coordinated spectroscopic follow-up with well-characterised instruments. 
Such observations will provide more quality data sets and allow the temporal evolution of coronal-line emission to be tracked in greater detail.

\section{Conclusions}\label{sec:conc}
We compiled optical spectra of $\approx30$ ECLEs reported in the literature and divided these into variable (TDE-linked) and non-variable (AGN-linked) systems.
Emission line widths were measured for both low-ionisation and coronal line transitions and these were used to estimate characteristic radial distances of the emission line gas, assuming virial motion.
The inferred radii indicate that coronal lines typically originate at sub-pc scales, intermediate between the low-ionisation NLR and the compact BLR.
Evidence for ionisation stratification was seen clearly in 14 objects, with higher-ionisation species generally located closer to the BH.
The incidence of stratification and the inferred gas distances are broadly similar for both the v-ECLEs and nv-ECLEs, which suggests that TDE flares illuminate a pre-existing circumnuclear gas distribution.
The absolute sizes of the emission regions scale non-linearly with BH mass, implying their scales are determined by photoionisation.

Coronal lines provide a powerful diagnostic of the structure and kinematics of nuclear gas in galactic nuclei, 
with v-ECLEs offering a unique opportunity to map this material during transient illumination.
However, the population of currently known ECLEs remains small, which limits the statistical power of studies such as this one.
We aim to continue our systematic search for additional ECLEs within DESI data using the \textsc{sleipnir} pipeline, 
thereby substantially expanding the sample size and enabling a more robust testing of distance-mass relations and the comparative properties of TDE- and AGN-linked systems.

\section*{Acknowledgements}\label{sec:ack}
DK, OG and PC acknowledge support from the UK Science and Technology Facilities Council (STFC) through grants ST/S000550/1, ST/W001225/1 (DK \& OG) and ST/Y001850/1 (PC).
SP is supported by the international Gemini Observatory, a program of NSF NOIRLab, which is managed by the Association of Universities for Research in Astronomy (AURA) under a cooperative agreement with the U.S.\ National Science Foundation, on behalf of the Gemini partnership of Argentina, Brazil, Canada, Chile, the Republic of Korea, and the United States of America.

The authors wish to thank Jason Hinkle for providing the Magellan spectrum of AT\,2021dms; 
Francesca Onori for providing the spectra of AT\,2017gge; 
Yibo Wang for providing the spectra of AT\,2018gn; 
Zhong-Xiang Wang for providing the spectra of AT\,2021acak;
and Chenwei Yang for providing the 2011 MMT spectra of SDSS\,J0748, SDSS\,J0952, SDSS\,J1241, SDSS\,J1342 and SDSS\,J1350.

The authors also thank the DESI reviewers, Matthew Temple and Ragadeepika Pucha, for their careful reading of an early draft of the paper and their recommendations for improvements.

This material is based upon work supported by the U.S.\ Department of Energy (DOE), Office of Science, Office of High-Energy Physics, under Contract No.\ DE–AC02–05CH11231, and by the National Energy Research Scientific Computing Center, a DOE Office of Science User Facility under the same contract. 
Additional support for DESI was provided by the U.S.\ National Science Foundation (NSF), Division of Astronomical Sciences under Contract No.\ AST-0950945 to the NSF’s National Optical-Infrared Astronomy Research Laboratory; the Science and Technology Facilities Council of the United Kingdom; the Gordon and Betty Moore Foundation; the Heising-Simons Foundation; the French Alternative Energies and Atomic Energy Commission (CEA); the National Council of Humanities, Science and Technology of Mexico (CONAHCYT); the Ministry of Science, Innovation and Universities of Spain (MICIU/AEI/10.13039/501100011033), and by the DESI Member Institutions: \url{https://www.desi.lbl.gov/collaborating-institutions}. 
Any opinions, findings, and conclusions or recommendations expressed in this material are those of the author(s) and do not necessarily reflect the views of the U.S.\ National Science Foundation, the U.S.\ Department of Energy, or any of the listed funding agencies.

The authors are honored to be permitted to conduct scientific research on I'oligam Du'ag (Kitt Peak), a mountain with particular significance to the Tohono O’odham Nation.

Funding for the Sloan Digital Sky Survey V has been provided by the Alfred P.\ Sloan Foundation, the Heising-Simons Foundation, the National Science Foundation, and the Participating Institutions. SDSS acknowledges support and resources from the Center for High-Performance Computing at the University of Utah. SDSS telescopes are located at Apache Point Observatory, funded by the Astrophysical Research Consortium and operated by New Mexico State University, and at Las Campanas Observatory, operated by the Carnegie Institution for Science. The SDSS web site is \url{www.sdss.org}.

SDSS is managed by the Astrophysical Research Consortium for the Participating Institutions of the SDSS Collaboration, including the Carnegie Institution for Science, Chilean National Time Allocation Committee (CNTAC) ratified researchers, Caltech, the Gotham Participation Group, Harvard University, Heidelberg University, The Flatiron Institute, The Johns Hopkins University, L'Ecole polytechnique f\'{e}d\'{e}rale de Lausanne (EPFL), Leibniz-Institut f\"{u}r Astrophysik Potsdam (AIP), Max-Planck-Institut f\"{u}r Astronomie (MPIA Heidelberg), Max-Planck-Institut f\"{u}r Extraterrestrische Physik (MPE), Nanjing University, National Astronomical Observatories of China (NAOC), New Mexico State University, The Ohio State University, Pennsylvania State University, Smithsonian Astrophysical Observatory, Space Telescope Science Institute (STScI), the Stellar Astrophysics Participation Group, Universidad Nacional Aut\'{o}noma de M\'{e}xico, University of Arizona, University of Colorado Boulder, University of Illinois at Urbana-Champaign, University of Toronto, University of Utah, University of Virginia, Yale University, and Yunnan University. 

For the purpose of open access, the authors have applied a Creative Commons Attribution (CC BY) licence to any Author Accepted Manuscript version arising.

\section*{Data Availability}
In compliance with the DESI data management policy, the data required to reproduce all of the figures in this paper are available in machine-readable format through Zenodo \citep[doi:10.5281/zenodo.20445180]{Kynoch26}.
This work includes data from the DESI EDR and DR1; these are detailed in the papers \cite{DESI-EDR} and \cite{DESI-DR1}, and data access instructions can be found \href{https://data.desi.lbl.gov/doc/releases/}{online}.
More recent DESI data (since 2022 June) will be made public in due course.

\href{https://sdss.org/}{SDSS} and \href{https://archive.eso.org/}{ESO VLT/X-shooter} spectra can be obtained from their public archives.
Additional archival spectra were retrieved from the Weizmann Interactive Supernova Data Repository (\href{https://www.wiserep.org/}{WISeREP}; \citealt{Yaron12}) and the Transient Name Server (\href{https://www.wis-tns.org/}{TNS}).
Private data from previous studies in the literature were obtained from the corresponding authors of those works, as detailed in the Acknowledgements.



\bibliographystyle{mnras}
\bibliography{refs} 



\appendix
\section{Notes on individual sources}\label{sec:appendix}
\subsection{Variable ECLEs}
\subsubsection{SDSS\,J0113}
SDSS\,J011306.68$+$093712.2 (SDSS J0113) was identified as a candidate ECLE in the systematic search for ECLEs in the SDSS Baryon Oscillation Spectroscopic Survey (BOSS: \citealt{Dawson13}; \citealt{Smee13}) low-redshift galaxy sample by \cite{Callow25}. 
Its 2012 BOSS spectrum exhibits strong coronal and low-ionisation optical emission lines.
A follow-up spectrum taken in 2023 with the Gemini-North Multi-Object Spectrograph (GMOS: \citealt{Hook04}) showed that both the coronal lines and low-ionisation lines had faded away, and the blue continuum ($\lambda\lesssim5000$~\si\angstrom) was also reduced.
Analysis of the archival MIR data obtained with \wise\ appear to show a light echo of the (unobserved) TDE;
the W1- and W2-band fluxes peak around mid-2010 and then drop sharply by $\approx1$~mag by 2013, followed by a gradual long-term decline (\citealt{Callow25}).
This MIR behaviour is consistent with that of other TDE-linked v-ECLEs \cite{Clark24}.

\subsubsection{SDSS\,J0748}\label{sec:sample:sdssj0748}
SDSS\,J074820.67$+$471214.3 (SDSS\,J0748), first reported by \cite{Wang11}, was the second ECLE to be identified (following SDSS\,J0952 by \citealt{Komossa08}: see Section~\ref{sec:sample:sdssj0952}). 
\cite{Wang11} noted the presence of `superstrong' very high ionisation coronal lines including \fex~$\lambda6374$, \fexi~$\lambda7891$, \fexiv~$\lambda5302$, and \arxiv~$\lambda4414$, but no \fevii\ lines in the 2003 SDSS optical spectrum.
The coronal lines appeared to have faded in follow-up spectra taken by \cite{Wang11} 4--5~yr later,
with \oiii\ emission increasing by a factor of 10 in the same period.
Higher-S/N optical spectra obtained by \cite{Yang13} with the Blue Channel Spectrograph (BCS: \citealt{Schmidt89}) mounted on the 6.5-m MMT in 2011 confirmed the absence of coronal line emission, so the authors classified it as a v-ECLE.
Investigating its MIR evolution in \textit{WISE} lightcurves, \cite{Clark24} found its MIR luminosity decreased substantially until it plateaued around 2018.
Its $\mathrm{W1}-\mathrm{W2}$ colour index began at the AGN dividing line and moved progressively into the non-AGN region.
\cite{Clark24} confirmed the categorisation of SDSS\,J0748 as a v-ECLE.

\subsubsection{SDSS\,J0952}\label{sec:sample:sdssj0952}
SDSS\,J095209.56$+$214313.3 (SDSS\,J0952) was the first reported ECLE.
\cite{Komossa08} noted the presence and unusually high strength relative to \oiii~$\lambda5007$ of several iron CLs (including \fevii~$\lambda6087$, \fex~$\lambda6375$, and \fexiv~$\lambda5303$) in the SDSS spectrum taken on 2005 December 30.
Follow-up optical spectroscopy in 2007 December 4--5 revealed that the CLs had faded substantially (relative to \oiii~$\lambda5007$) over the intervening 2~yr. 
Further diminution in the relative strengths of the CLs was seen in spectra taken throughout 2008 \citep{Komossa09} and again in its 2011 MMT/BCS spectrum (\citealt{Yang13}).
\cite{Clark24} found that the coronal lines were absent in 2021 spectra taken with the New Techology Telescope (NTT: \citealt{Tarenghi89}) and DESI, and that the source showed a strong decline in MIR brightness, confirming its classification as a v-ECLE.

\subsubsection{SDSS\,J1241}\label{sec:sample:sdssj1241}
Following the discoveries of the ECLEs SDSS\,J0952 (Section~\ref{sec:sample:sdssj0952}) and SDSS\,J0748 (Section~\ref{sec:sample:sdssj0748}), 
\cite{Wang12} performed a systematic search for other ECLEs in SDSS DR7.
SDSS\,J124134.26$+$442639.2 (SDSS\,J1241) was one of the five additional ECLEs found, along with SDSS\,J1342 (Section~\ref{sec:sample:sdssj1342}), SDSS\,J1350 (Section~\ref{sec:sample:sdssj1350}), SDSS\,J0938 (\ref{sec:sample:sdssj0938}), and SDSS\,J1055 (\ref{sec:sample:sdssj1055}).
\textcolor{black}{We note that \cite{Rose15} included SDSS\,J1241 in their sample of CLiF AGN.}

\cite{Yang13} classified SDSS\,J1241 as a nv-ECLE, noting the persistence of the \fevii~$\lambda3758$ coronal line in the 2011 MMT follow-up spectrum, taken nearly 8~yr after the 2004 SDSS discovery spectrum.
Later, \cite{Clark24} reclassified it as a v-ECLE because optical coronal lines were not detected in the later follow-up 2021 spectrum (taken with the Kast spectrograph on the Shane Telescope at the Lick Observatory: \citealt{Miller94}), 
and its MIR evolution was similar to other v-ECLEs.
The W1 and W2 fluxes of SDSS\,J1241 were observed to diminish steadily over the \textit{WISE} observing period,
however the decline was much shallower than the four other v-ECLEs studied ($\approx+0.25$ and $+0.5$~mag, respectively).
Its $\mathrm{W1}-\mathrm{W2}$ colour index remained in the non-AGN region throughout the \textit{WISE} observations.
A DESI spectrum of this source was recorded in 2025 March, which we newly analyse as part of this study.

\subsubsection{Raichu (DESI\,J204.3990$+$03.8775)}\label{sec:sample:raichu}
Raichu is a TDE-linked v-ECLE that occurred in the galaxy DESI\,J204.3990$+$03.8775. 
It is one of three variable sources (including Raticate: Section~\ref{sec:sample:raticate} and Pidgeot: Section~\ref{sec:sample:pidgeot}) discovered in a systematic search through the DESI EDR for ECLEs using the purpose-built Spectroscopic Light Echo Identification Protocol Now In Real-time (\textsc{sleipnir}) pipeline (\citealt{Clark26}).  
Extreme coronal line emission was observed in its DESI spectrum (DESI target ID 39627884763023878), and \cite{Clark26} noted the long-term monotonic decline of its MIR brightness and non-AGN $\mathrm{W1}-\mathrm{W2}$ colour index, classifying the source as a v-ECLE.

\subsubsection{SDSS\,J1342}\label{sec:sample:sdssj1342}
Another of the five \cite{Wang12} ECLEs identified in the first systematic search for them, 
the 2002 SDSS spectrum of SDSS\,J1342 dispayed the highest-ionisation optical Fe coronal lines \fex, \fexi, and \fexiv\ but no \fevii\ emission.
\cite{Yang13} obtained an MMT spectrum in 2011 that showed the later appearance of \fevii\ emission and disappearance of the higher ionisation lines.
The DESI spectrum of SDSS\,J1342, taken in 2021, revealed that \fevii\ emission remained, but the higher ionisation lines did not reappear \citep{Clark24}.
Therefore, unlike most other v-ECLEs, SDSS\,J1342 was found to have optical coronal lines that persisted for nearly two decades following its discovery \citep{Yang13,Clark24}.
\textit{WISE} observations of the source since 2010 showed that its MIR evolution brightness decreased systematically, with its $\mathrm{W1}-\mathrm{W2}$ colour index transitioning from AGN-like to non-AGN-like. 
\cite{Clark24} therefore classified SDSS\,J1342 as a v-ECLE.

\subsubsection{SDSS\,J1350}\label{sec:sample:sdssj1350}
Similar to SDSS\,J1342, SDSS\,J1350 was another ECLE identified in the \cite{Wang12} search, initially showing \fex--\fexiv\ emission lines in its 2006 SDSS spectrum which had faded in the 2011 follow-up spectrum obtained on the MMT but were replaced by the appearance of \fevii\ lines \citep{Yang13}.
\cite{Clark24} assessed the long-term MIR evolution of the source since 2010;
as with SDSS\,J1342, SDSS\,J1350 decreased in brightness and transitioned to a non-AGN-like MIR colour.
SDSS\,J1350 was therefore confirmed as a v-ECLE.

\subsubsection{Raticate (DESI\,J216.5891$+$00.1436)}\label{sec:sample:raticate}
Raticate is a TDE-linked ECLE that occurred in the galaxy DESI\,J216.5891$+$00.1436.
Extreme coronal line emission was observed in its DESI spectrum (DESI target ID 39627794400938039) and reported by \cite{Clark26}.
The \textit{WISE} MIR lightcurves of Raticate reveal two strong outbursts with W1 and W2 changing by $>1$ and $>1.5$~mag, respectively.
Its $\mathrm{W1}-\mathrm{W2}$ colour index fluctuates between AGN-like values when bright and non-AGN-like when faint.
\cite{Clark26} classified Raticate among their v-ECLEs.

\subsubsection{Pidgeot (DESI\,J243.3798$+$55.7528)}\label{sec:sample:pidgeot}
Pidgeot is a TDE-linked ECLE that occurred in the galaxy DESI\,J243.3798$+$55.7528.
Extreme coronal line emission was observed in its DESI spectrum (DESI target ID 39633332819985805) and reported by \cite{Clark26}.
Although showing a long-term decline in MIR brightness, an outburst beginning in 2015 was identified in its \textit{WISE} lightcurves, in which W1 and W2 increased by $>0.5$~mag.
The $\mathrm{W1}-\mathrm{W2}$ colour index remained in the non-AGN region.
\cite{Clark26} classified the source as a v-ECLE.

\subsection{CL-TDEs}
\subsubsection{TDE\,2022upj}
The transient TDE\,2022upj was discovered by the Zwicky Transient Facility (ZTF: \citealt{Bellm19}), in which it is known as ZTF\,22abegjtx, on 2022 August 31 (\citealt{Fremling22}), with its optical peak occurring on 2022 November 1 (\citealt{Newsome24}).
An optical spectrum obtained with the Floyds spectrograph on the Faulkes Telescope of the Las Cumbres Observatory (LCO; \citealt{Brown13}) during the optical peak (on 2022 November 2) revealed extremely strong high-ionisation coronal lines \fex~$\lambda6374$ and \fexiv~$\lambda5302$, making TDE\,2022upj the first nuclear transient to show coronal lines at its peak (\citealt{Newsome24}).
\cite{Newsome24} presented a detailed multi-wavelength study of the source, including a map of the nuclear material illuminated by the outburst.
They found evidence of a stratified emission line region on $\lesssim0.1$~pc scales, surrounded by circumnuclear dust at $\gtrsim0.4$~pc. 

\subsubsection{AT\,2018gn (ASASSN-18ap)}
The transient AT\,2018gn was first reported by the All-Sky Automated Survey for Supernovae (ASAS-SN: \citealt{Shappee14}; \citealt{Kochanek17}) with the name ASASSN-18ap and initially classified as a supernova (SN) candidate (SN\,2018gn) by \cite{Falco18}.  
The classification was reexamined by \cite{Wang24} who noted a delayed flare in infrared luminosity and the detection of strong optical coronal emission lines.
The peak of the optical emission occurred around 2018 February 12.  
Two optical spectra of AT\,2018gn were obtained around this time (on January 15 and February 11) with the FAST spectrograph \citep{Mink21} on the Tillinghast telescope of the Fred L.\ Whipple Observatory (FLWO), the first having been reported by \cite{Falco18}.
\cite{Wang24} also obtained three spectra over 2022-23 with the Double Spectrograph (DBSP: \citealt{Oke82}) on the Hale 200 inch telescope at the Palomar Observatory (see Table~\ref{tab:spectra_1}).
Additionally, three DESI spectroscopic observations of the galaxy were made recently: one on 2021 October 16 and two on 2022 October 21.
Unfortunately, the fibre does not appear to have been centred on the galactic nucleus during any of these observations.

\subsubsection{AT\,2021dms}\label{sec:sample:at2021dms}
AT\,2021dms (ZTF\,20acsacog) was discovered by means of a ZTF \textit{g}-band optical flare in the galaxy MCG-02-09-033 on 2021 February 21 and was reported as a SN candidate by \cite{Forster21}. 
\cite{Hinkle24a} obtained an optical spectrum of the target 173~d later on 2021 August 13 with the Low Dispersion Survey Spectrograph (LDSS) on the Magellan Clay Telescope (the VPH-All grism was used, for which the resolution $R\approx860$).
The spectrum exhibited strong high-ionisation coronal lines including \fex~$\lambda6374$, \fexi~$\lambda7891$, and \fexiv~$\lambda5302$, and \cite{Hinkle24b} noted its similarity to other coronal line emitting galaxies.
The source has not yet been observed by DESI and we are not aware of any other post-flare optical spectra having been reported in the literature.

\subsubsection{TDE\,2019qiz}\label{sec:sample:at2019qiz}
Initially discovered on 2019 September 19 by the ZTF and classified as a SN candidate, TDE\,2019qiz (ZTF\,19abzrhgq) was identified as a rising TDE on 2019 September 25 by \cite{Siebert19} following inspection of the optical spectrum obtained using the Low-Resolution Imaging Spectrometer (LRIS: \citealt{Oke95}) at the 10-m Keck telescope on that date.
Its optical flux peaked the following month, on 2019 October 10 (MJD 58766: \citealt{Nicholl20}).
\citealt{Nicholl20} instigated regular spectroscopic observations of the target with various facilities on 2019 September 27.
This monitoring included several UV/optical/NIR spectra obtained with VLT/X-shooter that are included in this study (see Table~\ref{tab:spectra_1}).
Later, \cite{Short23} reported the delayed appearance of extremely strong coronal lines (fluxes $\approx2.33$--$14.14\times10^{-16}$~erg\,s$^{-1}$\,cm$^{-2}$, corresponding to $\approx7$--44~per cent the flux of \oiii~5007), making the source an ECLE. 

\subsubsection{AT\,2021acak}
AT\,2021acak was discovered by the Asteroid Terrestrial-impact Last Alert System (ATLAS: \citealt{Tonry18}) on 2021 October 22 and reported by \cite{Tonry21} (with the name ATLAS21bkkj) the following day.  
The transient occurred in the $z=0.136$ galaxy SDSS\,J103447.90$+$152922.4 (this galaxy has not been spectroscopically observed by the SDSS).
Noting the similarity of its ZTF lightcurves to other TDEs, \cite{Li23} obtained several follow-up optical spectra with the Yunnan
Faint Object Spectrograph and Camera on the Lijiang telescope (\citealt{Wang19}) in 2022 April and June.
The high-ionisation lines \fex~$\lambda6374$ and \fexiv~$\lambda5302$ were present, but the optical \oiii\ doublet was not (it was not measured by \citealt{Li23}) therefore we consider the source to be an ECLE.
A DESI spectrum of the galaxy was obtained approximately 1~yr later, on 2023 April 15.
We note that \cite{Li23} proposed that the host galaxy of AT\,2021acak contains an AGN, 
on the basis of a fit to its pre-outburst SED and AGN-like MIR colours and brightness.
 
\subsubsection{TDE\,2022fpx}\label{sec:sample:at2022fpx} 
The transient TDE\,2022fpx (ATLAS22kjn) was first reported by \cite{Tonry22} on 2022 March 31 and subsequently classified as a TDE by \cite{Perez-Fournon22} in 2022 June on the basis of the optical spectrum taken with the SPectrograph for the Rapid Acquisition of Transients (SPRAT: \citealt{Piascik14}) on the Liverpool Telescope (LT: \citealt{Steele04}).\footnote{\url{https://www.wis-tns.org/object/2022fpx}}
It was associated with the galaxy SDSS\,J153103.70$+$532419.3 (although unfortunately there is no SDSS optical spectrum of this source).
The luminosity peak of the optical ($g$- and \textit{r}-band) emission occurred around 2022 July 24.
\cite{Koljonen24} presented an optical spectrum obtained with the Alhambra Faint Object Spectrograph and Camera (ALFOSC) on the Nordic Optical Telescope (NOT: \citealt{Djupvik10}) 11~d after the peak, along with new and archival multiwavelength observations.
Several coronal lines were observed in the NOT spectrum and the \fexiv\ line was measured to have a flux 1.7 times greater than that of \oiii~$\lambda5007$; on this basis the authors classified TDE\,2022fpx as a v-ECLE.  

\subsubsection{AT\,2018dyk}\label{sec:sample:at2018dyk}
AT\,2018dyk (ZTF\,18aajupnt) was identified as an astronomical transient in the ZTF on 2018 May 31 \citep{Fremling18}.
In 2018 August \cite{Arcavi18} suggested that it may be a TDE because of the broad H and He emission lines seen in the LCO/Floyds optical spectrum.
Follow-up optical spectra obtained by \cite{Frederick19} showed the appearance of extremely strong coronal lines, although the authors interpreted AT\,2018dyk as a `changing-look' LINER: a galactic nucleus transitioning from a low-ionisation nuclear emission line region to a broad-line AGN.
\cite{Huang23} revisited the source, finding (i) a fast rise and power-law decline in its X-ray, UV, and optical lightcurves, and (ii) a lag of the X-ray peak brightness behind the UV/optical, 
consistent with a TDE origin of the transient.
\cite{Hinkle24b} included AT\,2018dyk in a sample of anomalous nuclear transients (ANTs: extreme transients in galactic nuclei of which the physical origin is presently ambiguous; e.g., \citealt{Wiseman25}).  
An in-depth investigation of the source by \cite{Clark25}, including analysis of follow-up optical spectroscopy and multiwavelength data, confirmed its TDE nature.

\subsubsection{AT\,2017gge}\label{sec:sample:at2017gge}
The $z=0.0655$ galaxy SDSS\,J162034.99$+$240726.5 was observed spectroscopically by the SDSS in 2004 and classified as star-forming.
The transient AT\,2017gge was detected in this galaxy by ATLAS on 2017 August 3 (ATLAS17jrp; \citealt{Tonry17}).
\cite{Fraser17} suggested a TDE origin of the event, based on the appearance of broad Balmer lines and blue continuum in an optical spectrum taken on 2017 September 13 with the EFOSC2 instrument on the NTT.
A MIR flare occurring $\sim200$~d after the transient's discovery was observed in the \textit{WISE} lightcurves (\citealt{Jiang21}; \citealt{Onori22}). 
\cite{Onori22} reported the late appearance ($>200$~d following discovery) of strong, transient, optical coronal lines in follow-up spectra.

\subsubsection{TDE\,2021qth}
TDE\,2021qth (ZTF\,21abhrchb) was not identified as a TDE in real time by optical surveys, but was later classified as such following its X-ray detection by SRG/eROSITA \citep{Yao23}. 
A TDE rather than SN origin of the transient was favoured by \cite{Yao23} on the basis of its high X-ray luminosity. 
Optical spectra taken 27 and 300~d after its optical peak revealed the emergence of strong coronal lines in the latter, and \cite{Yao23} noted the similarity to SDSS\,J0952, classifying TDE\,2021qth as an ECLE.

\subsubsection{AT\,2018bcb (ASASSN-18jd)}
Unfortunately, whilst useful optical spectra of AT\,2018bcb exist in the literature, the publicly available data are of insufficient spectral resolution for our purposes, and higher-resolution data were not available to us; we therefore do not include it in the following analyses but mention it here for completeness.
AT\,2018bcb was identified as a transient by ASAS-SN (its internal name is ASASSN-18jd) on 2018 April 9 in the nucleus of the $z=0.1192$ galaxy 2MASX\,J22434289$-$1659083 (\citealt{Bersier18}).
\cite{Neustadt20} reported the discovery and early-time evolution of the source in UV/optical and X-ray bands.  
In the year following its discovery, UV/optical photometry revealed a very slow decline in brightness, well-described by a $T\sim2.5\times10^4$~K blackbody.
X-ray observations revealed luminous, variable emission that faded (although not smoothly) over the subsequent year. 
Optical spectra showed persistent broad Balmer emission together with transient \heii, \oiii, and coronal Fe features.
\fex~$\lambda6374$ and possibly \fexiv~$\lambda5302$ were observed in early optical spectra, but \fevii\ emission was not seen.
At its brightest, \fex~$\lambda6374$ was comparable in strength to \oiii~$\lambda5007$, significantly stronger than typically observed in Seyfert nuclei.
The strength and transience of the high-ionisation Fe lines make AT\,2018bcb similar to v-ECLEs; however \cite{Neustadt20} noted similarities with AGN-like nuclear variability, leaving its precise classification somewhat ambiguous.

\subsection{Non-variable ECLEs}
\subsubsection{Charizard (DESI\,J027.9525$+$04.1951)}
Charizard is an AGN-linked ECLE candidate in the nucleus of the galaxy DESI\,J027.9525$+$04.1951. 
Extreme coronal line emission was observed in its DESI spectrum (DESI target ID 39627887837451274) and reported by \cite{Clark26}.
The follow-up Gemini spectrum obtained a few months later showed no change in the coronal line emission.
In the MIR, Charizard shows only minor variations and a consistently AGN-like colour index.

\subsubsection{Arbok (DESI\,J115.5730$+$39.4366)}
Arbok is associated with the nucleus of the $z=0.416$ galaxy DESI\,J115.5730$+$39.4366. 
Extreme coronal line emission was observed in its 2021 DESI spectrum (DESI target ID 39633066745924399) and reported by \cite{Clark26}.
The galaxy has an archival SDSS/BOSS spectrum taken in 2010 containing very strong and broad Balmer lines.
These appear to have confused the SDSS spectroscopic pipeline, which identified \ha\ as Ly\,$\upalpha$, resulting in an erroneous redshift ($z=6.625$) being recorded for the galaxy.
This error meant that the source was missed by a prior search for ECLEs among SDSS BOSS galaxies \citep{Callow25}.
The SDSS and DESI spectra are very similar, and the source shows little MIR evolution during \textit{WISE} observations, with its colour index remaining in the AGN region.
Because of its lack of spectroscopic or photometric evolution over a decade, \cite{Clark26} classified the source as a non-variable, AGN-linked ECLE candidate.

\subsubsection{SDSS\,J0807}
SDSS\,J0807 was identified as an ECLE by \cite{Callow24} who noted strong \fevii\ lines in its SDSS spectrum, as well as weak \fex\ and strong \oiii.
Its MIR evolution was more characteristic of an AGN than a TDE, showing a gradual brightening and maintaining an AGN-like $\mathrm{W1}-\mathrm{W2}$ colour index.  
\cite{Callow24} classified the source as a nv-ECLE.

\subsubsection{SDSS\,J0938}\label{sec:sample:sdssj0938}
SDSS\,J0938 was one of five ECLEs identified in the systematic search of \cite{Wang12}.
Its 2006 SDSS spectrum shows clear \fevii, \fex, \fexi, and \fexiv\ coronal lines that were still present in the 2011 MMT spectrum obtainted by \cite{Yang13}, 
who reclassified the source as a Seyfert 2 AGN rather than a TDE.
The long-term MIR behaviour of the source presented by \cite{Clark24} support this classification of SDSS\,J0938: its \textit{WISE} fluxes and AGN-like colour do not vary substantially.

\subsubsection{SDSS\,J1055}\label{sec:sample:sdssj1055}
Similar to SDSS\,J0938, SDSS\,J1055 is a \cite{Wang12} ECLE that was classified as an AGN by \cite{Yang13} because of the similarity between its 2002 and 2011 optical spectra.
Further, \cite{Clark24} reported little change in its long-term MIR luminosity, and AGN-like $\mathrm{W1}-\mathrm{W2}$ colour index, confirming SDSS\,J1055 as a nv-ECLE.
Unlike SDSS\,J0938, SDSS\,J1055 is a type 1 AGN having both broad and narrow permitted emission lines.

\subsubsection{Sandslash (DESI\,J172.6675$+$50.6179)}
Sandslash is an AGN-linked ECLE candidate in the galaxy DESI\,J172.6675$+$50.6179. 
Extreme coronal line emission was observed in its DESI spectrum (DESI target ID 39633255741260888) and reported by \cite{Clark26}.
The target also has an archival SDSS spectrum that is very similar in appearance to its DESI spectrum.
Baldwin-Phillips-Terlevich (BPT: \citealt{BPT81}) optical emission line diagnostics place the source in the Seyfert AGN region, although its MIR properties are similar to quiescent galaxies (its W1 and W2 magnitudes show barely any change over a decade, and its $\mathrm{W1}-\mathrm{W2}$ colour index remains in the non-AGN region).

\subsubsection{SDSS\,J1207}
SDSS\,J1207 was identified by \cite{Callow24} as an ECLE in the galaxy SDSS\,J120719.91$+$241155.8 (also known as Mrk\,648 and identified as a Seyfert 1 AGN by \citealt{VeronCetty10}) on the basis of strong \fevii\ emission lines.
Weak \fex\ and possibly \fexi\ lines were also noted following visual inspection of its SDSS spectrum.
Its MIR fluxes and colour were found to be constant and so the source was confidently identified as a nv-ECLE.
We present a new DESI spectrum of the source as part of this work.

\subsubsection{SDSS\,J1238}
SDSS\,J1238 was identified as an ECLE by \cite{Callow24} on the basis of strong \fevii\ emission lines, although no other strong coronal lines were apparent in its 2008 SDSS spectrum.
Comparing the 2023 DESI spectrum, \cite{Callow24} noted that the broad Balmer lines were much reduced in strength and width and overall the spectrum was redder, although the \fevii\ coronal lines were still present.  
The source's MIR fluxes showed a decline over the \textit{WISE} observing period; however, rather than the smooth decline of TDE-linked v-ECLEs the flux evolution was non-monotonic.
Its $\mathrm{W1}-\mathrm{W2}$ colour index remained in the AGN locus throughout.
\cite{Callow24} concluded that SDSS\,J1238 was not a TDE-like v-ECLE.

\subsubsection{SDSS\,J1247}
Strong \fevii~$\lambda6087$ emission in the 2006 SDSS spectrum of SDSS\,J1247 resulted in its ECLE selection by \cite{Callow24}, who also noted weak \fex~$\lambda6374$.
Its 2023 DESI spectrum was very similar, with the coronal lines still present.
Its MIR flux and colour evolution were consistent with an AGN.
Given the constancy of its optical and MIR characteristics, SDSS\,J1247 was classified as a nv-ECLE. 

\subsubsection{SDSS\,J1402}
SDSS\,J1402 was identified as an ECLE by \cite{Callow24} on the basis of strong \fevii\ emission lines in its 2007 SDSS spectrum.
The source also showed strong \oiii\ emission, broad Balmer lines and a blue continuum.  
All of these features remained present within its 2023 DESI spectrum.
Its MIR fluxes drop then recover during the \textit{WISE} observing period and its $\mathrm{W1}-\mathrm{W2}$ colour index approaches the AGN/non-AGN dividing line in its faint phase.
\cite{Callow24} judged that this behaviour was inconsistent with other TDE-like v-ECLEs, and classified SDSS\,J1402 as a nv-ECLE.

\subsubsection{Nidoqueen (DESI\,J212.9556$+$52.7676)}  
Nidoqueen is an AGN-linked ECLE with coronal line emission observed in its DESI spectrum (DESI target ID 39633290029695741) and reported by \cite{Clark26}.
The source has an archival SDSS spectrum (SDSS\,J141149.34$+$524603.4) from 2002, classified as that of a galaxy.
Considering its lack of optical spectroscopic and MIR photometric evolution, AGN-like MIR colour and Seyfert/LINER BPT diagnostics, the source was classified as a nv-ECLE.

\subsubsection{SDSS\,J1458}
The 2008 SDSS spectrum of SDSS\,J1458 includes strong \fevii~$\lambda6087$ emission in addition to prominent \oiii\ emission, broad Balmer lines and a blue continuum, all of which are still present in its 2021 DESI spectrum \citep{Callow24}.
Its MIR fluxes also show very little change over the 10~yr of \textit{WISE} observations and its MIR colour is well within the AGN region.
The source was therefore classified by \cite{Callow24} as a nv-ECLE.

\subsubsection{SDSS\,J1459}
Selected by \cite{Callow24} for having moderately strong \fevii\ lines, the 2004 SDSS spectrum of SDSS\,J1459 also contains broad Balmer lines.
The emission lines are of a similar strength in the 2022 DESI spectrum, although the UV/optical continuum is brigher and bluer.
Its MIR fluxes generally increased over the \textit{WISE} observing period, with its MIR colour index remaining indicative of an AGN.
SDSS\,J1459 was classified as a nv-ECLE. 

\subsubsection{SDSS\,J1715}
SDSS\,J1715 was selected by \cite{Callow24} for having strong \fevii~$\lambda6087$ emission in its SDSS spectrum from 2000.
Strong Balmer lines, \oiii\ emission and blue continuum all suggested that the source was an AGN.
Likewise, little evolution was seen in its \textit{WISE} MIR fluxes or colour, reinforcing its identification as an AGN.
The source has recently been observed by DESI (2025 April 22), and we present its spectrum in this paper.

\subsubsection{SDSS\,J2220}
\cite{Callow24} found SDSS\,J2220 to have moderately strong \fevii\ lines in its 2001 SDSS spectrum.
Similar to SDSS\,J1715, its optical spectrum strongly resembles that of typical type 1 AGNs, with broad permitted emission lines, strong \oiii\ emission and a blue UV/optical continuum.
However, its MIR fluxes showed a marked increase in brightness in the gap between \textit{WISE} and \textit{NEOWISE} operations ($\approx2011$--2013), 
along with an apparent transition of its $\mathrm{W1}-\mathrm{W2}$ colour index from AGN-like to non-AGN-like.
Throughout \textit{NEOWISE}, the MIR fluxes of SDSS\,J2220 slowly decreased. 
This MIR evolution may partly be attributed to contamination by a star within the \textit{WISE} point spread function: when corrected for, SDSS\,J2220's MIR flux increase is greatly reduced and its colour index stays mostly in the AGN-like region. 
\cite{Callow24} did not firmly define SDSS\,J2220 as either a variable or nv-ECLE but concluded its MIR evolution was more similar to nv-ECLEs.
Given the persistence of its coronal lines, broad emission lines and blue UV/optical continuum over $\approx20$~yr between SDSS and DESI spectra, 
we include SDSS\,J2220 with the nv-ECLEs, but note its unusual MIR behaviour.


\bsp	
\label{lastpage}
\end{document}